THE HONG KONG POLYTECHNIC UNIVERSITY

INSTITUTE OF TEXTILES AND CLOTHING

# Function-led Design of Multifunctional Stimuli-responsive Superhydrophobic Surface Based on Hierarchical Graphene-Titania Nanocoating

**LIU Yang**

**A Thesis**
**Submitted in Partial Fulfillment of the**
**Requirements for the Degree of**
**Doctor of Philosophy**

# Abstract

With the advancements in both nanotechnology and micro-fabrication techniques, the development of superhydrophobic surface (SHS) enters a new era which highlights the significance of multifunctionality, tunable surface property, and smart response. Novel multifunctional SHS may accomplish multiple tasks simultaneously in real-field applications while tunable surface property can effectively induce smart responses on the SHS, which further benefit the functional optimization, life cycle, and prompt the total system efficiency of SHS. Based on the clear understanding of the application requirement and the material property, function-led design of SHS can be realized so that high performance SHS are achieved with reduced cost and optimized functions. In order to fabricate SHS with highly interactive interfacial dynamics, graphene, one of the thinnest among the existing materials, was chosen and applied to the flexible textile fabrics. Extremely strong bonding was observed between graphene and the fibers, and the thin sheets of graphene were also observed to tightly wrap according to the surface profiles of the fibers. Surprisingly, the hydrophilic cotton fabric became hydrophobic with a graphene loading of only 0.5 wt%. By further upgrading the monolayer graphene coating on the fabric, a nanofilm of titania ($TiO_2$) was subsequently applied onto the graphene surface which created a hierarchical surface structure showing superhydrophobicity. As a result of the photoactivity of $TiO_2$, UV-triggered superhydrophilic conversion can be readily achieved on this novel SHS. Enhanced control over the conversion rate and the wettability were also achieved due to the interfacial charge transfer interaction between $TiO_2$ and graphene. Tunable adhesion, spreading, and transport of water droplets can be realized on the as-obtained hierarchical graphene-$TiO_2$ (GT) SHS, which can further transform the GT-SHS into a versatile platform for micro-droplet/fluid manipulation. Multiple functions of application significance, i.e. directional water transport, micro-droplet storage and transfer, self-cleaning and gas sensing, are readily demonstrated on GT-SHS; exceptional oil/water separation performance is also shown by the membrane based on GT-SHS. Therefore, potent potential in the sweat and moisture management of quick-dry garment, micro-volume droplet storage, and oily water treatment can be anticipated for the novel stimuli-responsive GT-SHS. To speak further, being perceived as an essential bridging step between the fundamental research and practical application, the function-led design criterion can be envisaged as the new guiding code for the research and development of SHS.

# TABLE OF CONTENTS

# List of Symbols and Abbreviations

| | |
|---|---|
| SHS | Superhydrophobic Surface |
| 3D | Three Dimensional |
| GT | Hierarchical Graphene/Titania Surface |
| GO | Graphene Oxide |
| SEM | Scanning Electron Microscopy |
| TEM | Transmission Electron Microscopy |
| HRTEM | High Resolution Transmission Electron Microscopy |
| SAED | Selected Area Electron Diffraction |
| AFM | Atomic Force Microscopy |
| XRD | X-ray Diffractometry |
| EDX | Energy Dispersive X-ray Spectroscopy |
| XPS | X-ray Photoelectron Spectroscopy |
| CCD | Charged Coupled Device |
| UV | Ultra-Violet |
| $\theta$, $\theta_Y$ | Young's Contact Angle |
| $\theta^*$ | Apparent Contact Angle |
| $\theta_c$ | Critical Contact Angle between Wenzel and Cassie States |
| $\varphi_s$ | Solid Fraction at the three phase contact line |
| $\gamma$ | Surface Tension |
| WCA | Water Contact Angle |
| SA | Sliding Angle |
| HF | Hydrofluoric Acid |
| RF-PECVD | Radio Frequency Plasma Enhanced Chemical Vapor Deposition |
| PMMA | Poly(methyl methacrylate) |
| RIE | Reactive Ion Etching |
| SMP | Shape Memory Polymer |
| PDMS | Polydimethylsiloxane |
| AR | Aspect Ratio |
| PTFE | Poly(tetrafluoroethylene) |
| FAS | Fluoroalkylsilane |
| SLIPS | Slippery Liquid Infused Porous Surfaces |
| PS | Polystyrene |
| PP | Polypropylene |
| NP | Nanoparticle |
| AATCC | American Association of Textile Chemists and Colorists |

# Chapter 1. Introduction

## 1.1 Research background

Superhydrophobic surface (SHS) is referred to the surface of solids which has a water contact angle larger than 150° and sliding angle smaller than 5°.[1] The SHS exists both in nature and can be engineered artificially. Intriguingly, a large number of species, including insects, plants and birds, possess the SHS characteristics, as the surface of their shells, leaves and feathers clearly exhibits high functional performance of water-repellent and stimuli-responsive microfluidic dynamics.[2] Gaining insights from the great nature, the scientific community has devoted to the development of SHS for several decades, inspiring by the extreme efficiency, durability and sustainability of the master-pieces of natural SHS. Since the first reveal of "lotus effect" by Barthlott et al. in 1992, extensive research effort has been contributed to the understanding of the internal mechanism of this phenomenon and transfer the essential knowledge into real artificial SHS.[3] Inherited from natural SHS, the artificial SHS also possess the characteristics of high water contact angle and low sliding angle, demonstrating both low adhesion to water droplets and high repellency to flowing water streams. Based on the "lotus effect", the artificial SHS can be realized by sufficing two dominating factors: the formation of micro- and nano-scale dual roughness surface texture and low surface energy coating.[4] To speak further, the dual roughness factor minimizes the contact between solid and liquid as the cohesion force between the water molecules overcomes the adhesion force between water and solid surface; while the low surface energy coating maximizes the contact angle between the solid and water, which synergistically induces the super water-repellent property.

Started from the SHS that impedes the contact of static water droplets, the development of artificial SHS experiences tremendous growth in recent decades. Multivariant processing techniques, such as lithography, etching and 3D printing, have been utilized to fabricate artificial SHS; and the interdisciplinary manipulation of nanotechnology in constructing the hierarchical surface micro-textures significantly enhances the functional performance of the artificial SHS to unprecedented level.[5-10] Being moved from experimental to practical, microscopic to macroscopic, and static to dynamic, the state-of-the-art artificial SHS is now possessing the new characteristics of: (i) extremely anti-wetting property even to liquids with substantially low surface tension (i.e. fluorine liquids); (ii) stimuli-responsive and tunable wetting pattern on the surface regions; (iii) multifunctionality and high durability suited for the real-field application.

In order to obtain the targeted functions to fulfill the application requirement, function-led design is necessary for realizing the form and structure of SHS in a cost-effective manner.[11-13] Based on the end-user requirements, the targeted functions of SHS are properly identified and selected, which are subsequently realized by the integration of different materials and structures. High-end applications may require advanced engineering over the forms and structures of the materials, where the nanotechnology and advanced fabrication techniques can find their places. Given that the exact knowledge regarding the chemical, physical, and mechanical properties of the materials is acquired, one can easily fit the materials or their combinations to the functional requirements of the end product, and also predicts their potential performance. On the other hand, by designing the microscopic and macroscopic structures of the device, different material components can be extensively integrated, where high intensity of interfacial area is formed. And enhanced functional performance can be anticipated from these synergistically interfacial interactions at the versatile material interfaces.

Along with the innovations in the fields of information technology and the rise of interdisciplinary research, new opportunities and challenges are emerging for the science community which require the discovery of new knowledge and solutions. A broad perspective from engineering to commercialization is necessary to fulfill the advanced missions in science and technology in the coming era of open science and boost the technology transfer. Therefore, the function-led design of multifunctional stimuli-responsive SHS is of both present and future significance as conceptual methodology to inspire the related research.

## 1.2 Objectives

(i) Design and development of a stimuli-responsive multifunctional SHS which is capable of tuning the water spreading and adhesion on its surface, as well as switches between superhydrophilicity and superhydrophobicity.

(ii) Understand the mechanisms involved in the interfacial interactions and their influences on the surface wettability.

(iii) Develop a smart textile coating based on the stimuli-responsive SHS, manipulate the microfluidic behavior within the fabric matrix and evaluate the functional performance of the coating in various novel applications.

(iv) Apply the function-led design criterion in the research process of the SHS, compared the research output between the non-function-led design and function-led design, and generate informative suggestions for conducting the research in the function-led process.

## 1.3 Framework of the thesis

The thesis comprehensively demonstrates the engineering approaches used to obtain the full-spectrum wettability tuning on cotton fabric and the underlying mechanisms involved were also dedicatedly discussed.

Chapter 1 introduces the basic concept of superhydrophobic surface and its brief research history.

Chapter 2 reviews the theories developed for superhydrophobic surface and the core concepts involved in the design of superhydrophobic surface. State-of-the-art research on the superhydrophobic surface and new innovative concepts are also highlighted.

Chapter 3 describes the synthesis approaches used to fabricate the advanced nanostructured materials and the research methodologies involved in the characterization of the surface properties of the as-obtained nanomaterials.

Chapter 4 depicts the fabrication process of the hydrophobic graphene coated cotton fabric (G-fabric) and the related characterizations. The tuning of surface wettability of G-fabric is also demonstrated.

Chapter 5 demonstrates the fabrication process of the superhydrophobic cotton fabric with hierarchically nanostructured graphene/titania surface coating (GT-fabric). The surface morphology, chemical composition and full-spectrum tuning of surface wettability on GT-fabric are also characterized and described.

Chapter 6 presents the advanced applications of GT-fabric as a result of its full-spectrum tunable surface wettability. Multiple novel functions with practical significance, such as directional water transport, microfluid manipulation and oil-water separation, are realized using the GT-fabric.

Chapter 7 depicts the mechanism study on the superhydrophilic conversion of $TiO_2$ surface during ultra-violet (UV) radiation. New perspectives on the conversion mechanism are provided by investigating the interfacial interaction between graphene and $TiO_2$, and the structural reconfiguration on $TiO_2$ from the atomic scale to macro scale.

Chapter 8 summarizes the research and findings in the thesis and provides insights into the function-led design of superhydrophobic surface. Suggestions for the future developmental trends of superhydrophobic surface are also provided.

# Chapter 2. Literature review

## 2.1 The Young's equation and superhydrophobic state

In 1805, Sir Thomas Young firstly published the discovery on the theory of the solid-liquid-gas interfaces, which was eventually contained in an apparently simple equation: $\gamma_{sg}=\gamma_{sl}+\gamma_{lg}\cos\theta$, namely the Young's equation.[14] The Young's equation defines the equilibrium between the surface tensions acting on the three-phase contact point of solid, liquid and gas interfaces. The symbol $\gamma_{sg}$ represents the surface tension of the solid-gas interface, while $\gamma_{sl}$ and $\gamma_{lg}$ represent the surface tension of solid-liquid and liquid-gas interfaces, respectively. Given a small volume of droplet (gravity is negligible) sitting on an ideally flat surface (continuous solid-liquid interface), the contact angle θ between the solid surface and the liquid droplet is given by:

$$\cos\theta = \frac{\gamma_{sg} - \gamma_{sl}}{\gamma_{\mathrm{lg}}} \quad (1)$$

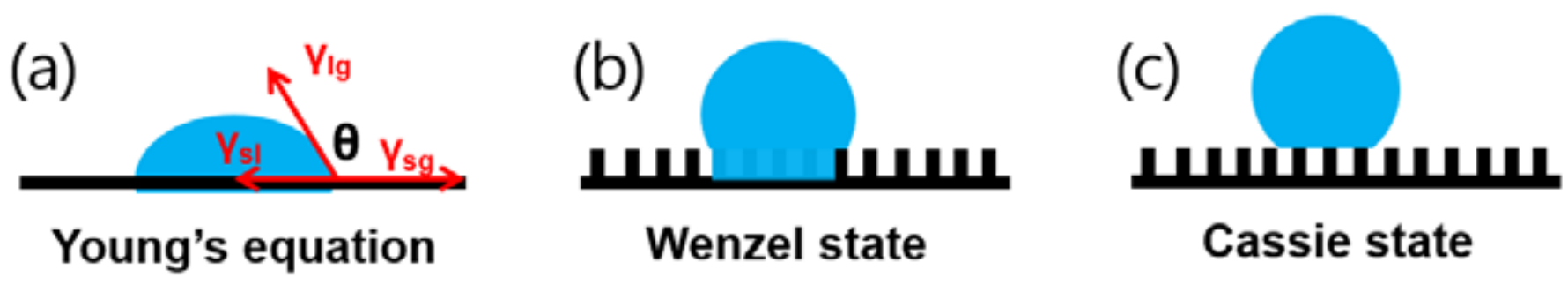

**Figure 1** Schematic illustrations of Young's equation (a) and the hydrophobic states (b, c). The blue meniscus represents the water droplet while the flat solid line and solid line with pillars represent the smooth and rough surfaces, respectively.

As schematically illustrated in Figure 1a, the Young's equation is only best for the equilibrium state of the contact between homogeneous solid surface and liquid, which is a rare case in real. In 1936, Wenzel researched the contact of liquid on heterogeneous surface, and subsequently discovered the effect of the surface texture on the formation of solid-liquid interface.[15] The roughness factor r, which indicates the intensity of surface texture, has been introduced to the Young's equation to formulate the Wenzel model. In the Wenzel model, the cosine of the apparent contact angle $\theta^*$ equals to the product of r and $\cos\theta$: $\cos\theta^*=r\cos\theta$ (2). As r is calculated from the actual surface area over the apparent surface area, and there is always $r>1$, the apparent contact angle $\theta^*$ decreases compared to $\theta$ in the case that $0°<\theta<90°$; on the contrary, $\theta^*$ increases compared to $\theta$ as $90°<\theta<180°$. As demonstrated in Figure 1b, the Wenzel model considers the effect of surface texture on the contact between the liquid and solid surface, and the space between the textures is filled with liquid. However, air may be still entrapped in the space between the textures, which is usually the case of SHS, and the Wenzel model does no longer effectively hold. To address this issue, the Cassie-Baxter model was proposed in 1944, and the apparent contact angle was defined as: $\cos\theta^*=f_1\cos\theta_1+f_2\cos\theta_2+\dots f_i\cos\theta_i$ (3), where 1 and 2 are indexed to different surface components with the area fraction of $f_i$ (i=1,2,3,…,n) and a contact angle of $\theta_i$ (i=1,2,3, …,n), respectively.[16] While there is air presented between the contact of solid and liquid, equation (3) can be written as: $\cos\theta^*=f_1\cos\theta_1-f_2$ (4). Provided that ($f_1+f_2=1$), equation (4) can be transformed into: $\cos\theta^*=-1+\varphi_s(1+\cos\theta)$ (5), while $\varphi_s$ represents the solid fraction at the solid-liquid interface.[17] As in the hydrophobic region, the ranges of equation (2) and (5) overlap, which means both of them should be hold in the region from moderate hydrophobicity to superhydrophobicity.[18] By equating (2) and (5), the threshold $\theta_c$ for this region can be obtained as:

$$\cos\theta_c = \frac{\phi_s - 1}{r - \phi_s} \quad (6)$$

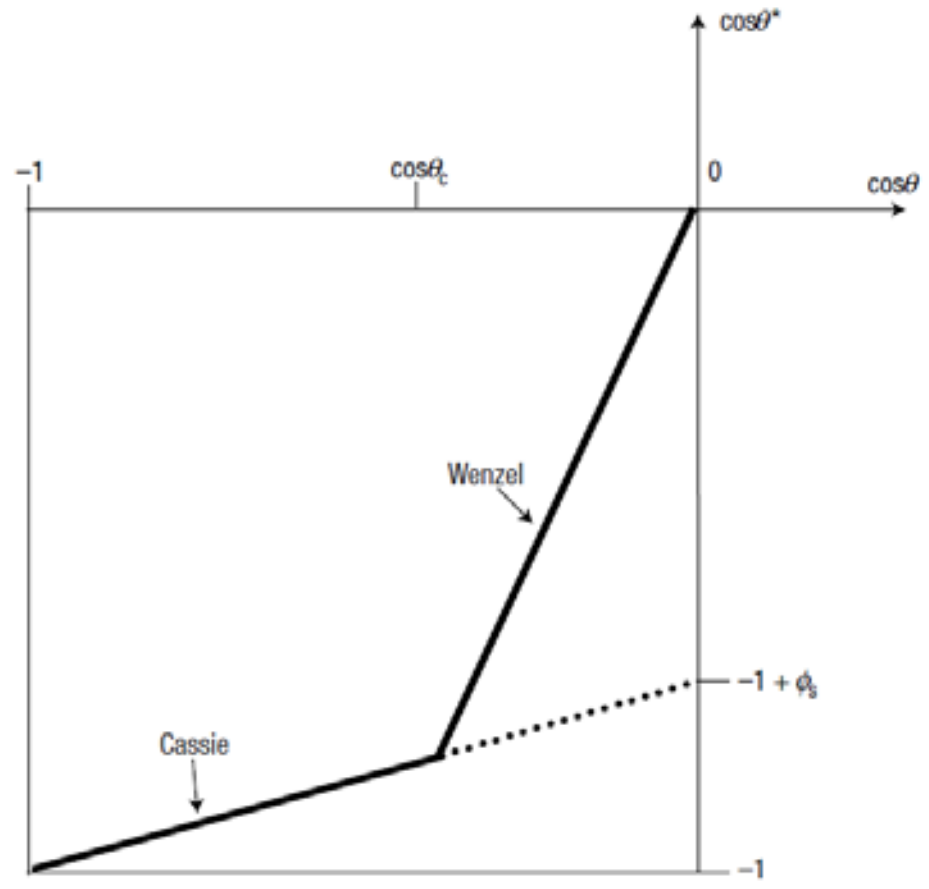

**Figure 2** Diagram shows the mathematical regions of Cassie and Wenzel states. For 90° < θ < $\theta_c$, the apparent contact angle θ* is given by the Wenzel equation. However, for θ > $\theta_c$, θ* is given by the Cassie-Baxter equation. It is also noticed that the Cassie state can still exist as θ < $\theta_c$ in certain circumstances. This metastable region is represented by the dot line.[17]

Therefore the Wenzel model still holds for $90°<\theta^*<\theta_c$, and the Cassie-Baxter model is also valid in the region of $90°<\theta<180°$, ideally from the point of $\cos\theta^*=\varphi_s-1$.[19] Figure 2 summarizes the relationship between the Wenzel and Cassie-Baxter models and their effective regions, presumably in the hydrophobic state ($90° < \theta^* < 180°$).

Even though their theoretical implications are different, the apparent contact angles between the Wenzel and Cassie states differ slightly.[20] The essential dissimilarity lies in the adhesion of liquid, as the hysteresis is significantly smaller in the Cassie state as compared to the Wenzel state, which can reach a factor of 20.[21] Being both hydrophobic states, a transition can be induced from the Cassie to the Wenzel state, by using proper external pressure.[22] The process of the transition is accomplished by replacing the air-liquid interface with the solid-liquid interface, as the entrapped air is dissipated from the textured surface. The work required for the process is normally done by applying an external pressure or by the gravity of the liquid.

Through the research of the hydrophobic states, the underlying mechanisms involved in the interfacial solid-liquid interactions can be progressively unfolded, which can not only benefit the prediction of surface hydrophobicity by simple parameters measured from the static solid-liquid contact, but also promote the precise engineering control over the nature of the solid-liquid interactions, e.g. sliding, adhesion and spreading.

## 2.2 Development in the forms and functions of superhydrophobic surface

### 2.2.1 “Lotus effect” and dual roughness

The developments of SHS towards real-field applications started at the introduction of "lotus effect" in 1992. W. Barthlott et al. elaborately studied the surface morphology of the hydrophobic plants using scanning electron microscope (SEM), and subsequently proposed the self-cleaning "lotus effect" derived from the superhydrophobicity.[23] Based on the SEM characterization of the lotus surface, a distinct hierarchical structure consisting of micro-papillae and wax tubules is revealed.[24] On the upper epidermis of lotus leaf, the micro-papillae can be clearly observed, while a dense layer of wax tubules is coated on the surface of and between the papillae, as shown in Figure 3. The wax tubules are majorly composed of $C_{29}$-dols in the case of lotus, which cover the surface area of the epidermis with high porosity as a result of their nanoscale dimensions and random orientation. At the same time, the micro-papillae provide solid supports to the water droplet which then favors the formation of Cassie state. This distinct structural configuration of order micro-papillae and low surface energy nano-scale wax coating thus constitutes the "dual roughness" of lotus leaf, resulting in outstanding water-repellent and self-cleaning properties.

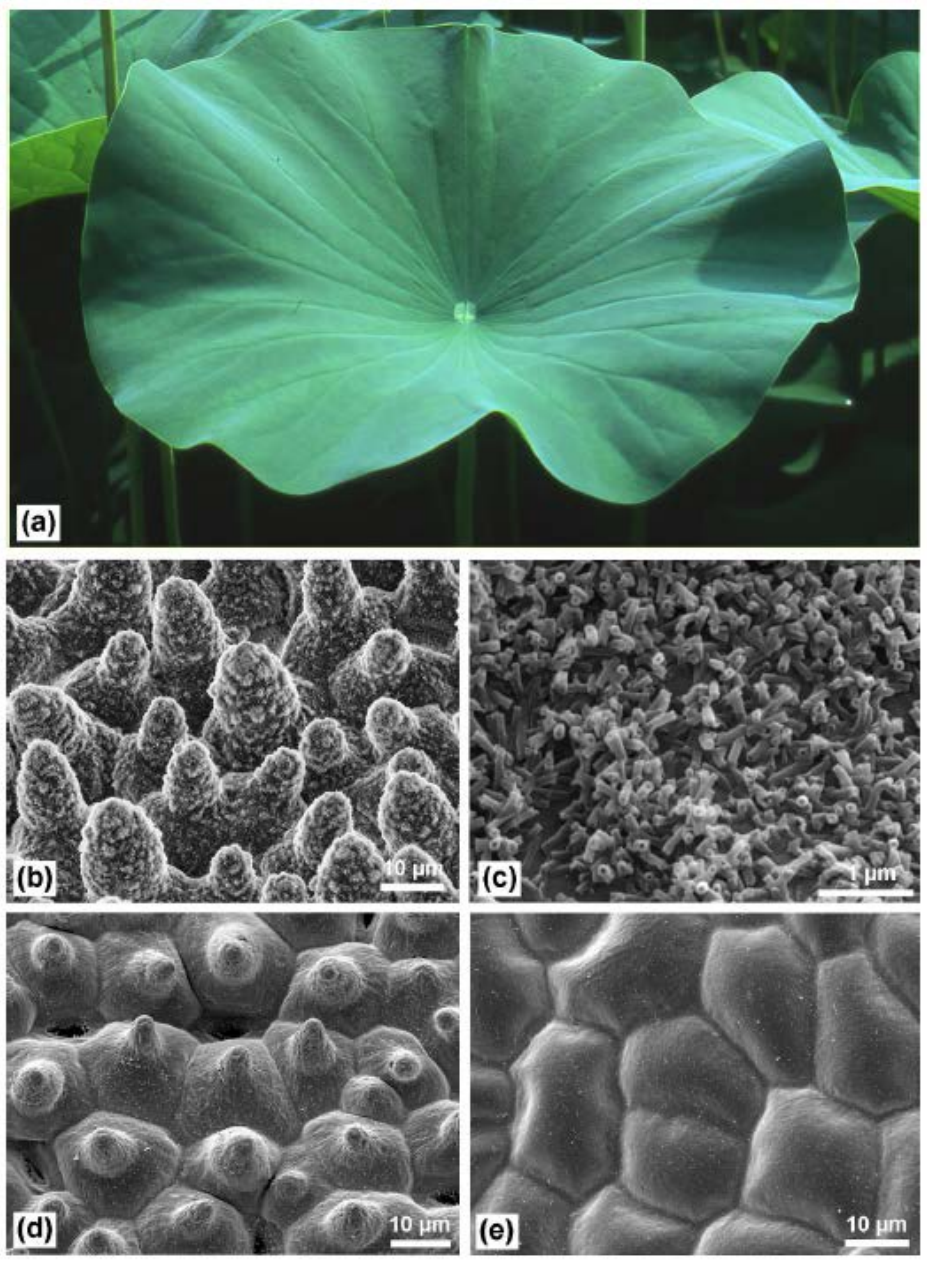

**Figure 3** (a) Photograph of a lotus leaf. (b) SEM image of the lotus leaf surface showing the hierarchical structures of micro-papillae, wax cluster coating and wax tubules. (c) Magnified view of the wax tubules showing the nano-dimensional feature. (d) SEM of the lotus leaf surface after wax removal, where the stomata are clearly observed. (e) SEM of the underside leaf surface after wax removal, which is composed of convex cells without stomata.[24]

### 2.2.2 Fabrication of SHS on flat solid surface: chemical etching

Foundation on the concept of "dual roughness" inspired by the "lotus effect", various flat surfaces are engineered and transformed into high performance SHS, including metal, plastic, ceramic, silicon, carbon, and glass.[25-30] In order to create the desired roughness on the flat surface, the chemical etching was found to be an effective method to render the surfaces rough with both simplicity and processability.[31] In a typical process, flat metal surfaces, i.e. steel & copper, can be turned into SHS by simple dipping-drying-coating processes; the chemical etching, as the essential step to generate surface roughness, was done by immersing the metal pieces into the etching solution (40 wt% HF/30 wt% $H_2O_2$, v/v=1:1), and subsequently dried the etched piece in elevated temperature.[32] Low surface energy coating was then applied by dipping the etched metal pieces in the hydrolyzed fluorosilane solution [$CF_3(CF_2)_7CH_2CH_2Si(OCH_3)_3$], which were eventually dried at 140°C to obtain the SHS. The resulting SHS on either steel or copper alloy showed remarkable superhydrophobicity as water droplets of relatively large sizes can stand still with high contact angles, as shown in Figure 4. The high superhydrophobicity of the as-obtained SHS was attributed to the hierarchically rough surface structures formed by the chemical etching and silane coating processes, which can be observed as the disoriented micro-textures with nanoscale sub-structures, as shown in Figure 5.

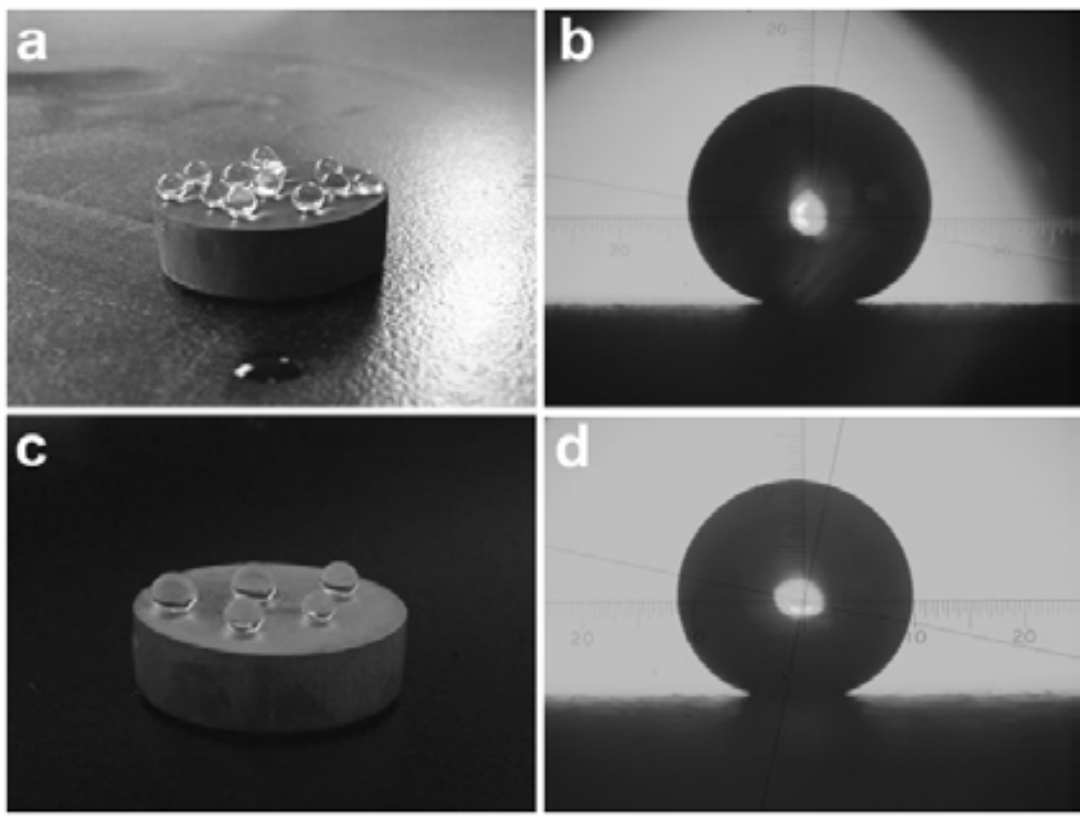

**Figure 4** (a) Image showing water droplets with different sizes resting on the superhydrophobic steel surface prepared by chemical etching. (b) Magnified view of a water droplet resting on the steel surface with a contact angle of 161°±1°. (c) Image showing water droplets with different sizes on the superhydrophobic copper alloy prepared by chemical etching. (d) Image of a water droplet on the copper alloy surface with a contact angle of 158°±1°.[32]

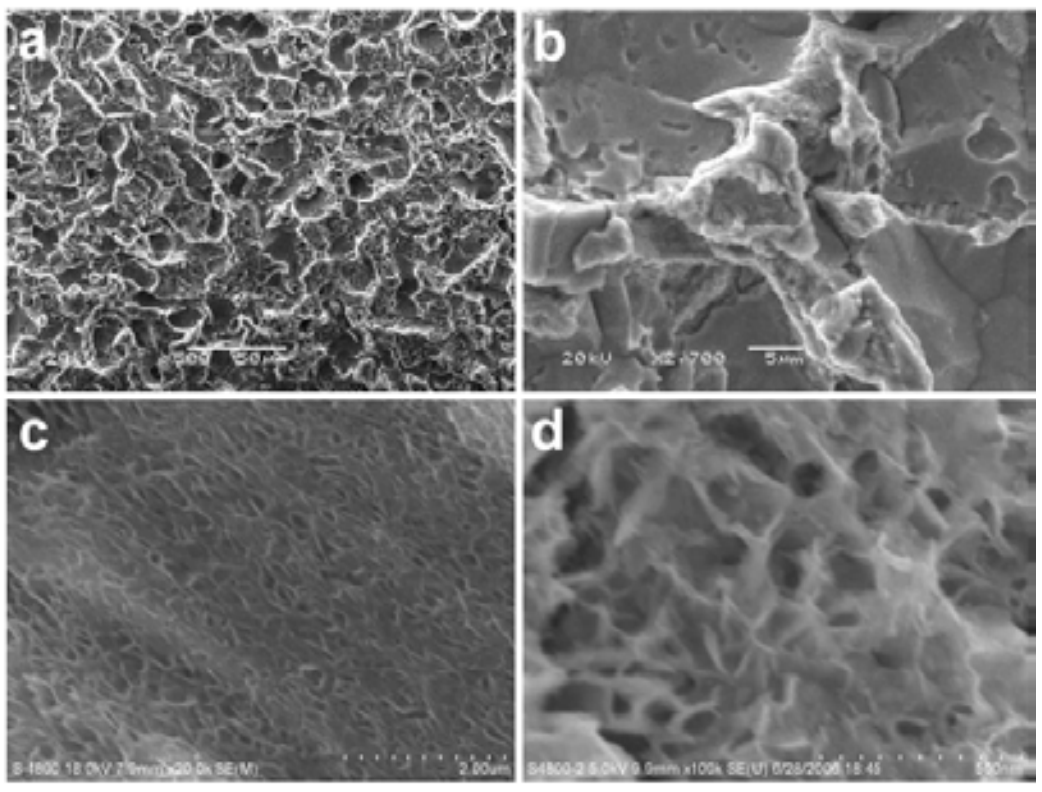

**Figure 5** SEM (a) and the sequentially magnified SEM images (b, c, d) of the chemical etched steel surface coated with fluoroalkylsilane.[32]

### 2.2.3 Fabrication of SHS on flat solid surface: physical etching

Besides chemical etching, "dual roughness" can also be generated on flat surface through physical etching processes, such as plasma etching, laser ablation, and micromachining.[33-35] By using radio frequency plasma enhanced chemical vapor deposition (RF-PECVD), SHS can be obtained on the flat poly(methyl methacrylate) (PMMA) surface.[36] In the first step, PMMA was etched by a glow discharge of $CF_4$ gas to create the surface texture of micro-roughness. After a short hydrolysis process in water to remove the additional solid fraction, the as-obtained PMMA was subsequently coated with the low surface energy organosiloxane to achieve superhydrophobicity, as depicted in Figure 6. The superhydrophobic PMMA surface exhibited both high water contact angle and low hysteresis, as the impinged water drops can be elastically bounced upon impact, as shown in Figure 7. Nonetheless, it is of ease to perceive that the hierarchical roughness consisting of the micro-pillars and the siloxane coating is the essential factors to induce the excellent superhydrophobicity.

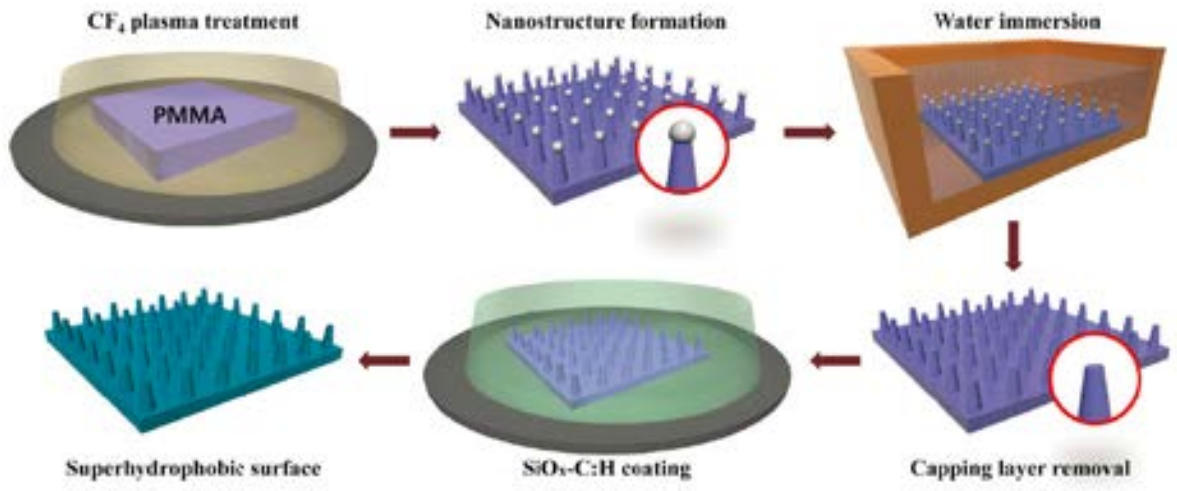

**Figure 6** Schematic illustration of the plasma-etching process to create superhydrophobic surface on PMMA substrate.[36]

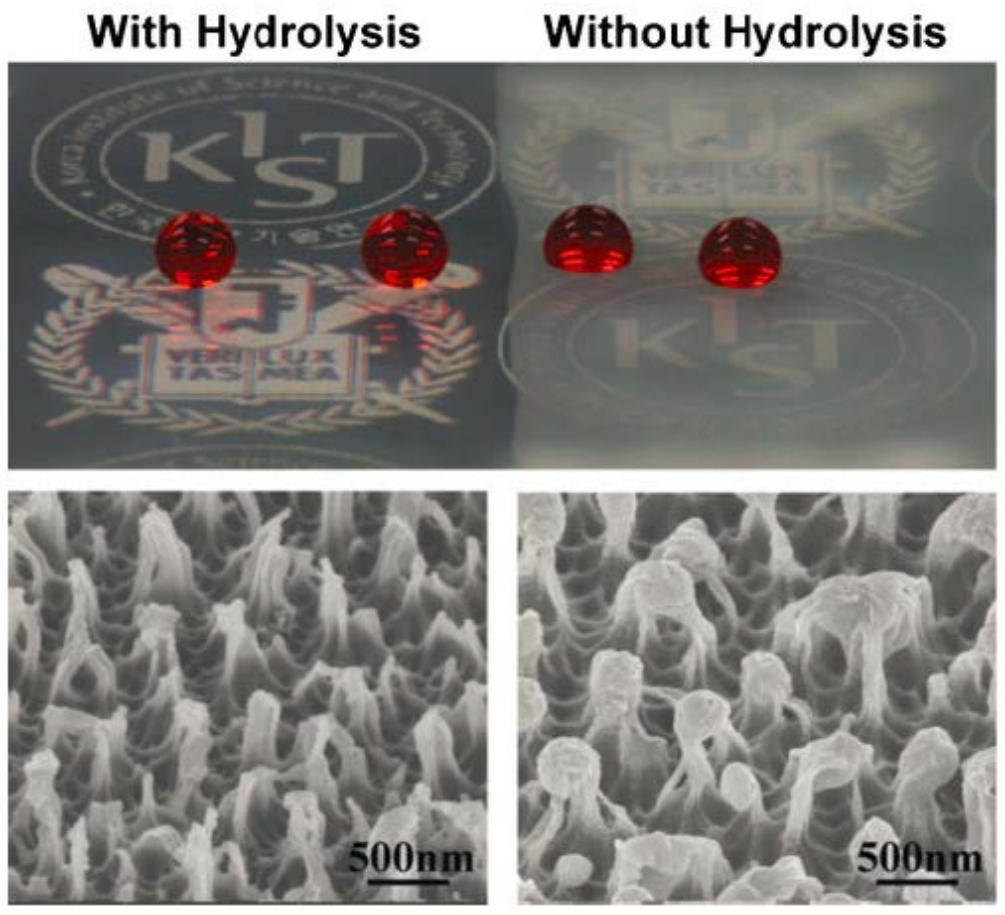

**Figure 7** Photograph and SEM image of the plasma etched PMMA surface with additional hydrolysis processing (left images) and the plasma etched PMMA surface without hydrolysis (right images). Both higher water contact angle and transparency are observed on the hydrolyzed PMMA surface, which can be attributed to the enhanced surface roughness after hydrolysis.[36]

### 2.2.4 Fabrication of ordered SHS on flat solid surface: photolithography

The creation of "dual roughness" by chemical etching is effective in obtaining SHS on flat surface, which is also suitable for large scale processing. However, the chemical etching method also associates with intrinsic drawbacks as only randomly oriented surface texture can be formed, and the subtle morphology of the texture is created without precise control. These drawbacks thus reduce the potential of optimization for the SHS obtained by chemical etching. The SHS with heterogeneous surface texture may also exhibit lower resistance to drop impact, as the region with low degree of roughness cannot provide substantial air pockets to support the Cassie state. Aiming to obtain optimized superhydrophobic property on flat surface, the technique of photolithography has been adopted to fabricate the SHS with ordered arrays of micropillars & posts.[37] Step-wise processes are developed for photolithography so that fine control over the microstructures can be achieved. During the photolithography process, sequential steps including cleaning of the wafer, coating of the barrier layer, coating of photoresist, soft baking, mask alignment, UV exposure, and hard baking are involved, as shown in Figure 8.[38] On the other hand, different post-lithography processes, such as etching, lift-off, and molding, can be applied to the as-obtained molds to sculpture the desired patterns. Upon using different combination of the processes, such as alternating the photoresist, changing the mask alignment, and applying different post-lithography techniques, multivariant fine morphologies on the micropost array can be generated, which exhibit intriguing superhydrophobic properties such as low hysteresis, super-repellent, and pressure resistance.[39-41] By combining the photolithography with the reactive ion etching (RIE) technique, high performance SHS can be created on Si wafers.[42] The novel SHS possesses unique surface texture, which is consisting of the microhoodoo arrays with defined geometric parameters, as shown in Figure 9. Special reentrant structure is designed for the individual microhoodoo which can induce substantial vertical components of surface tension to support the Cassie drops. For this order surface texture, enhanced liquid repellent property was clearly observed as even low surface energy hydrocarbons (e.g. heptane, methanol) can form Cassie drops on it, as shown in Figure 10.[43]

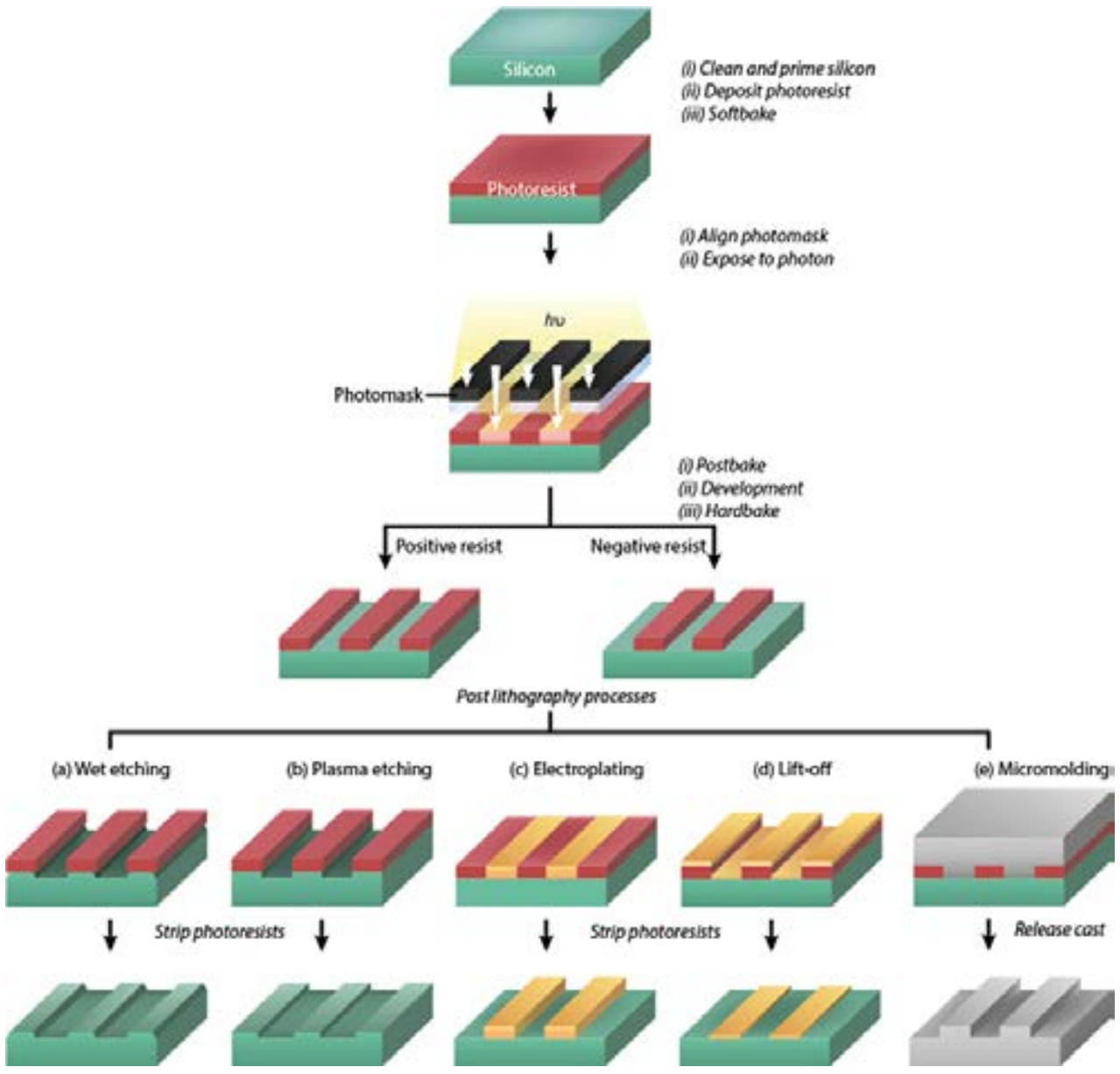

**Figure 8** Schematic illustration of a typical photolithography process.[38]

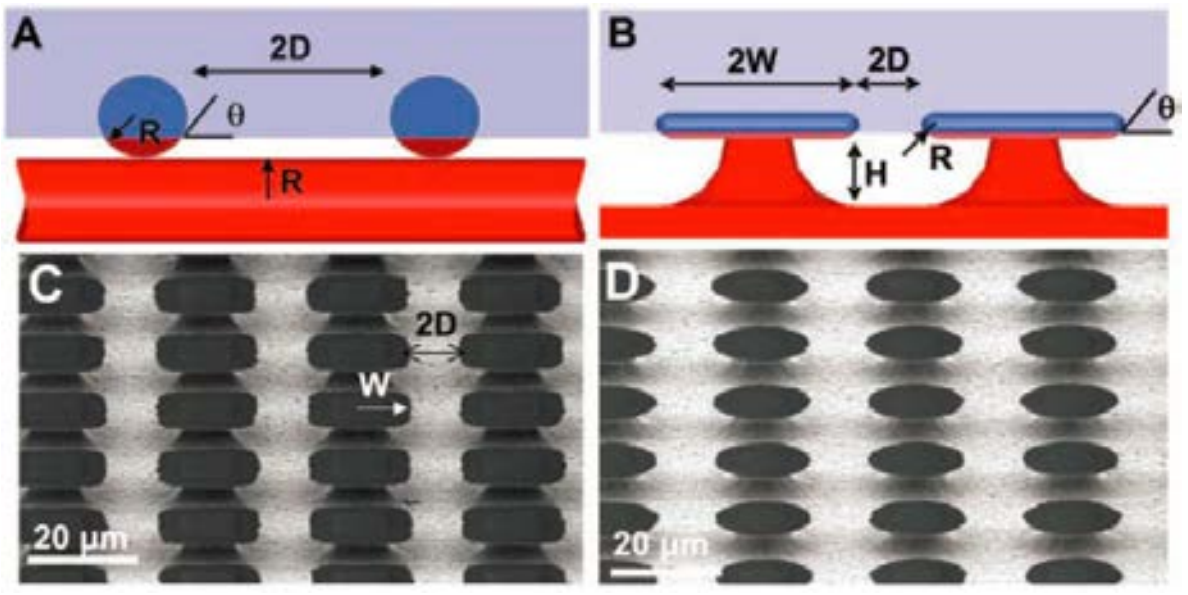

**Figure 9** (A, B) Schematics showing the formation of a composite interface on the surface with reentrant microtextures, including the nanofiber surface (A) and microhoodoo surface (B). The blue color indicates the wetted area while the red color indicates the non-wetted area, corresponding to the case that the surface is contacting with a liquid that has an equilibrium contact angle θ < 90°. (C, D) SEM images of the microhoodoo surfaces with square (C) and circular (D) flat caps, obtained at a tilting angle of 30°.[42]

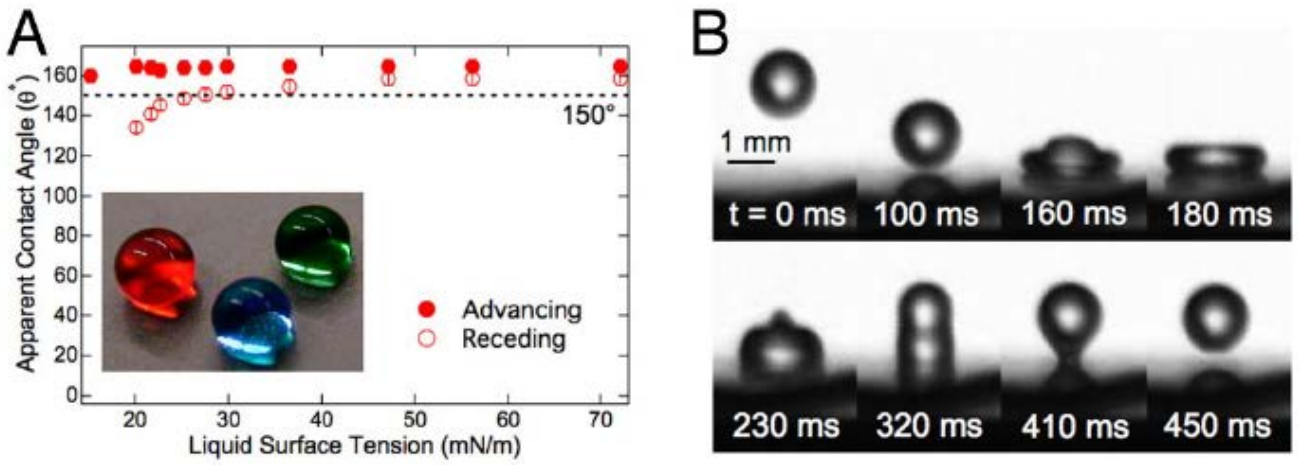

**Figure 10** (A) The advancing and receding contact angles of liquids with different surface tensions on the silanized microhoodoo surface (reentrant structure). Inset shows the heptane (red), methanol (green) and water (blue) droplets resting on the microhoodoo surface. (B) Sequential frames showing the bouncing of a hexadecane droplet on the silanized microhoodoo surface.[43]

### 2.2.5 Fabrication of ordered SHS on flat solid surface: replica molding &

**chemical vapor deposition**

Besides photolithography, other cost-effective techniques can also be utilized to engineer order micro-textures on flat surface and achieve superhydrophobicity, such as replica-molding and chemical vapor deposition.[44-45] Polymer-based SHS with order micropillar surface texture can be obtained using the replica-molding process. Compared with photolithography, the replica-molding possesses have the advantages such as simplified procedures and deformable surface textures. Using a shape memory polymer (SMP) as the replicating resin, the SMP-based SHS can be produced from the polydimethylsiloxane (PDMS) mold.[46] Micropillar arrays with defined pillar spacing, height and pillar diameter can be readily fabricated on the SMP surface. The fabrication process is simple and involved only pouring the SMP precursors in the PDMS mold, curing, and then peeling off (Figure 11). The as-obtained SMP with micropillar arrays can effectively maintain Cassie drops with high water contact angle of 155.6° and low sliding angle of 7.5° (for 10 μL water droplet).[47] The shape memory property of the SMP also enables controlling of the hydrophobic state on its surface. The original Cassie state can be transformed into Wenzel state as the surface texture of the SMP was deformed by a shear force, and the micropillars were bended according to the force direction. The surface can be subsequently restored to the Cassie state by heating to release the strain, and the micropillars also returned to their original shapes, as shown in Figure 12. This novel reversible process, however, can realized the fine control over pinning and de-pinning of droplets on SHS, which is suited for the microfluidic application.

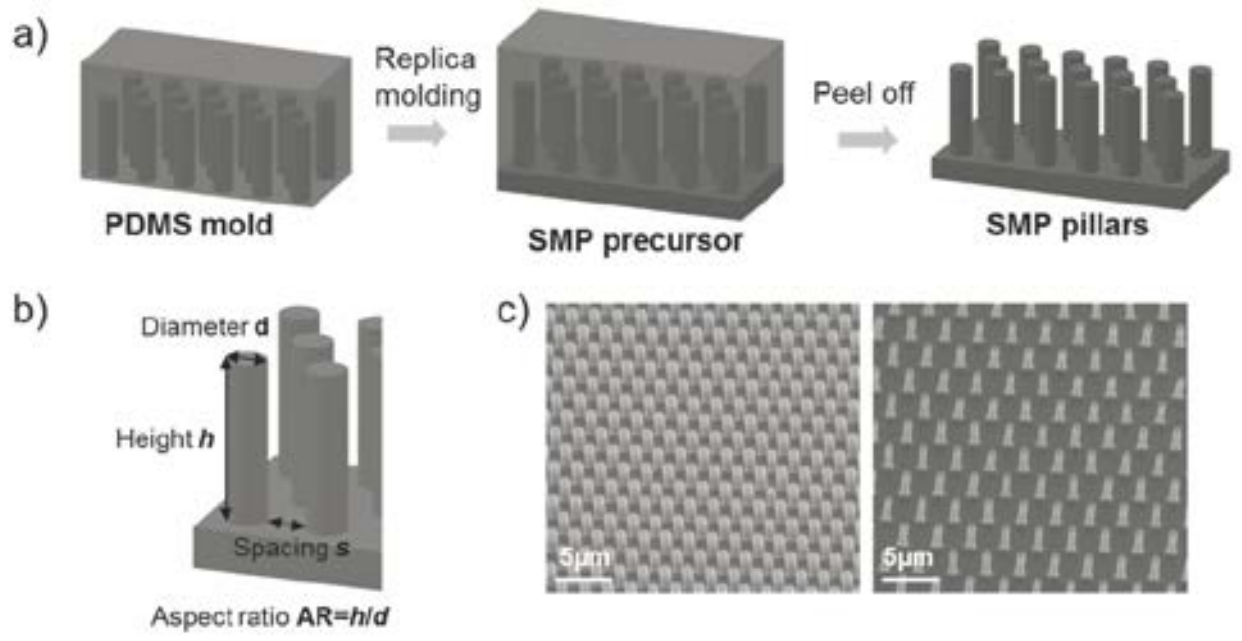

**Figure 11** (a) Schematic illustration of the replica molding process used to fabricate the shape memory polymer micropillars. (b) Characteristic parameters of the micropillars and (c) SEM images of a hexagonal SMP micropillar array with d=1 μm, s=1 μm (left) and s=2 μm (right); AR=4.[46]

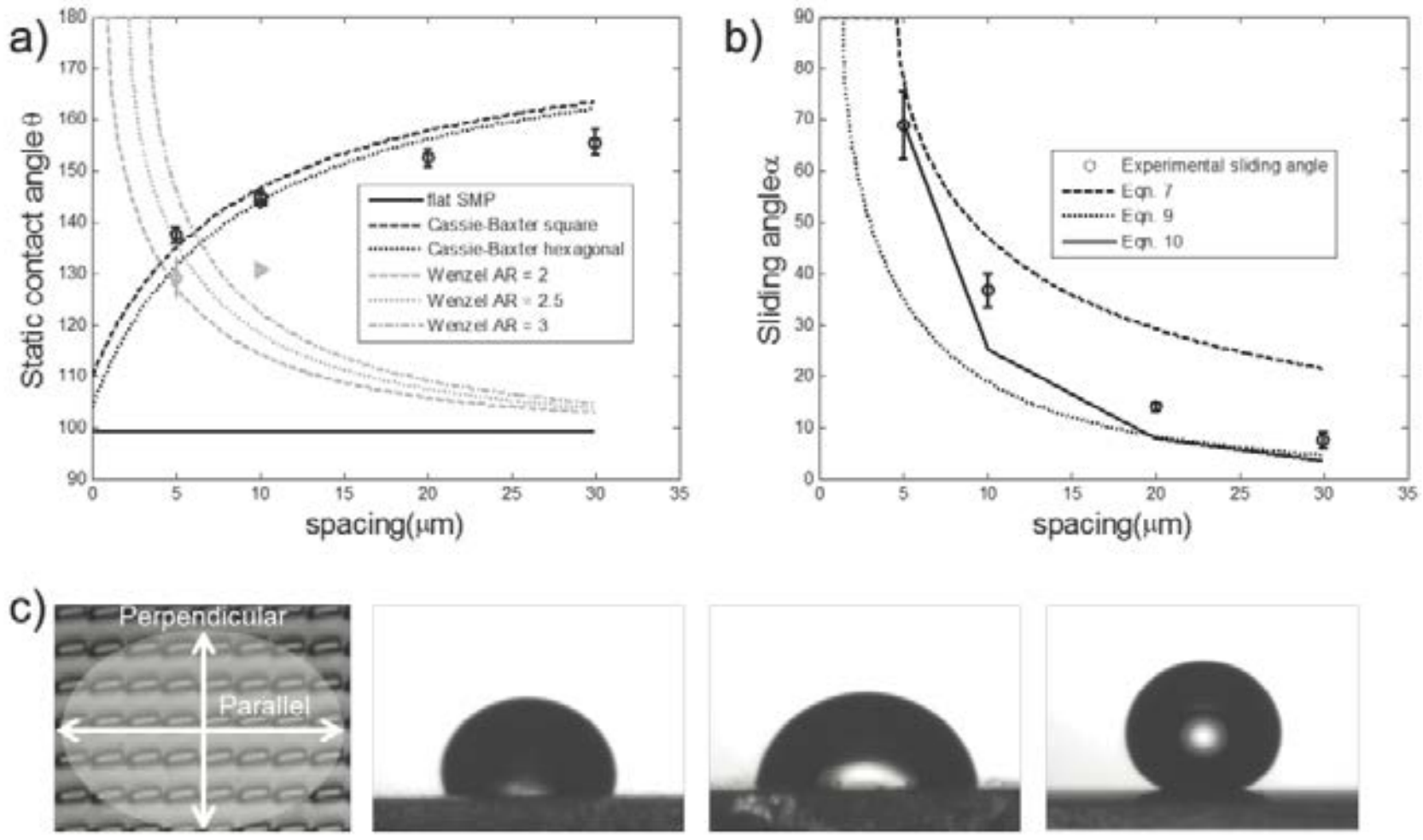

**Figure 12** Static water contact angles of SMP micropillars with different structural parameters (solid dots). The circle dots represent the square array and the square dots represent the hexagonal array. The WCAs of the deformed pillars are also shown, as the diamond dots represent the case of AR=2 and the triangle dots represent the case of AR=3. The lines represent the theoretically predicted

values by Cassie and Wenzel model. (b) The experimental sliding angles (circles) vs. theoretical predictions based on different equations on a SMP micropillar surface. (c) Demonstration of the anisotropic wetting on the deformed SMP micropillar surface. From left to right: Top-view of the deformed SMP pillar array with the white arrows indicating the viewing direction of the droplet; the perpendicular view of the droplet; the parallel view of the droplet; and the droplet on the un-deformed SMP pillars, respectively.[47]

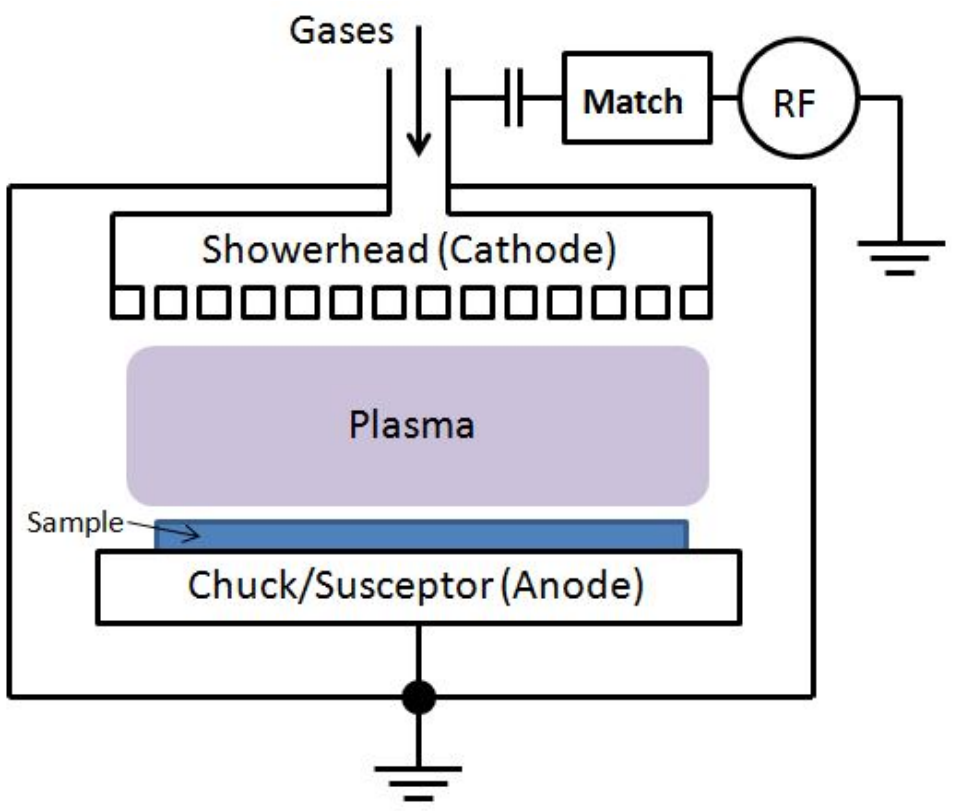

**Figure 13** Schematic illustration of a plasma enhanced chemical vapor deposition (PECVD) set-up.[64]

Order nanostructured arrays can also be produced on flat surfaces using chemical vapor deposition (CVD). Different from photolithography, which prints the pattern on the substrate using external device, and categorized as the "Top-down" approach, CVD directly grows the nanostructures on the substrate surface, as the chemical vapors of the precursors interact with the catalyst particles pre-impregnated substrate, and continuously deposit on the surface of the catalysts via either vapor-liquid solid (VLS) or vapor-solid (VS) mechanisms.[48-50] In this regard, CVD is most recognized as the "Bottom-up" approach. Compared with photolithography and replicate-molding, CVD is more versatile in selecting the precursor and substrate. Precursors such as metals, oxides, sulfides, nitrides, phosphides, carbides,[51-56] and substrates such as silicon, metal, polymer, and electrospun fiber mats,[57-60] are suitable for the CVD processing. To produce SHS, order patterns of nanostructures are firstly grown on the desired substrate by CVD, which are subsequently coated with low surface energy molecules, i.e. fluoroalkylsilane, to achieve superhydrophobicity. Morphology control over the as-obtained nanostructures can be implemented by changing the CVD time, precursors, and catalyst crystallinity.[61-63] Due to the high density of nanostructures grown by CVD, which can be depicted as "Forest", robust superhydrophobicity can be readily achieved, as a result of the relatively homogeneous air-solid interface with low solid fraction formed atop the order nanostructured “forest”. A schematic illustration of the CVD process is shown in Figure 13.[64] For example, hexagonally patterned zinc oxide (ZnO) nanorod forest can be produced by the CVD of ZnO powder and graphite powder.[65] Large area of hexagonal ZnO nanorod forest was grown on the sapphire substrate, following the hexagonal pattern of the pre-deposited gold particles, as shown in Figure 14. The lengths of the nanorods can be changed from hundreds of nanometers to a few microns by controlling the CVD time. Utilizing the versatility of CVD process, fascinating ZnO nanostructures with unique superhydrophobic property can be produced. Novel sea urchin-like nanostructured ZnO was produced on a porous anodic alumina (PAA) substrate by CVD.[66] Firstly alumina nanowire pyramids were grown on the PAA substrate by electrodeposition, and the ZnO nanowires were then nucleated and grown radially on top of them, as shown in Figure 15. The as-obtained sea urchin-like ZnO nanowires exhibited both high water contact angle (151°) and high water adhesion, as shown in Figure 16. It thus indicates water droplets can form the so-called "partial impregnated Cassie state" on the sea urchin-like ZnO surface, which is regarded as one of the metastable intermediate state between the Cassie and Wenzel states.[67]

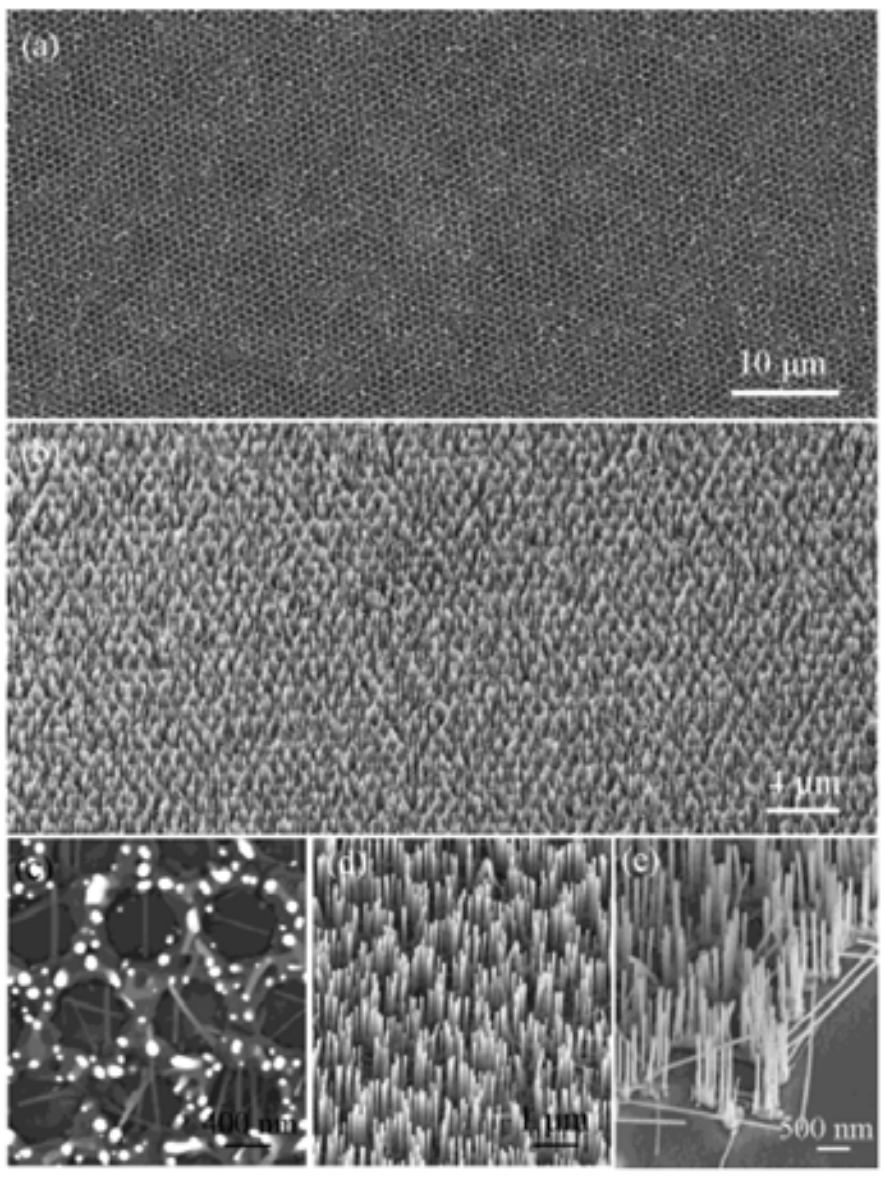

**Figure 14** Order hexagonal ZnO nanorod array produced by CVD. (a, b) Top-view and side view (tilting angle=30°) SEM images of the ZnO nanorod array grown on the honeycomb catalyst pattern, respectively. (c, d) Magnified top-view and side-view SEM images showing the hexagonal arrangement of the nanorod array. (e) The hexagonal order is maintained at the edge of the catalyst pre-impregnated sapphire surface.[65]

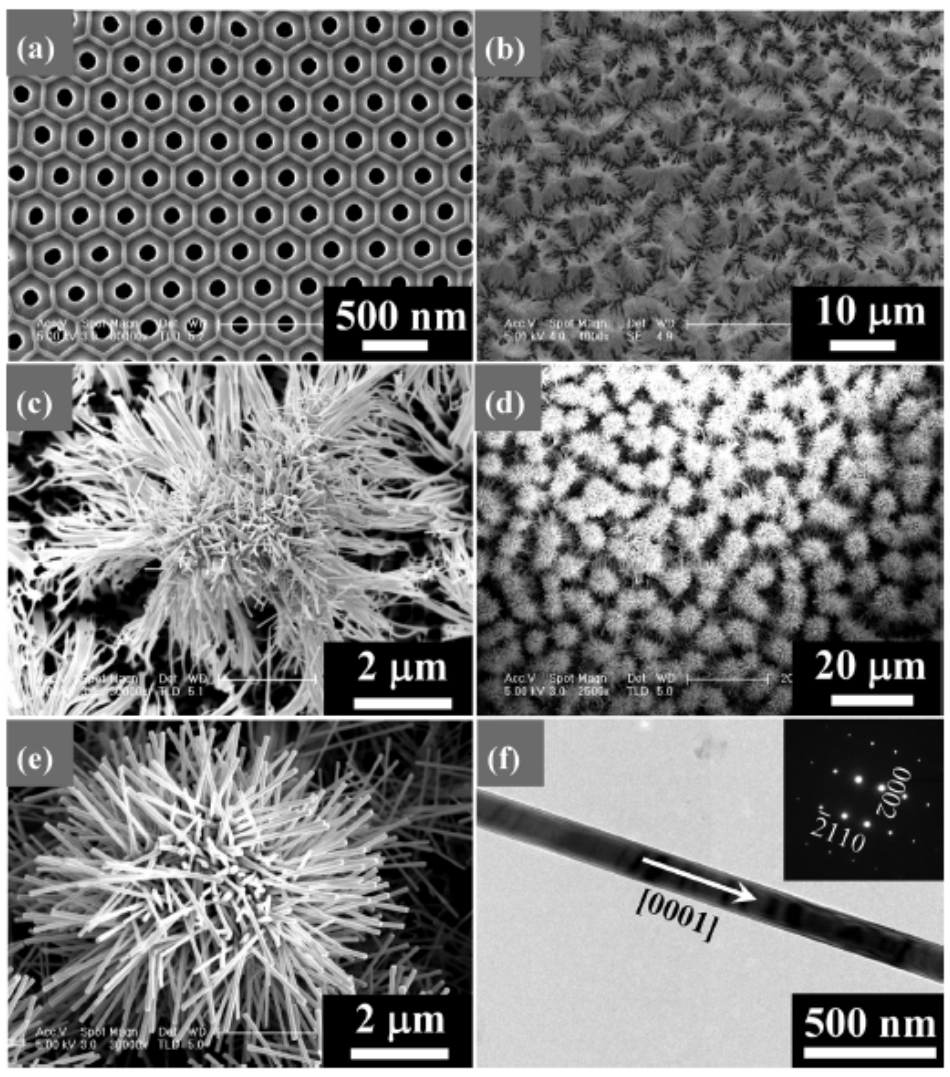

**Figure 15** (a, b) SEM images of the porous anodic alumina (PAA) film (a) and the alumina nanowire pyramids (b) grown on its surface by chemical etching. (c) SEM image showing the ZnO nuclei on top of the alumina nanowire pyramid. (d, e) SEM and magnified SEM images of the lotus-leaf like ZnO microstructures showing both the micro- and nano-dimension roughness. (f) TEM image of a ZnO nanorod in the lotus-leaf like structures. Inset shows the corresponding SAED pattern of the nanorods.[66]

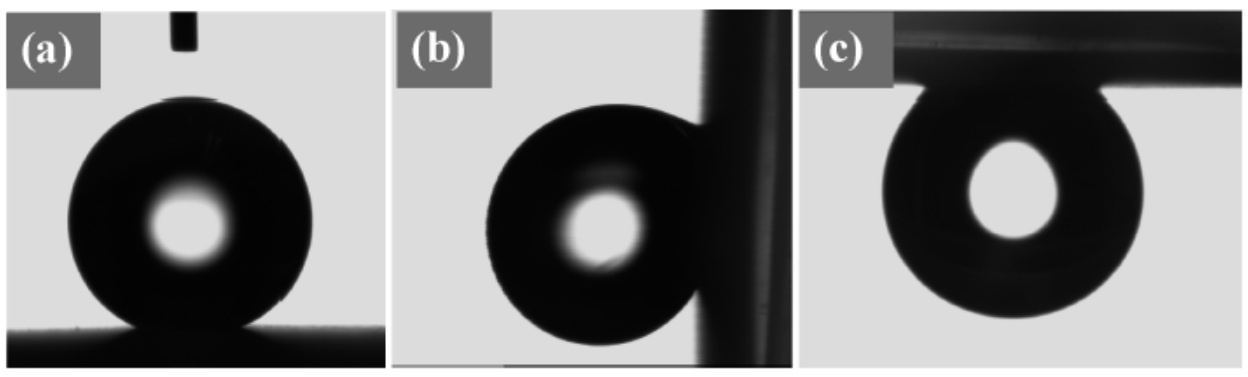

**Figure 16** Photographs showing a water droplet resting on the lotus-leaf like ZnO microstructures with a tilting angle of 0°, 90° and 180°. High water

adhesion is demonstrated by the lotus-leaf like ZnO surface.[66]

Besides ZnO, carbon nanotubes (CNTs) are also widely-used for the generation of SHS via CVD.[68] Different from ZnO, which spontaneously obtains superhydrophobic characteristics on the as-formed nanostructures, the CNT forests generally reach superhydrophobicity by the additional low surface energy coating.[69] Excellent water repellency with high water contact angle and low hysteresis thus can be achieved simultaneously. In a typical process, CNT forest was produced on the silicon substrate impregnated with Ni catalyst using plasma enhanced CVD.[70] Control over the CNT morphology can also be implemented by changing the CVD parameters and the catalyst layer thickness. To achieve the superhydrophobicity, a poly(tetrafluoroethylene) (PTFE) coating was subsequently applied onto the CNT forest using hot-filament CVD. Both high contact angle and low hysteresis were observed on the CNT forest after the coating process, as shown in Figure 17. And it was also demonstrated that the coating was effective in stabilizing the morphology of CNT during water evaporation. Intriguingly, the PTFE coated CNT forest can remain superhydrophobic even to the tiny water droplets down to micrometer size, which demonstrated robust hydrophobic property to water condensation, as shown in Figure 18. This novel phenomenon thus unmasks the importance of the nanoscale size and density of order microstructures in achieving superhydrophobicity to the molecular level.

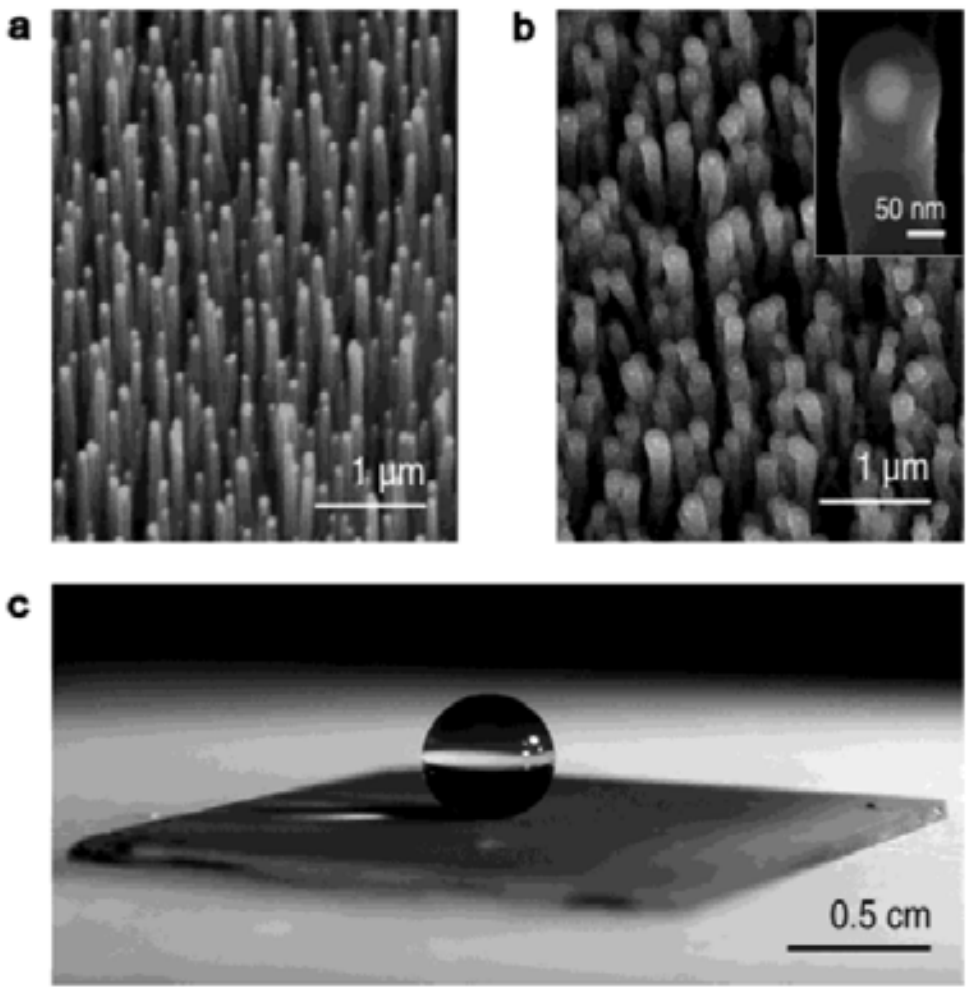

**Figure 17** (a) SEM image of the aligned CNT forest grown by PECVD with a diameter of 50 µm and a height of 2 µm. (b) SEM image of the CNT forest after the PTFE coating. And (c) Photograph showing a spherical water droplet with high contact angle sitting on the surface of PTFE coated CNT forest.[70]

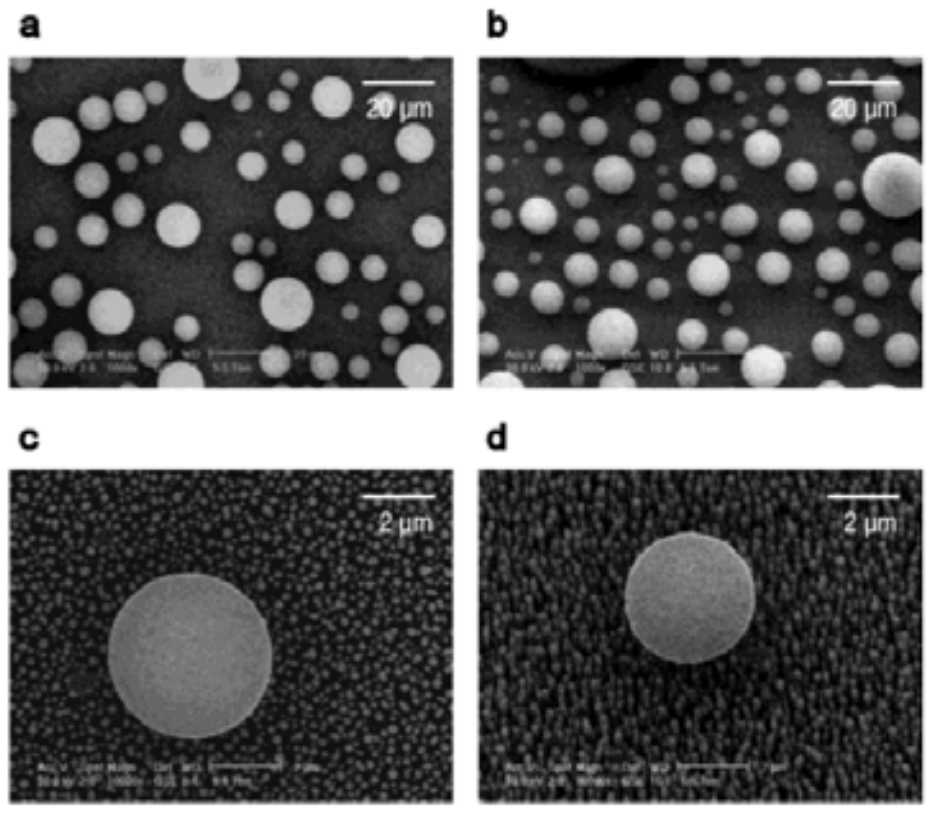

**Figure 18** Environmental SEM images showing the micron-sized water droplets sitting on the PTFE coated CNT forest. (a, b) Top-view and side-view (tilting angle=15°) of the micro-droplets on the surface of CNT forest, respectively. (c, d) Magnified top-view and side-view images of a single micro-droplets on the CNT forest, the tips of CNTs underneath can be clearly observed.[70]

### 2.2.6 Fabrication of SHS on non-flat solid surface: low surface energy coating

To create "dual roughness" and achieve superhydrophobicity, "Top down" or "Bottom up" approaches are in general required to fabricate the micro-textures on flat surfaces. In this case, artificial control can be implemented on the surface texture based on the precise microfabrication techniques. However, for the intrinsically non-flat surface, direct low surface energy organic coating can be applied to reach superhydrophobicity. Generally speaking, non-flat surfaces of engineering importance, i.e. textiles, papers, foams, are having intrinsic microscale surface textures which are induced during the manufacturing processes.[71-73] Taking advantages of this built-in roughness, high performance SHS can be realized on these engineering surfaces cost-effectively. Using the fluoroalkyl functionalized water-born siloxane (FAS) as the coating agent, plain-weave cotton fabric can be transformed from hydrophilic to superhydrophobic after a simple pad-dry-cure process.[74] Relatively high water contact angle (150°) and low sliding angle (15°) were observed for the FAS coated cotton. However, after applying the plasma pretreatment to the fabric surface before the FAS coating, the superhydrophobicity of the fabric can be further enhanced with a WCA of 154° and SA of 7°. Very good self-cleaning property was observed on the FAS coated and plasma pretreated cotton fabric [CO(PT)-FAS], as demonstrated in Figure 19. The FAS coating also showed good wash fastness up to ten standard washing cycles, indicating by the preservation of high contact angles on the fabric surface (Figure 20). On the surface of the textile fabric, the engineered networks of fibers may form the intrinsic roughness consisting of the surface curvatures existed between the fibers and yarns.

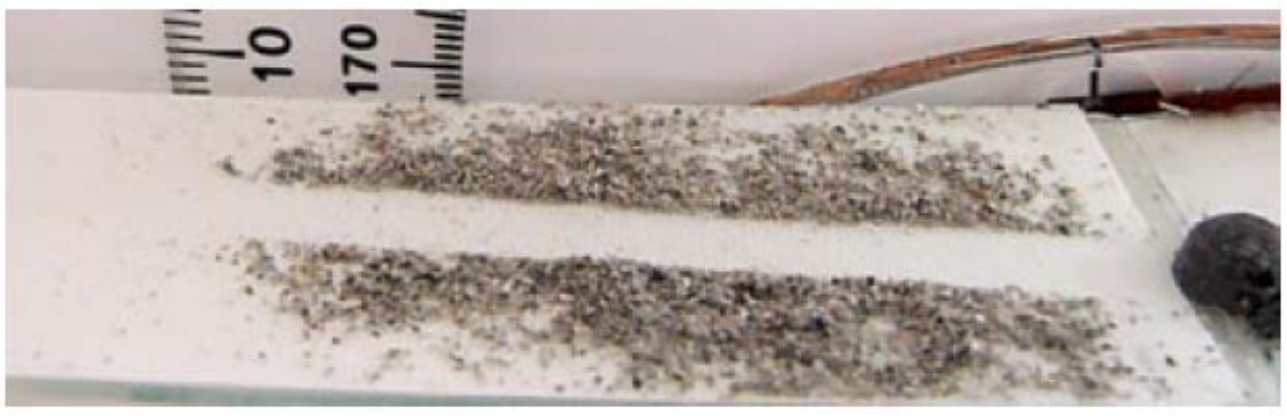

**Figure 19** The self-cleaning property of the CO(PT)-FAS sample surface is demonstrated by rolling a water droplet through the area contaminated by solid particles.[74]

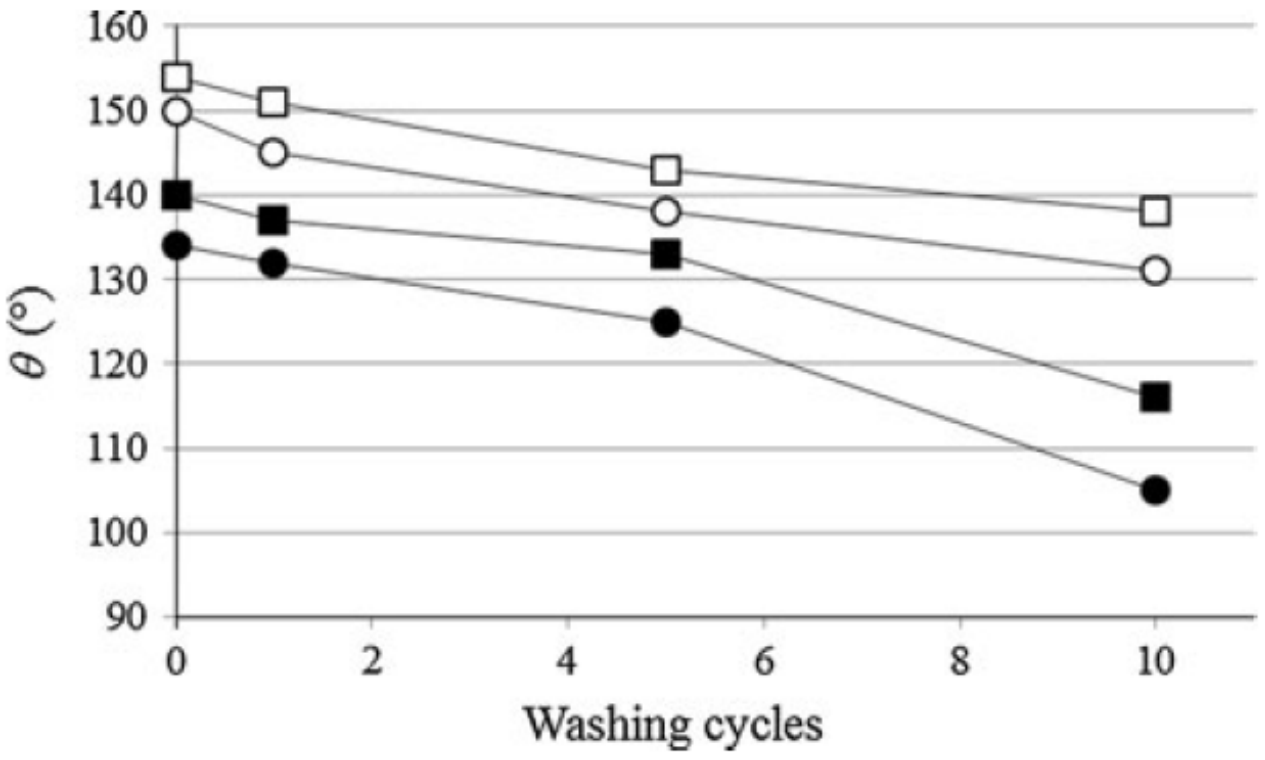

**Figure 20** The contact angle vs. washing cycle for water and n-hexadecane on the CO(PT)-FAS and CO(UN)-FAS (without plasma treatment) surface. The open circle and open square represent water droplets on CO(UN)-FAS and CO(PT)-FAS, respectively. The filled circle and filled square represent hexadecane droplets on CO(UN)-FAS and CO(PT)-FAS.[74]

Similarly, roughness can also be created on the paper surface during manufacturing, as the short paper fibers entangled and stacked to form the fiber mat. In this regard, the paper surface can also be transformed from hydrophilic to superhydrophobic using simple low surface energy coating. By a facile vapor-phase coating process, Whatman chromatography paper was made into superhydrophobic in a fast fabrication process.[75] In this case, the paper reacted with the vapors of fluoroalkyltrichlorosilane, which subsequently formed a low surface energy coating on its surface. The vapor phase silanization generates notable superhydrophobicity with WCA as high as 160° and hysteresis ranging from 7° to 20° (Figure 21). As a result of the low surface energy organosilane coating, the silanized paper also exhibits substantial oil-repellent property, which makes it impede the wetting of both water and oil, and reaches the so-called omniphobicity.

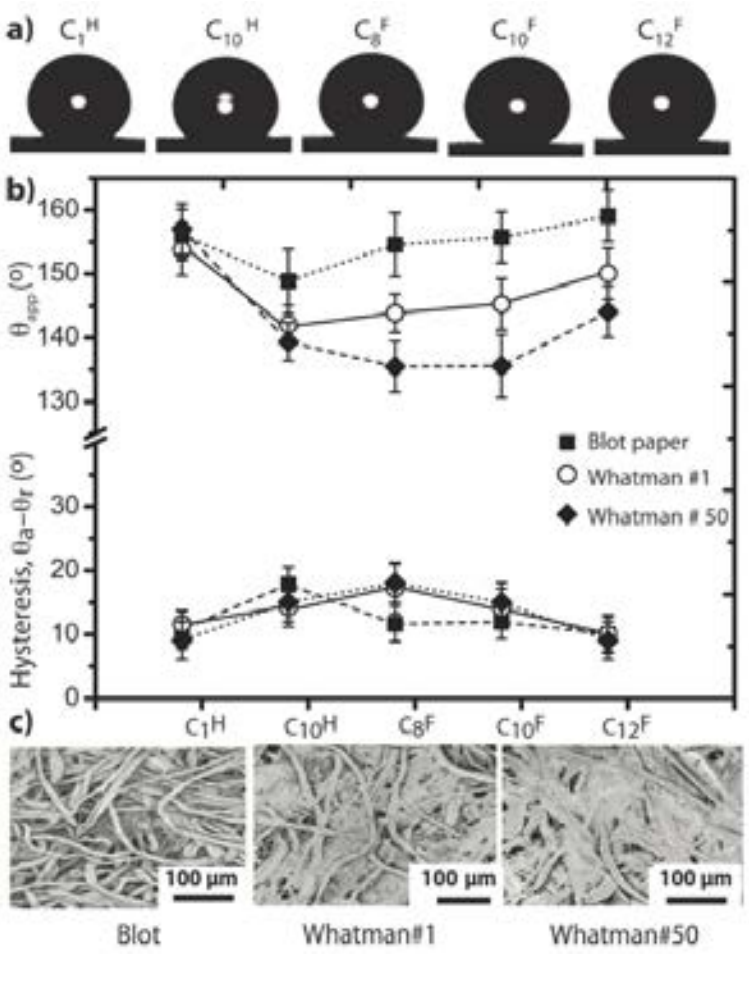

**Figure 21** (a) Photographs showing a water droplet resting on the paper silanized by fluoroalkyltrichlorosilanes with different alkyl chain lengths. (b) Plots showing the correlation of the static water contact angle and contact angle hysteresis to different silanization reagents. Three types of paper with different roughness and porosity are used as the substrate for silanization, including blot paper, Whatman #1 and Whatman #50 paper. (c) SEM images showing the different surface textures on these three types of paper.[75]

Besides fabrics and papers, SHS can also be prepared on the foam surface with ease. Foams consisting of different materials, such as metals, polymers, and carbons, have been successfully fabricated into high performance SHS as a result of their porous microtextures which provide substantial air fraction to support the Cassie state.[76-78] Upon using a simple dip-coating process, SHS can be realized on the copper foam, and the ethanol solution of n-tetradecanoic acid was used as the coating agent.[79] This process is also versatile as the copper foam with different pore density can be transformed into superhydrophobicity (Figure 22). By combining both the characteristics of SHS and foam, the superhydrophobic Cu foam becomes "super-floatable" on water surface, which can be utilized in the load conveyance in water (Figure 23).

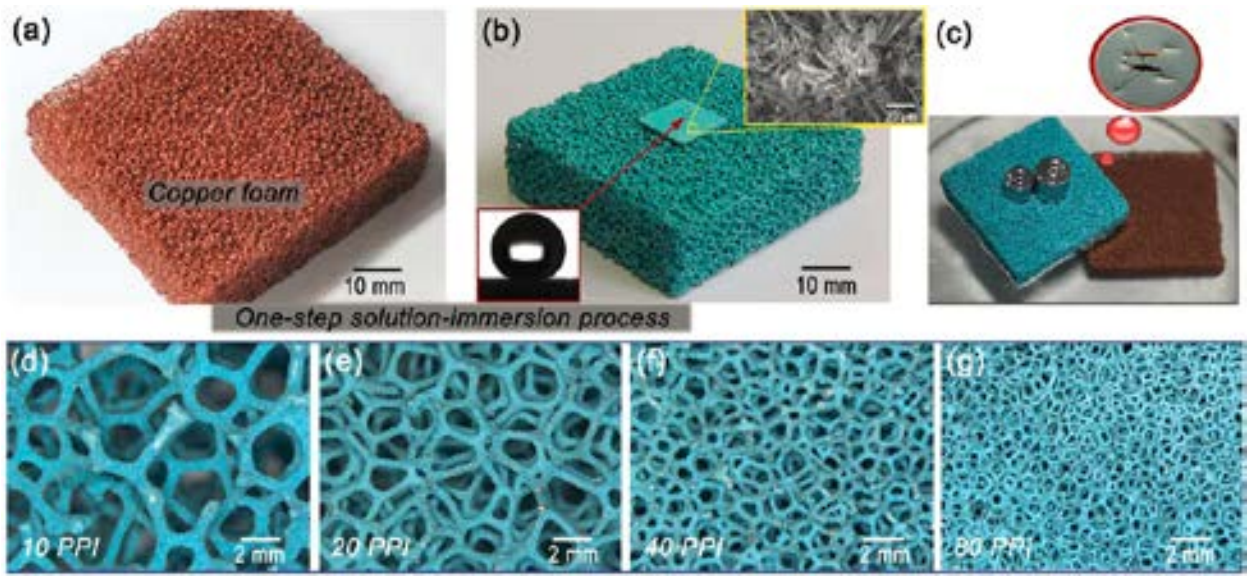

**Figure 22** (a, b) Optical images of a copper foam with 40 pores per inch (PPI) before (a) and after (b) the superhydrophobic coating by one-step solution immersion. Insets in (b) show the as-formed dual roughness structure on the foam surface and a water droplet resting on such a surface with high contact angle. (c) Photograph showing that the superhydrophobic foam can float on the surface of water with weight loading, assembling the water strider. (d-g) Optical images of the as-prepared superhydrophobic foams with different PPI (10-80).[79]

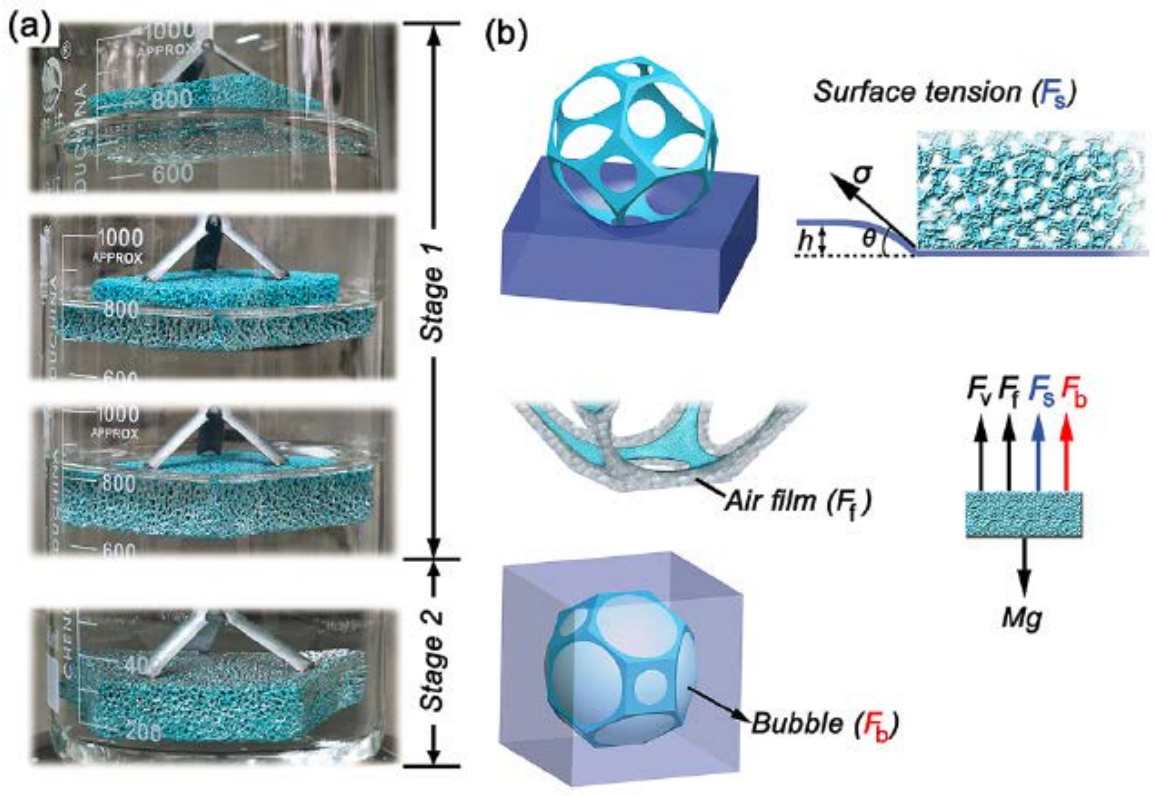

**Figure 23** (a) Analysis of the floating performance of a 40 PPI superhydrophobic foam in a two-stage approach: Partial immersion (stage 1) and completed immersion (stage 2). (b) A tetrakaidecahedron-like cell with eight regular hexagons and six square faces is used to model the structural unit of the foam.[79]

## 2.3 New concept based advanced superhydrophobic surface with

## substantial robustness

Foundation on the "dual roughness" concept, different materials and processes have been combined to obtain high performance SHS with low hysteresis and high WCA. With increasing demands over material and structural functions, the industries urgently need new SHS with enhanced hydrophobic properties, improved cost-effectiveness, and environmental friendliness. To achieve these goals, new concepts of SHS have been introduced, accompanied with the innovative processing techniques to bring about the SHS of present significances.

Different from the solid low surface energy coating used in the fabrication of SHS, a new concept of slippery liquid infused porous surfaces (SLIPS) have been proposed.[80] In this case, the porous surfaces, i.e. silanized epoxy nanopost arrays & Teflon nanofiber networks, were lubricated with the low surface energy liquids, such as 3M Fluorinert FC-70 ($\gamma$=17.1 mN $m^{-1}$). As the porous surfaces were completely wetted by the lubricating liquid, a robust and defect-free lubricant liquid surface could be formed on the porous surface. Unique omniphobic property was observed for the SLIPS as both water and organic solvent can be effectively prevented from wetting the underlying solid surface (Figure 24). For SLIPS to demonstrate its full potential, a surface silanization treatment to lower the surface energy is required. As shown in Figure 24b, the dyed pentane droplet replaced the lubricating film on the non-silanized epoxy within 10 s while remained stable on the silanized epoxy. More importantly, the SLIPS exhibited extremely high pressure stability which was equivalent to a hydrostatic pressure around 7 kilometers. This is probably the highest reported value for SHS since the liquid-air-solid interface of normal solid hydrophobic surfaces cannot be stable under such a high pressure. Due to the mobility of the lubricants at the solid-liquid interface, the SLIPS is also self-healable, while value-added multifunctionalities, including anti-staining, anti-icing, anti-adhesion, and optical transparency, are also demonstrated.

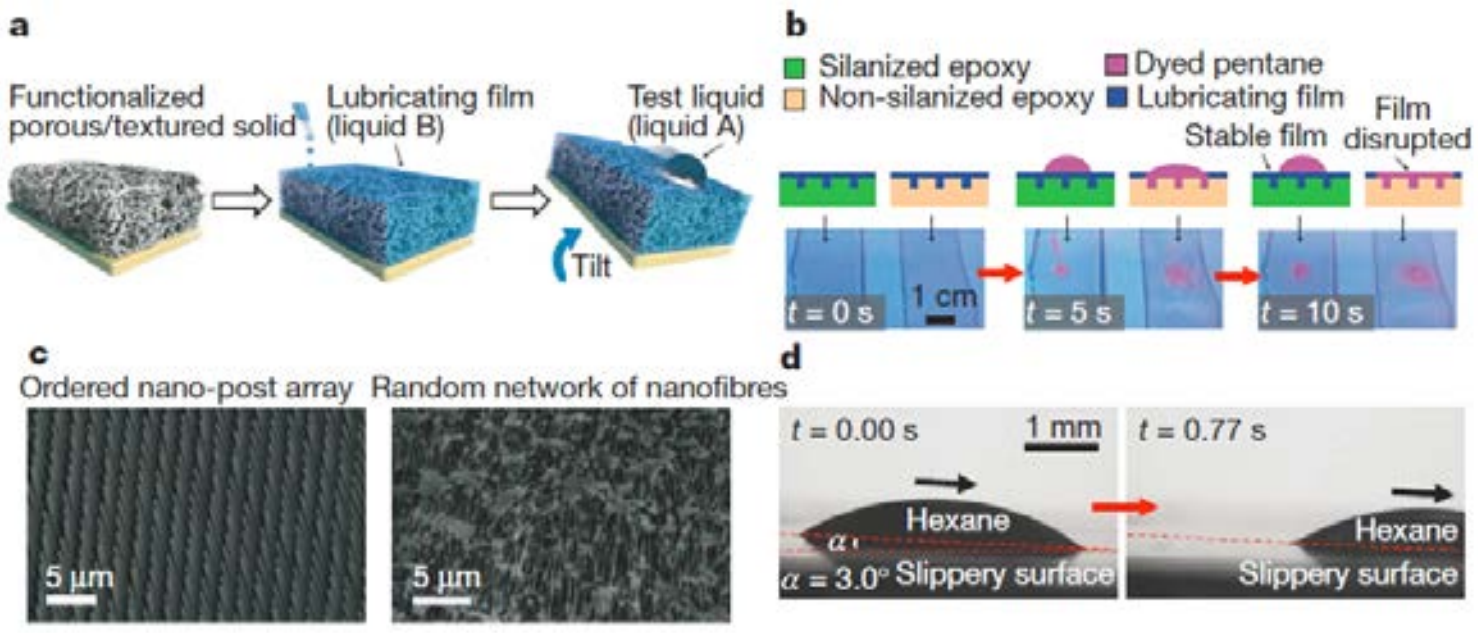

**Figure 24** (a) Schematic illustration of the fabrication process of SLIPS. A low surface energy liquid is imbibed into the functionalized porous or textured solid, which forms a physically smooth and chemically homogeneous lubricating film on the substrate surface. (b) Demonstration of the difference in the lubricating film stability for the silanized and non-silanized textured epoxy substrates. (c) SEM images of two representative porous/textured solid substrates, including the epoxy-based ordered nanopost array and the Teflon-based random nanofiber network. (d) Photographs demonstrating the sliding of a low surface energy liquid (hexane) on the SLIPS with a tilt angle as small as 3.0°.[80]

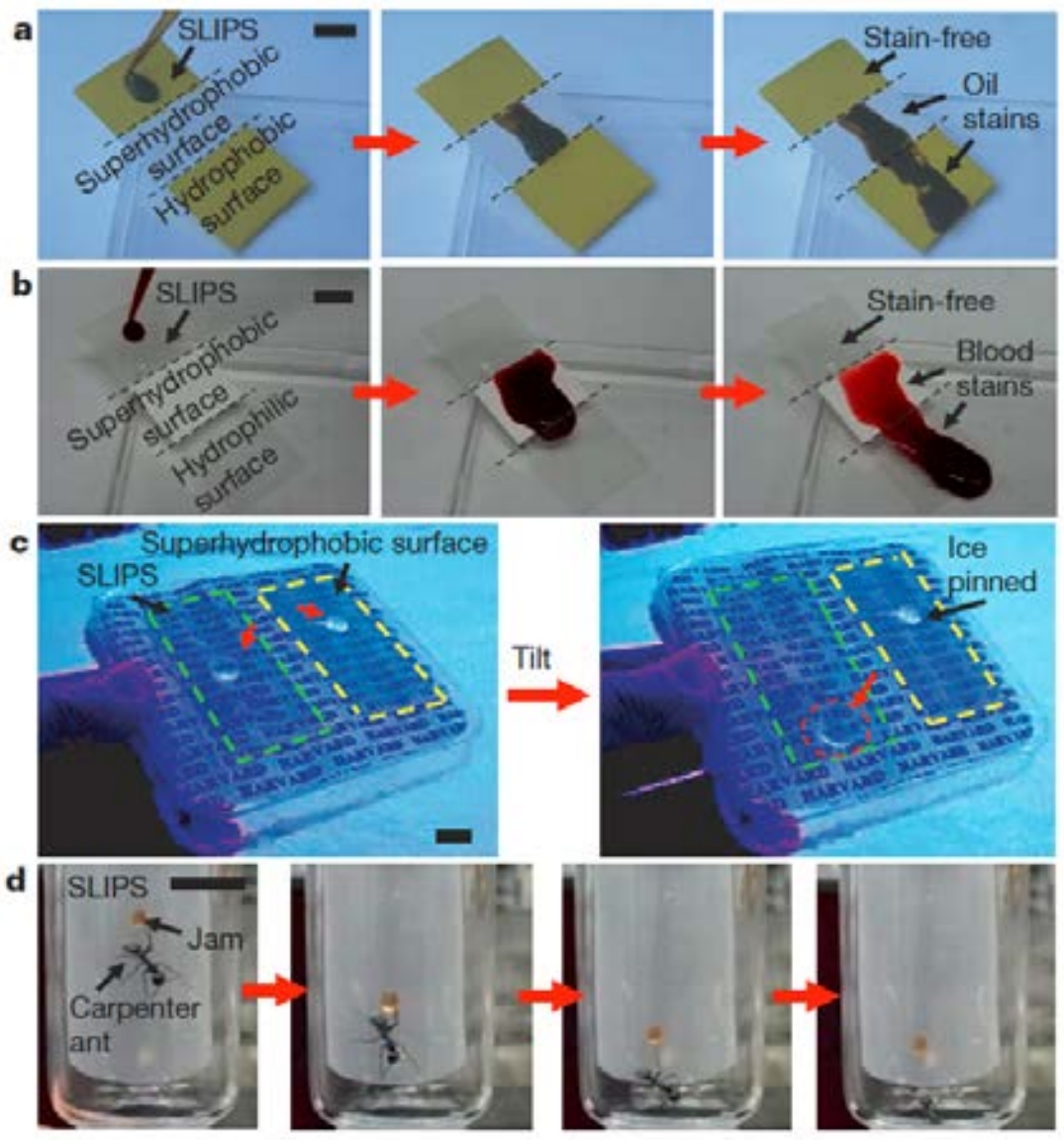

**Figure 25** (a, b) Photographs showing the movement of a crude oil drop (a) and a blood drop (b) on a substrate with integrated SLIPS, superhydrophobic, and hydrophobic regions. The SLIPS surpassed the other two surfaces in terms of the drop flow speed and anti-staining property. (c) Comparison between SLIPS and epoxy-based superhydrophobic surface on anti-icing. The ice particle swiftly slid off SLIPS while pinned on the epoxy-based SHS. (d) Sequential frames showing the sliding of a carpenter ant on the SLIPS upon tilting.[80]

The concept of SLIPS is definitely innovative and unprecedented super-repellent performance can be achieved by the formation of the highly mobile lubricant layer on the porous solid surface. However, it may still require the solid substrate to possess substantially low surface energy in order to maintain a stable lubricant layer without the replacement of other impinging liquids. By precisely manipulating the structures of surface texture, innovative super-repellent surface can even be obtained on the intrinsically hydrophilic substrate, e.g. silicon dioxide ($SiO_2$). Using a reactive ion etching (RIE) process, doubly reentrant structure was fabricated on $SiO_2$ surface.[81] Different from the reentrant structure, the doubly reentrant structure consisting of microposts with vertical nano-overhangs as side walls on the post top. The structural parameters, such as the post-top diameter (D), centre-to-centre distance (P), the nano-overhang thickness (t) and length ($\delta$), are engineered precisely to obtain the minimum solid fraction $f_s$ and the maximum liquid suspension ability (Figure 26). It was speculated that the doubly reentrant structure can suspend liquids even with a Young's contact angle $\theta_Y \approx 0°$, as substantial suspension force can be provided by the vertical components of the surface tension acted on the solid-liquid interface, which is generated by the unique nano-overhang structure. At the mean time, the simple micropost structure and the reentrant structure can only suspend liquids with $\theta_Y \approx 120°$ and $\theta_Y \approx 30°$, respectively (Figure 27). As a result, the doubly reentrant $SiO_2$ micropost arrays showed excellent super-repellent property, which can even resist the wetting of low surface energy liquids including 3M FC-72. On the other hand, as the super-repellency is solely based on the physical structure, and there is no need of polymer coating, the $SiO_2$ surface with doubly reentrant microposts also shows omniphobicity to non-volatile liquids at temperatures larger than 320°C, and demonstrates extremely temperature stability to even 1000°C storage.

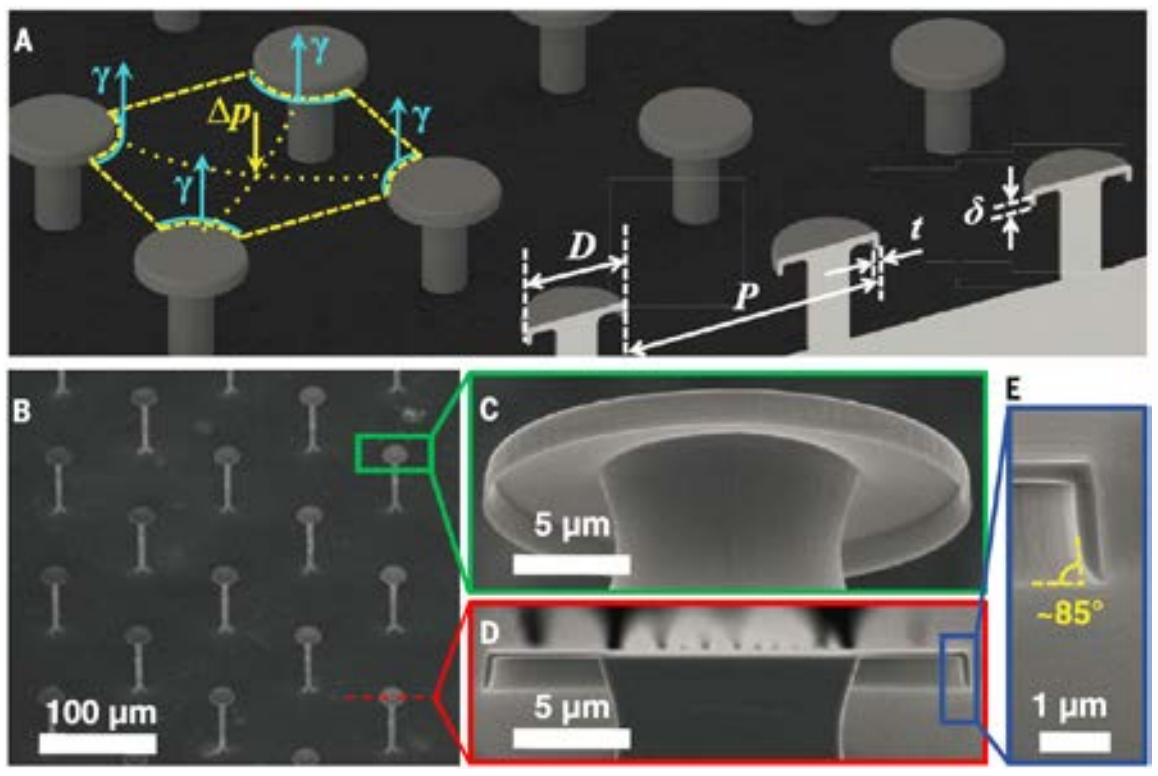

**Figure 26** (A) 3D model simulating the doubly reentrant surface structure design. The critical parameters are indexed in the model, including the post top diameter (D), center-to-center distance (P), and the length (δ) and thickness (t) of the vertical overhangs. (B, C, D) SEM images of the top-view (B), bottom-view (C), and cross-sectional view (D) of a micropost with doubly reentrant structure. (E) Magnified cross-sectional view of the overhang structure on the post edge.[81]

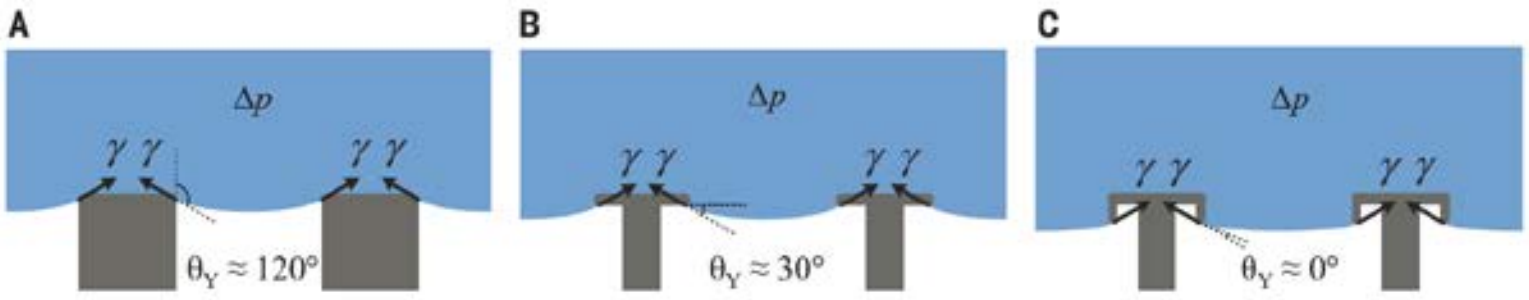

**Figure 27** Schematic illustration of liquid suspended on different surface structures, the pressure difference between the liquid and air is denoted as Δp. The simple micropost structure (A) can only suspend liquids with $\theta_Y > 90°$, while the reentrant structure can suspend liquid with $0° < \theta_Y < 90°$, but it will fail if $\theta_Y \approx 0°$. The doubly reentrant structure can suspend liquids with $\theta_Y \approx 0°$ as the substantially upward forces is provided by the surface tension acting on the vertical overhangs.[81]

The realization of the omniphobic surface which can remain super-repellent under extreme conditions has unraveled the functional potential of engineering the microscopic structures and compositions on solid substrates by nanotechnology. However, new insights to extend the functional performance of SHS beyond theoretical prediction have been given by manipulating the macrotextures of SHS.[82] As the microscopic roughness often exhibits macroscopic smoothness in the case of SHS, the creation of macroscopic surface texture might boost the super-repellent performance to a new height. By using the laser-ablation technique, SHS with "dual length scale textures" can be created, which shows remarkable de-wetting performance as the contact time of the impinging drop has been reduced significantly to below the theoretical limit.[83] The innovative macrotexture was introduced on a microtextured silicon surface coated with fluorosilane using laser ablation. The as-produced macrotextures obtain the characteristic full width at half maximum of 300 μm, heights comparable to or less than the impinging liquid film thickness, and lengths up to several millimeters. As the liquid drops impinge on the macrotextured superhydrophobic silicon surface, the ridge of the macrotexture can trigger the redistribution of the droplet mass and momentum, initiating a so-called center-assisted retraction process (Figure 28). Different from the axisymmetric retraction on smooth SHS, the non-axisymmetric center-assisted retraction on the macrotextured SHS results in significantly shortened retraction process, which in turn accelerates the bouncing process of the impinging drops considerably. To speak further, a 27% reduction in contact time compared to the theoretical limit on smooth SHS is obtained for the macrotextured silicone SHS. Nonetheless, this enhanced de-wetting process against impinging drops can effectively reduce the liquid residues remained on the SHS after dynamic contact, which is of great application potential in anti-freezing and prevents the ice formation.

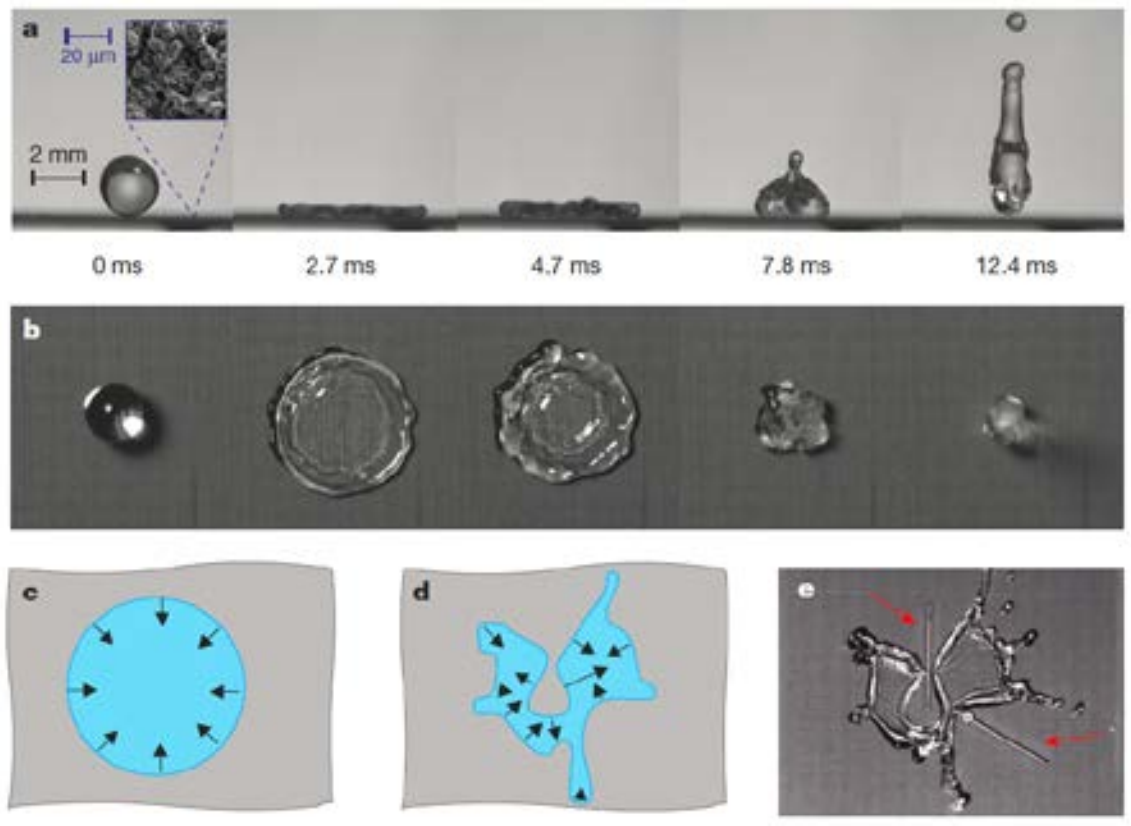

**Figure 28** (a, b) Sequential high speed frames showing the bouncing of a water droplet from a superhydrophobic silicon surface at the side-view (a) and top-view (b). Inset (a) shows the SEM image of the microtextured silicon surface. The axisymmetric recoil with nearly uniform retraction speed at the rim is observed in (b) and illustrated in (c). The non-axisymmetric recoil with divided retraction rims and additional retraction momentum in the center area is also illustrated in (d) and experimentally realized by the macrotextures in (e).[83]

However, the surface textures, which are the key to achieve SHS, may also become the first weakness to break it down. It is reasonable to envisage that the tiny surface textures on SHS can be easily damaged or removed by the external forces, which posts a critical drawback for the real-field applications of SHS. In order to address this issue, the concept of "robust" SHS has been proposed recently, which highlights the durability of SHS under repeatedly mechanical abrasion, stability under environmental contamination, and resistance to unfriendly external factors, i.e. elevated temperature, high pressure, and high humidity.[84-88] By using raspberry-like porous silica capsules, robust SHS with notable mechanical stability can be formed on the glass substrate.[89] In the first place, nano-silica was coated on the polystyrene (PS) microsphere template via the Stober method, forming the silica@PS capsules. Afterwards the silica@PS capsules were deposited on the amine-terminated glass plate by the electrostatic charge interaction. The porous silica capsules were then formed on the glass plate by the thermal degradation of the PS phase. Eventually a semi-silane coating was applied on to the porous capsules by CVD which not only made the capsules superhydrophobic, but also formed the silica bridges that cross-linked the adjacent capsules and the glass substrate (Figure 29). As a result of the chemical cross-linking provided by the silica bridges, the glass surface with the porous silica capsule coating exhibited robust superhydrophobicity which could withstand the peeling of the adhesive tape at a pressure of ~10 kPa, and remained stable after sand abrasion from an impact height of 30 cm. This novel robust SHS thus can be used in organic solar cells to expand their life cycles, which also shows high transparency and improves the total cell efficiency (Figure 30).

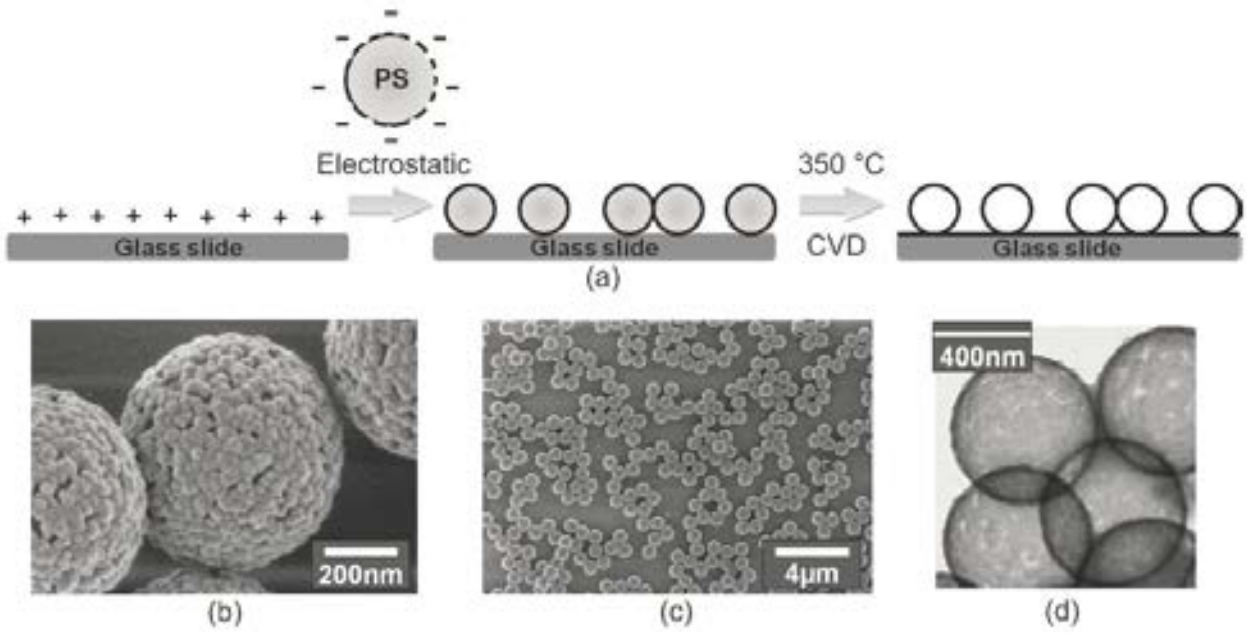

**Figure 29** (a) Schematic illustration of the preparation process for the transparent superhydrophobic surface composed of porous silica capsules on the glass substrate. (b, c) SEM images of the silica nanospheres coated polystyrene particles (b) and the glass slide surface covered with raspberry-like particles after the dip-coating process (c). (d) TEM image of the porous silica capsules after removal of the PS core.[89]

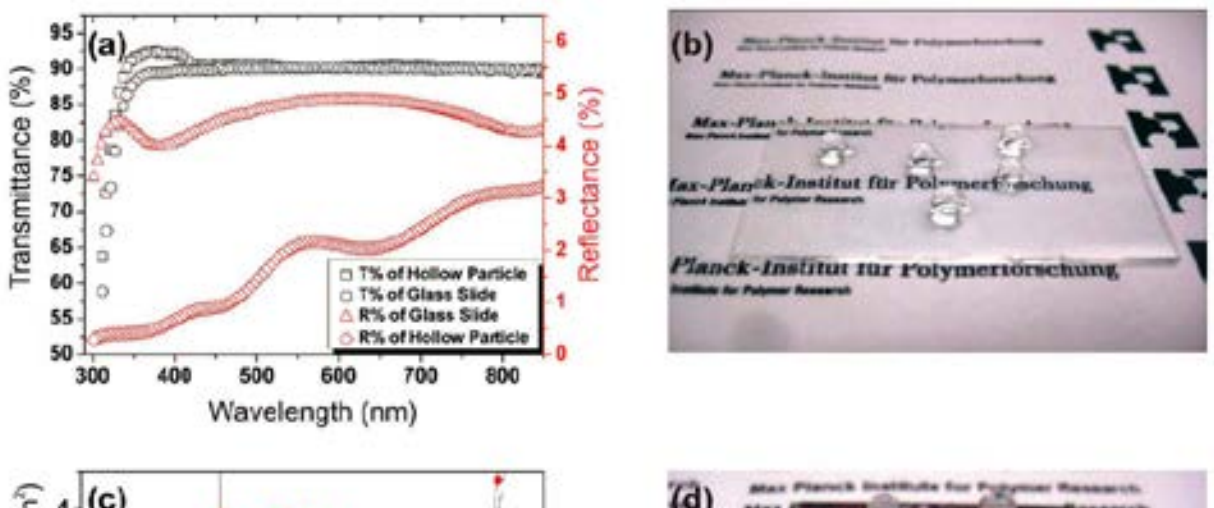

**Figure 30** (a) Comparison of the transmittance and reflectance spectra of the glass slide with and without the transparent superhydrophobic porous silica coating. (b) Photograph showing water droplets deposited on the glass substrate with the transparent superhydrophobic coating. (c) Comparison of the current-voltage curves between the organic solar cells with normal glass plate and glass plate with the superhydrophobic porous silica capsule coating. The inset shows the configuration of the organic solar cell. (d) Photograph showing the water droplet resting on the organic solar cell after the efficiency test, as the test did not affect the transparent superhydrophobic surface.[89]

Besides establishing the chemical bonding between the nanostructured coating and the substrate, robust SHS can also be achieved through engineering the adhesion of the coating to the substrate. Different processing techniques, such as PDMS composite coating, thermal compression, and surface pretreatment have been developed to create SHS with strongly adhering features.[90-92] A representative case is the creation of highly durable SHS on the cotton fabric surface. Using a simple dip-coating-curing process, FAS functionalized silica nanoparticles (NPs) were immobilized on the cotton fabric surface by the cross-linking networks of PDMS and FAS (Figure 31).[93] The novel composite coating exhibited excellent superhydrophobicity as well as superior mechanical durability. In fact, the coated fabric maintained high contact angles (>160°) and low sliding angles (~5°) even after 500 washing cycles and 10000 abrasion cycles under standard AATCC testing environment; and it also demonstrated good stability against boiling for 5 hours (Figure 32). It was speculated that a tire rubber-like resilient composite film was formed by the coating process, which provided substantial mechanical strength and substrate adhesion to resist the repeatedly mechanical deformation.

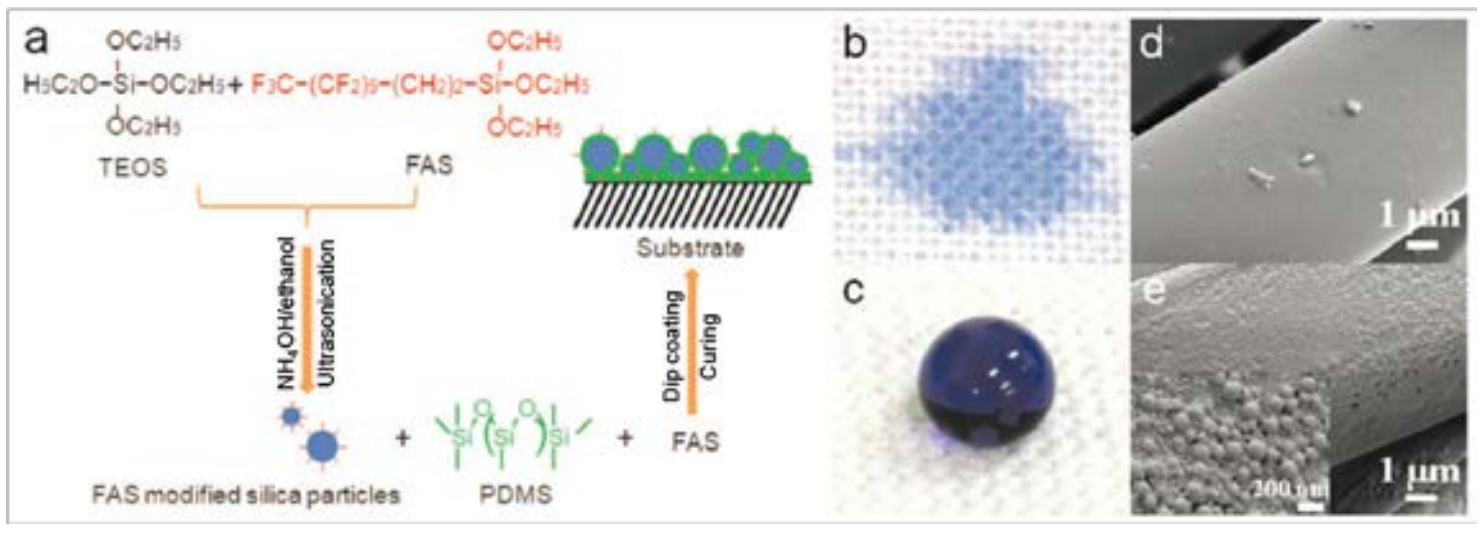

**Figure 31** (a) Schematic illustration of the fabrication process of applying the robust superhydrophobic coating on the fabric surface. (b, c) Photographs of the spreading (b) and stable (c) water droplets on the untreated (b) and superhydrophobic (c) fabrics' surface. (d, e) SEM images of a bare polyester fabric surface (d) and a silica/PDMS/FAS coated polyester fabric surface (e).[93]

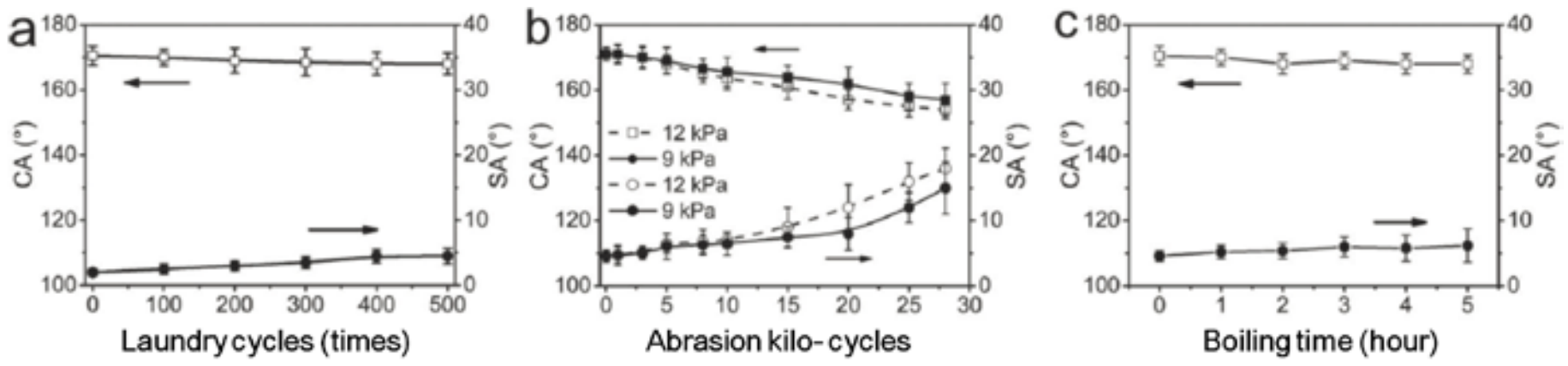

**Figure 32** The changes of water contact angle and sliding angle on the silica/PDMS/FAS coated superhydrophobic fabric surface as a function of washing cycles (a), abrasion cycles (b) and boiling time (c).[93]

Aside from improving the strength and adhesion of the microtextures to the substrate surface by chemical and physical bonds, robust SHS with mechanical stability can also be fabricated by using the self-healing or sacrificial surface components.[94-96] By combining a microfabrication and injection molding methods, robust SHS can be obtained on the polymer surface in a cost-effectively manner.[97] In this approach, microtextures were generated on the aluminum foil by a micro-working robot. The foil was subsequently made into the mold for injection molding and the microtextures were transferred onto the forming polypropylene (PP) surface. Different from homogeneous microtextured SHS, protective micropillars were designed and incorporated into the as-formed PP surface as periodic, large microhumps (Figure 33). Being higher and larger than the other microtextures, the microhumps would be more susceptible to the external compression and abrasion, which served as the sacrificial protection for the major microtextures. As a result of the presence of the protective micropillars, the microtextured PP surface showed robust superhydrophobicity with high WCA and low SA even under a compression pressure of 20 MPa and also survived in the wearing test up to 120 kPa, demonstrating notable potential for the real-field application.

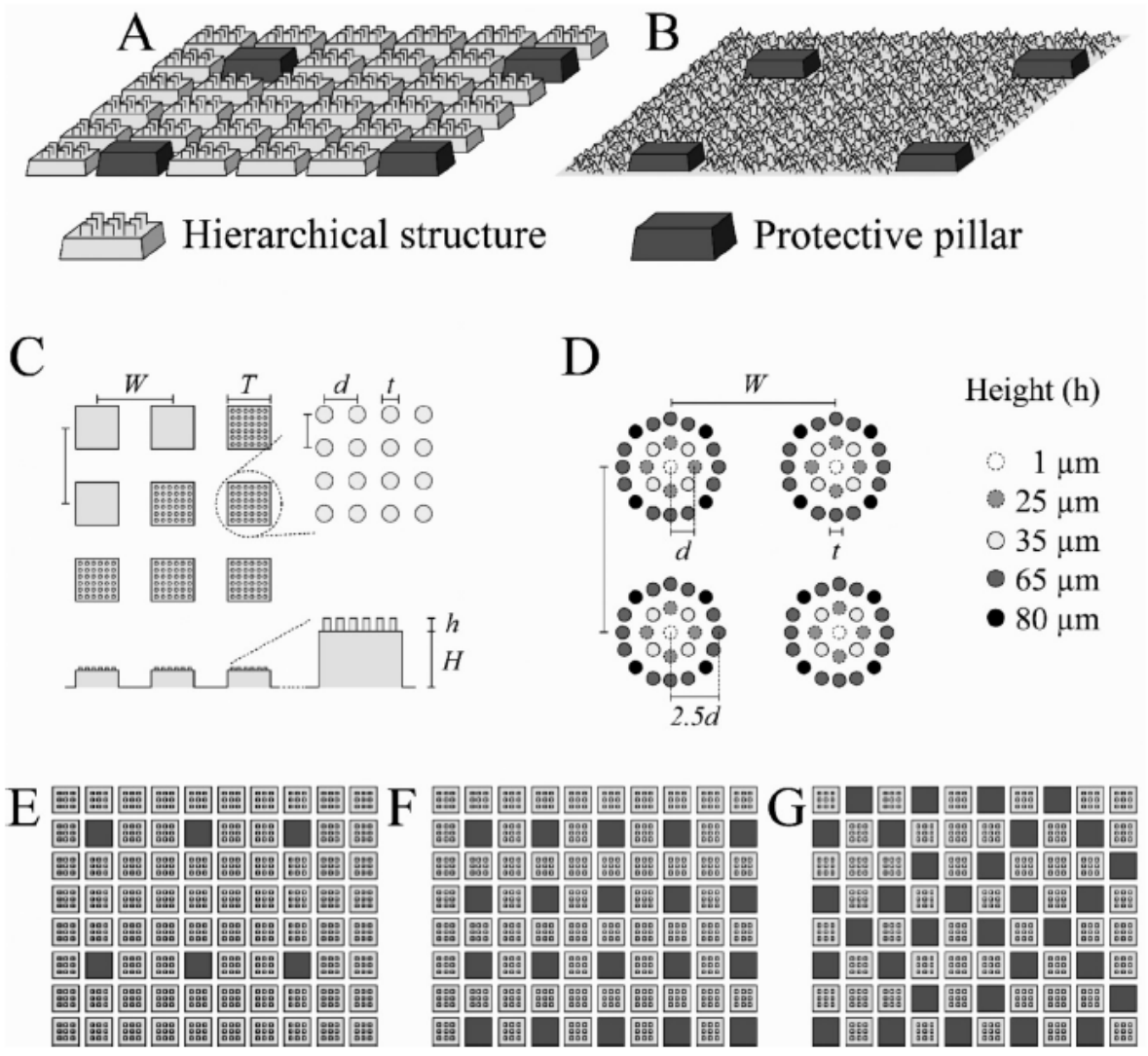

**Figure 33** (A, B) Schematic illustration of the designed superhydrophobic

structures on polymer surface with (A) the textures consisting of hierarchically nanostructured micropillar and protective pillar; and (B) the textures consisting of multiscale structure and protective pillar. (C, D) Design parameters of the hierarchical micropillar (C) and the multiscale structure (D). (E, F, G) Schematics showing the number and positions of the sacrificial protective pillars denoted as (E) p5, (F) p15 and (G) p25.[97]

All in all, there is no universal protocol of SHS which would fit in all the application requirements. By learning from the extensive research ouputs on the SHS with various hydrophobic characteristics, one should realize that the design and fabrication of SHS are rather end-use specified. Based on the current understanding on the relationship between SHS and microtextures, and to obtain SHS with optimized engineering, economic, and environmental impacts, the designer of SHS should closely link the core functions with the real-field application requirements, and reversely engineers the structure and property of the SHS through a function-led approach. Nonetheless, it is also essential to discover new applications for the SHS, where new functional potential of the artificial super-repellent surface can be unravelled.

## 2.4 Significance of the research

(i) Easy tunable wettability. Different from the SHS with fixed hydrophobicity, the wettability of the as-obtained graphene-titania (GT) SHS can be tuned from superhydrophilicity to superhydrophobicity repeatedly in a short time, using UV light as the stimulus. The versatility in wettability thus can serve as the basis for new functions and applications.

(ii) Interactive mechanism study. The mechanism of the UV-triggered superhydrophilic conversion on the anatase $TiO_2$ surface was investigated in the microscopic level. Observation on the conversion process was monitored from the lattice re-configuration to the surface reconstruction. The role of graphene in the conversion and its interaction with $TiO_2$ were also studied, which provided new theory for the functional optimization on the $TiO_2$ superhydrophilic conversion.

(iii) New applications for SHS. The potential of the as-obtained GT-SHS in the microfluidic manipulation was unravelled, which showed promising functional performance in droplet storage, transfer, and microfluidic sensing. Due to the easy-tunable wettability, surface energy engineering can also be applied to the GT-SHS to realize the unique one-way water transport process, as well as high efficient oily water separation.

# Chapter 3. Methodology

## 3.1 Materials

The graphite powder (98.0%), phosphorus (V) oxide (98.0%), and hydrogen peroxide (30%) were purchased from Uni-Chem. Potassium persulfate (99+%) was purchased from Acros Organics. Hydrazine monohydrate (99+%) was purchased from Alfa Aesar. And titanium tetraisopropoxide was purchased from Tokyo Chemical Industry (product number T0133). All chemicals were used as received without further purification. Cotton twill fabrics were used as the substrate for coating during the experimental processes.

## 3.2 Synthesis of graphene oxide

The graphene oxide was synthesized using a modified Hummers' method.[98] In a typical process, 25 g of potassium persulfate ($K_2S_2O_8$) and 2.5 g of phosphorus pentoxide ($P_2O_5$) were carefully weighted and subsequently added into the 12 mL 95% concentrated sulfuric acid ($H_2SO_4$) at 80°C under magnetic stirring. The mixture was continued stirring for half an hour to achieve homogeneous and afterwards 3 g of graphite powder was slowly added into the mixture. The graphite powder reacted strongly with the oxidative mixture and a black slurry was formed shortly. The black slurry was continued stirring at 80°C for 4.5 hours and became viscose. The slurry was then transferred to a 1000 mL glass beaker and subsequently 500 mL deionized (D. I.) water was added under magnetic stirring. After achieving homogeneous, the dispersion was filtered and washed with copious amount of D. I. water for several times in order to remove the unreacted impurities. The as-obtained carbon cake was dried in ambient overnight.

The carbon cake was then exfoliated to form graphene oxide using the Hummers' method. In a typical process, the carbon cake was dispersed in 120 mL concentrated sulfuric acid (95%) at 0°C ice bath under magnetic stirring. Afterwards 15 g potassium permanganate ($KMnO_4$) was slowly added into the dispersion and the reaction temperature was maintained below 20°C. The dispersion was then continued stirring at 35°C for 2 hours and 250 mL D.I. water was subsequently added to dilute the viscose dispersion. During the addition of D.I. water, the reaction temperature was maintained below 25°C. The dispersion was continued stirring for 2 hours and then 700 mL D. I. water was added. After 2 minutes of the water addition and as the dispersion became homogeneous, 20 mL 30% hydrogen peroxide ($H_2O_2$) was added and the dispersion abruptly turned brilliant yellow indicating the completion of the reaction. Intensive bubbling was also observed when $H_2O_2$ was added, in this case, the addition of $H_2O_2$ should be slow to keep the reaction stable.

The as-obtained yellowish dispersion was filtered and washed with copious amount (~1L) of 10% hydrochloric acid (HCl) and D. I. water. The as-obtained solid product was dried in ambient and re-dispersed in D. I. water to form a 0.5 wt% graphene oxide (GO) dispersion. The GO dispersion was subsequently dialyzed for one week to remove the remained impurities.

## 3.3 Synthesis of titania nanosol

The titania nanosol-gel was synthesized according to the previous reported method.[99] In a typical process, 300 mL titanium tetraisopropoxide ($Ti[OCH(CH_3)_2]_4$) was drop-wisely added into 700 mL aqueous solution containing 0.2% HCl and 0.2% acetic acid under magnetic stirring. Afterwards, the solution temperature was raised to 60°C and the hydrolysis reaction was proceeded for additional 16 hours. The nanosol was formed as a result of the generation of $TiO_2$ nanoparticle and the evaporation of water. Subsequently the as-obtained titania nanosol was aged at ambient for one week prior to use.

## 3.4 Microstructure characterization

The microstructures of the as-synthesized graphene and $TiO_2$ nanoparticles (NPs) were characterized by scanning electron microscope (SEM, JSM-6490, JEOL) and transmission electron microscope (TEM, JEM-2100F, JEOL). The surface morphologies of the as-synthesized graphene and $TiO_2$ NPs were investigated using SEM and their corresponding particle size, crystalline structure and lattice information were analyzed by TEM. To prepare the TEM sample, the graphene or $TiO_2$ nanosol dispersions was drop-casted onto the copper grid and dried under ambient condition.

The surface topography of the as-prepared graphene and $TiO_2$ thin films was measured by atomic force microscope (AFM, Nanoscope IV, Digital Instrument). To prepare the samples for AFM characterization, a 1×1 cm silicon wafer was pre-cleaned by sequentially sonicating in D.I. water and ethanol, respectively. Afterwards, the diluted dispersions of graphene or $TiO_2$ nanosol were drop-casted on the silicon wafer and dried in ambient, forming a relatively homogeneous thin film on the wafer's surface.

The crystalline structure and phase information of the samples were measured using X-ray diffractometry (XRD, Rigaku SmartLab). The sample in powder form was placed in the center of the holder and pressed to be flat using a glass slide.

## 3.5 Surface property characterization

The surface elemental composition was characterized using energy dispersive X-ray spectroscopy (EDX). High resolution SEM image was firstly obtained and then the EDX mapping was performed on the imaging area of the sample. The type and abundance of the elements on the surface area of the sample were subsequently obtained, which were shown as the colored pixels distributed on the imaging area.

The chemical bonding and energy state of the surface elements were further characterized by X-ray photoelectron spectroscopy (XPS, SKL-12, Leybold Heraeus). The sample for XPS characterization was prepared by drop-casting the diluted sampling dispersion onto a pre-cleaned silicon wafer, which subsequently formed a homogeneous thin film on the wafer surface.

## 3.6 Contact angle measurement

The water contact angle (WCA) was measured by a contact angle measurement and contour analysis system (OCA-15 plus, Dataphysics). A droplet volume of 2 μL was used for the WCA measurement. Each reported value of WCA was the average obtained from measuring at least five different locations on the solid surface. In order to measure the WCA accurately, the measurements were conducted after the as-deposited water droplet stabilized on the surface. For the samples without ultraviolet (UV) treatment, the stabilizing time is set as 5 minutes; for the samples with UV treatment, the stabilizing time is set as 5 seconds.

The water spreading and transport measurements were conducted on the hierarchical graphene/titania coated cotton fabric (GT-fabric). The time required for spreading and transport was defined as the completely disappearance of a 2 μL water droplet on the GT-fabric surface, despite the different capillary processes that may take place on the droplet. The life time of the water droplet on GT-fabric surface was monitored by a charge-coupled device (CCD) camera and each reported value was the average of three parallel measurements.

## 3.7 UV-induced superhydrophilic conversion on GT fabric

The as-prepared GT-fabric was mounted onto the center of a pre-cleaned glass slide and placed right beneath the UV lamp in a UV exposure box. Subsequently the GT-fabric was radiated by the UV lamp with a UV wavelength of 365 nm and intensity of 0.457 mW $cm^{-2}$. To control the wettability on the GT-fabric, different durations of UV radiation were applied, i.e. 1, 2 and 3 hours. After the completion of the UV treatment, the GT-fabric was taken out of the UV box and cooled in the ambient condition.

## 3.8 Fabrication of GT fabric based micro-droplet sensor arrays for gas sensing

The GT-fabric was curled and attached to the end of a glass rod by double-sided bonding tape. Afterwards, the droplet array with pre-defined spacing was deposited on the fabric surface using a micro-needle. In a typical process, a 3×3 water droplet array was fabricated and the volume of individual droplet was controlled to be 10 μL. In order to realize the sensory characteristic, different indicators were loaded into the as-deposited water droplets, which could exhibit special chromatic changes upon exposing to specific gas molecules. In this case, 0.05 M $CuCl_2$, 0.1 wt% Congo red, and 0.01 M $PdCl_2$ were used as the indicator solutions, which showed distinct chromatic responses to ammonia ($NH_3$), HCl, and ethylene diamine molecules, respectively. Quantitative sensing analysis can also be achieved on the micro-droplet sensing arrays by exposing them to different concentrations of gas, which may show a sequential chromatic change from low to high gas concentrations.

## 3.9 Oil/water separation test on GT-fabric

The GT-fabric was mounted onto the filter head of a vacuum filtration device (Jin Teng sand core filter with portable core) and used as the filtration membrane. Different oil/water mixtures, i.e. chloroform/water, hexane/water, dodecane/water, were fed into the filtration device and separated by the GT-fabric membrane using gravity. The volume ratio between the oil and water phase in the oil/water mixture was set as 1:1 and the time required for the completion of separation was also recorded using an electrical chronograph. The separation efficiency (R) was calculated using the following equation: $R=(W_i/W_0)\times100\%$, while $W_0$ and $W_i$ are the weight of the originally fed and finally separated oil, respectively.

# Chapter 4. Surface energy engineering on cotton fabric using graphene

## 4.1 Introduction

Graphene, the recent "superstar" of materials, has drawn extensive research interests as a result of its superior mechanical, electrical, and thermal properties. Furthermore, graphene has also been recognized as one of the most active materials in terms of surface chemistry. And the surface wettability of graphene can be easily tuned at either the molecular or atomic levels by controlling the crumble distribution on the graphene sheet and the absorption of molecules with different polarities. The combination of the intrinsically hydrophilic cotton fabric and hydrophobic graphene is expected to generate enormous functional possibilities. As enhanced adhesion would be established between the micro-meter sized cotton fibers and the extremely flexible graphene sheets with nanodimensional thickness, the wettability of the underlying fiber matrix can be strongly affected by the change of surface composition on the thin graphene coating. Furthermore, functional upgrades can be achieved by the nanoengineering of the graphene surface via nanoparticle deposition or chemical functionalization.

## 4.2 Fabrication of graphene nanocoating on cotton fabric

The as-obtained GO was dispersed in D.I. water and sonicated to make a 3 mg $mL^{-1}$ dispersion. Cotton fabric with defined dimensions (2×2 cm) was immersed in the GO dispersion for a short duration (~30 s). Subsequently the GO coated fabric was transferred to an aqueous solution of hydrazine and the temperature was elevated to 80°C to trigger the reduction process. Mild stirring was applied during the reduction in order to achieve homogeneous. As the reduction continued, it could be clearly observed that the yellowish brown GO absorbed on the fabric turned into deep black, indicating the formation of graphene. The reduction was proceeded for additional 1 hour as the temperature reached 80°C, then the as-obtained graphene coated cotton fabric (G-fabric) was taken out from the solution and rinsed with copious amount of D.I. water. The preparation process of G-fabric is schematically illustrated in Figure 34.

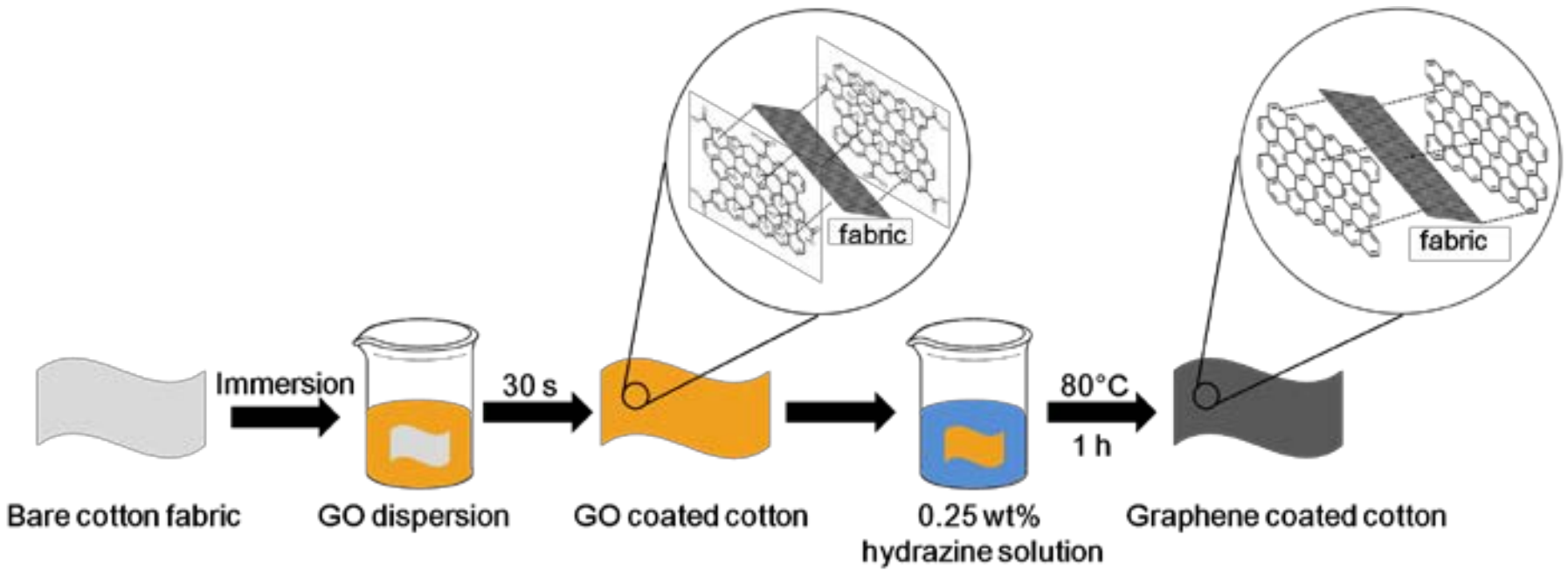

**Figure 34** Schematic illustration of the G-fabric fabrication process.

To further characterize the as-obtained graphene, the black dispersion remained in solution was taken out and re-dispersed in D.I. water. Approximately 1 mL of the dispersion was acquired using a syringe and re-dispersed in 4 mL D. I. water. Afterwards, the diluted dispersion was drop-casted onto a copper grid for the transmission electron microscope (TEM) characterization. As a result, graphene sheets with well-defined morphology can be observed in the TEM image (Figure 35); the selected-area electron diffraction (SAED) pattern of the graphene sheets also reveals good crystallinity of the as-obtained graphene, as typical hexagonal diffraction rings of the $sp^2$ carbon basal plane is preserved (Figure 35: inset).[100] The characteristic atomically thick edges of graphene can also be observed, as indicated by the semi-transparent contrast along the edge profile (Figure 36). The as-obtained graphene exhibited a few-layer characteristic with a thickness below 10 nm, indicating the quantum size effect was well-preserved regardless of the synthesis simplicity (Figure 37).

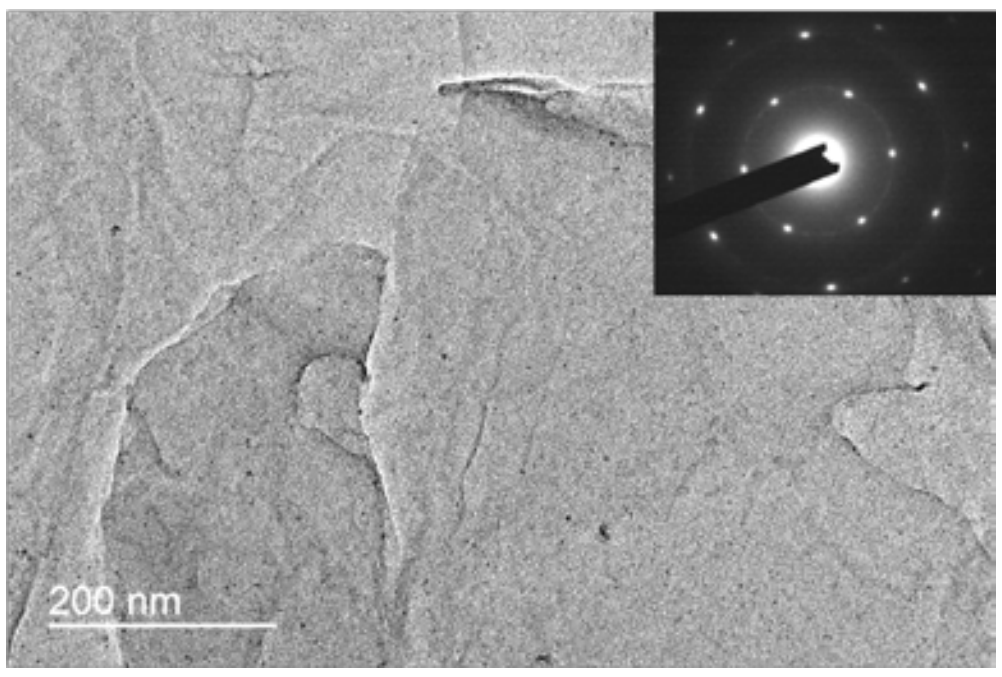

**Figure 35** TEM image of the as-obtained graphene. Inset shows the correspond SAED pattern.

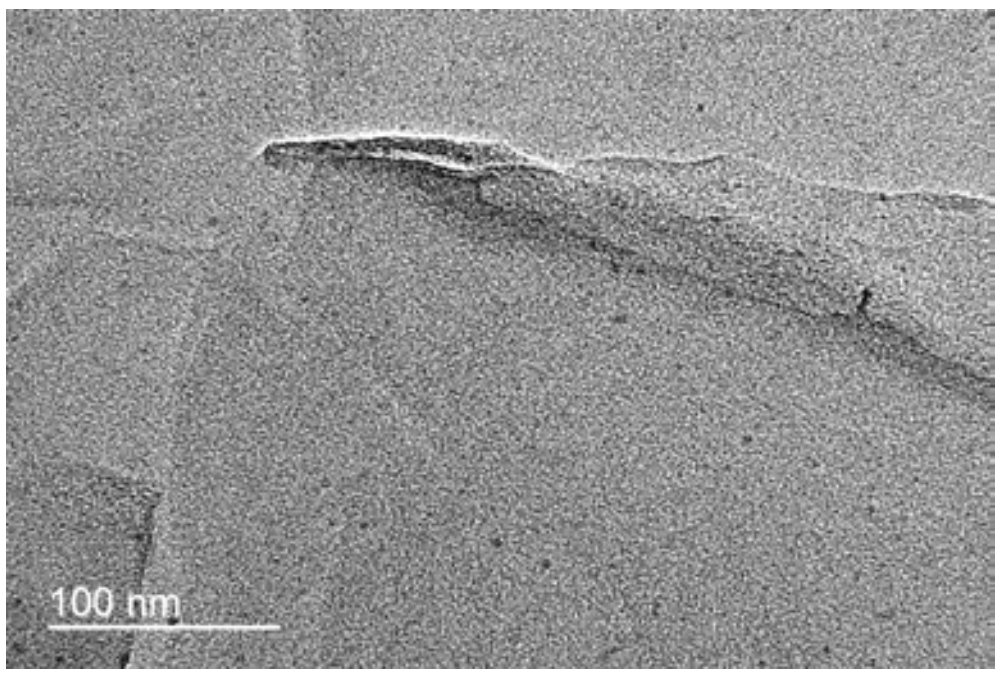

**Figure 36** Magnified TEM image of the as-obtained graphene with well-defined atomically morphology at the edge.

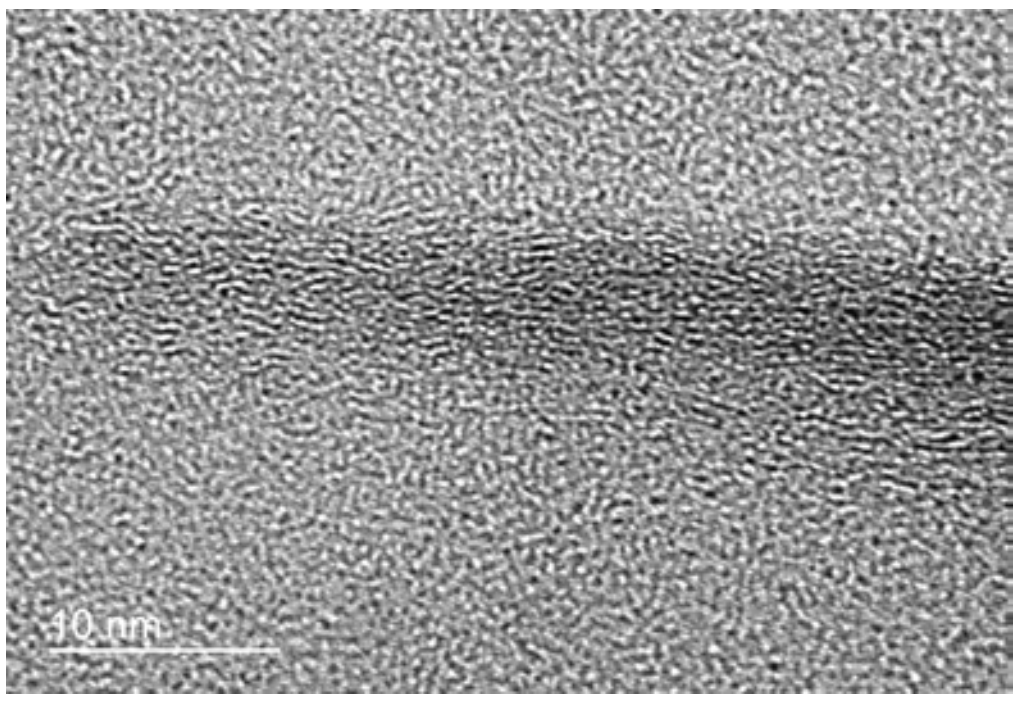

**Figure 37** High resolution TEM (HRTEM) image of the edge region in graphene. The edge is observed to be composed of less than ten stacked graphene layers, indicating the extremely small thickness of the as-obtained graphene.

The surface topography of the as-obtained graphene was also characterized by atomic force microscope (AFM). The AFM sample was prepared by drop-casting of 0.05 mL of the graphene dispersion obtained after the reduction process. The AFM image showed the topological structure of randomly oriented wrinkles on smooth thin sheets, which was in good accordance with the TEM observation (Figure 38). By re-constructing the AFM image, a 3D model on the graphene sheet can be obtained, by which the shape of graphene was clearly visualized (Figure 39). The X-ray diffraction (XRD) pattern of the as-obtained graphene showed two characteristic peaks at 11.22° and 22.6°, which can be indexed to the oxygen-intercalated graphene with a d-spacing of 0.8 nm and graphene with a d-spacing of 0.4 nm, respectively (Figure 40).[101] It thus indicated that the reduction process did not fully-reduced the GO into graphene. However, this phenomenon may not affect the wettability of the as-obtained graphene, which is discussed in section 4.3.

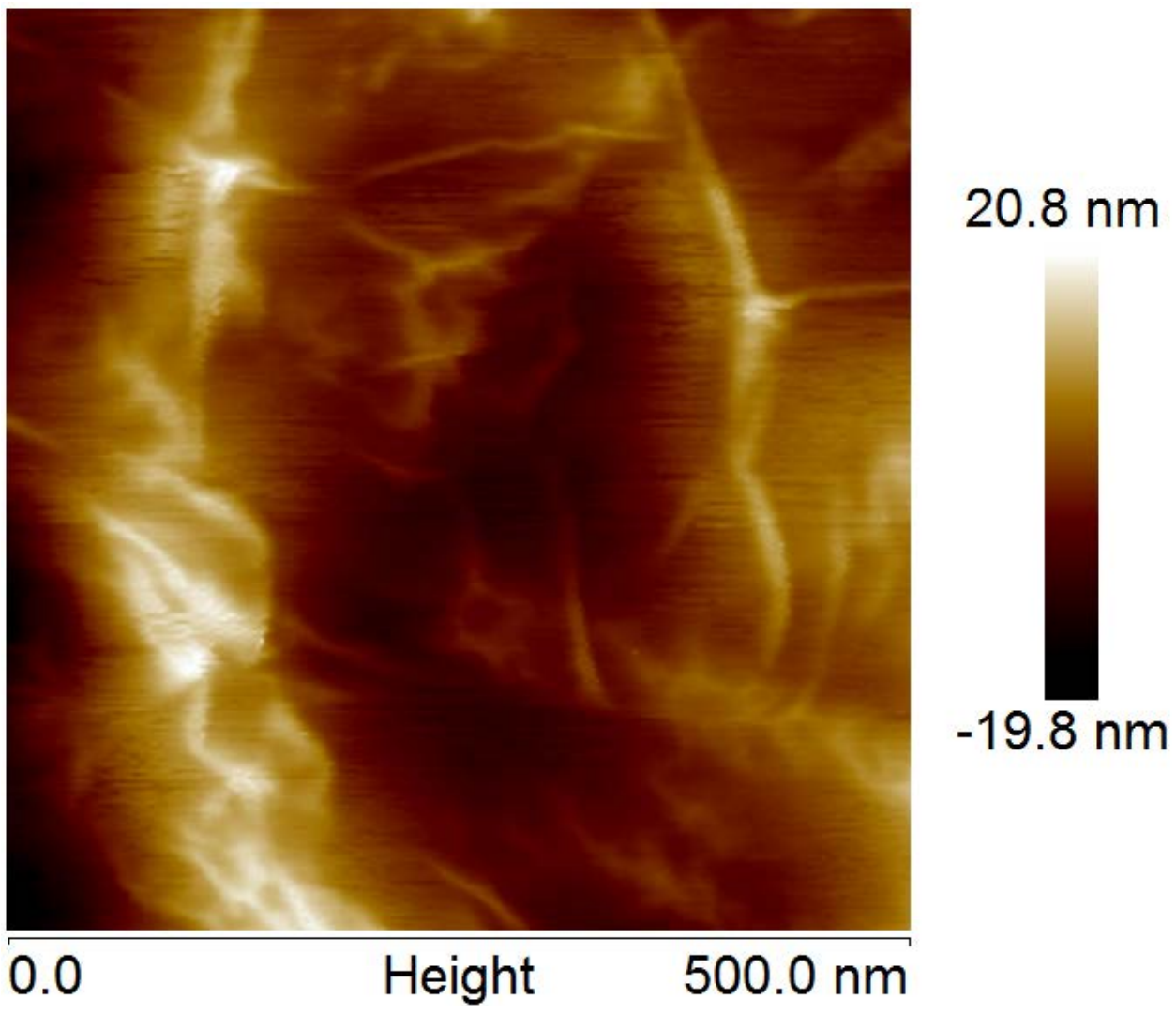

**Figure 38** AFM image of the as-obtained graphene, where the characteristic crumples of graphene sheets can be observed in the nanoscale level.

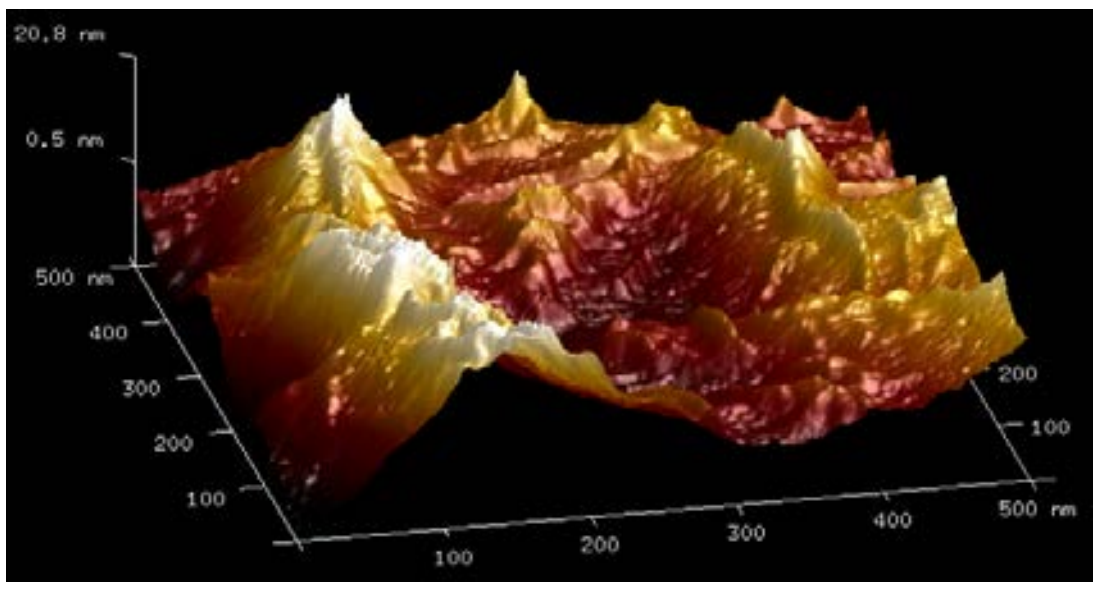

**Figure 39** 3D model of the graphene sheet constructed from the AFM image.

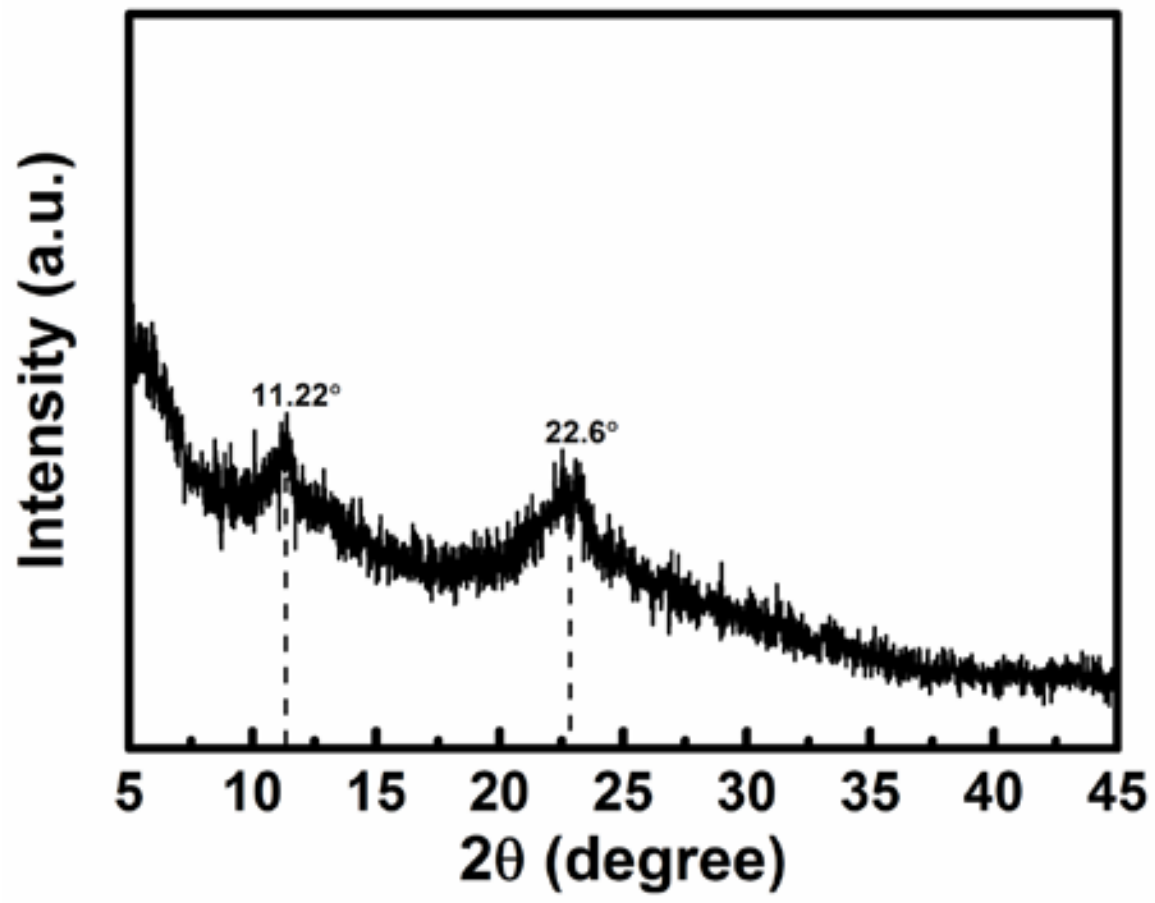

**Figure 40** XRD spectrum of the as-obtained graphene, showing two characteristic diffraction peaks at 11.22° and 22.6°.

The GO coating on the cotton fabric surface was directly observed using scanning electron microscope (SEM). It was found that the cotton fiber surface was covered by a single layer composed of large GO sheets which can span over tens of microns (Figure 41). The GO layer was found strongly adhere to the fiber surface which followed and duplicated the surface texture of the fibers (Figure 42). The surface morphology of the G-fabric was subsequently characterized by SEM. From the SEM image of G-fabric, the graphene coating can be observed as the rough "raincoat" layer on the cotton fiber surface (Figure 43). The as-formed graphene coating adhered strongly to the fiber surface which also followed the surface curvature of the fiber. The coating was found to exist on most of the fibers, indicating the effectively wrapping of graphene on the fabric during the reduction. From the magnified view of the coating, it was found that the fish-scale-like structure was consisting of nanoscale wrinkles which seemed to orient along the longitudinal direction of the fibers (Figure 44). It was speculated that this unique microstructure was formed during the reduction, as the oxygen atoms were released from the carbon basal planes of GO, the plane contracted and induced the formation of the wrinkle-like microstructure. The surface morphology of bare cotton fabric was also shown, which was rather smooth and flat as compared to the G-fabric (Figure 45).

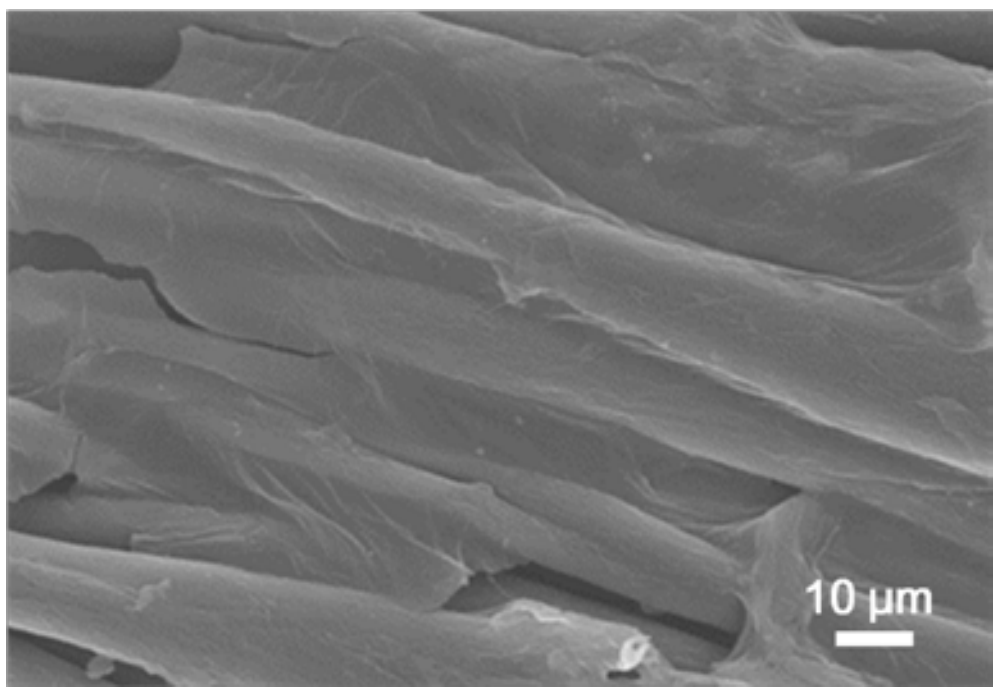

**Figure 41** SEM image of the graphene oxide (GO) coated cotton fabric.

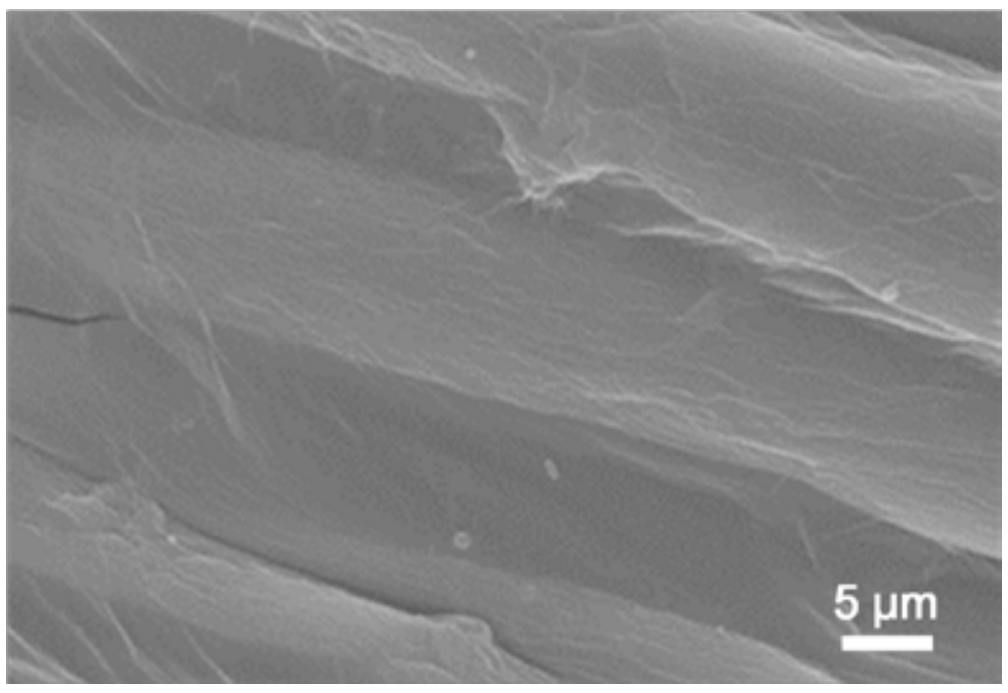

**Figure 42** Magnified SEM image of the GO coated cotton fabric, showing the strong adhesion of the GO sheet to the fiber surface.

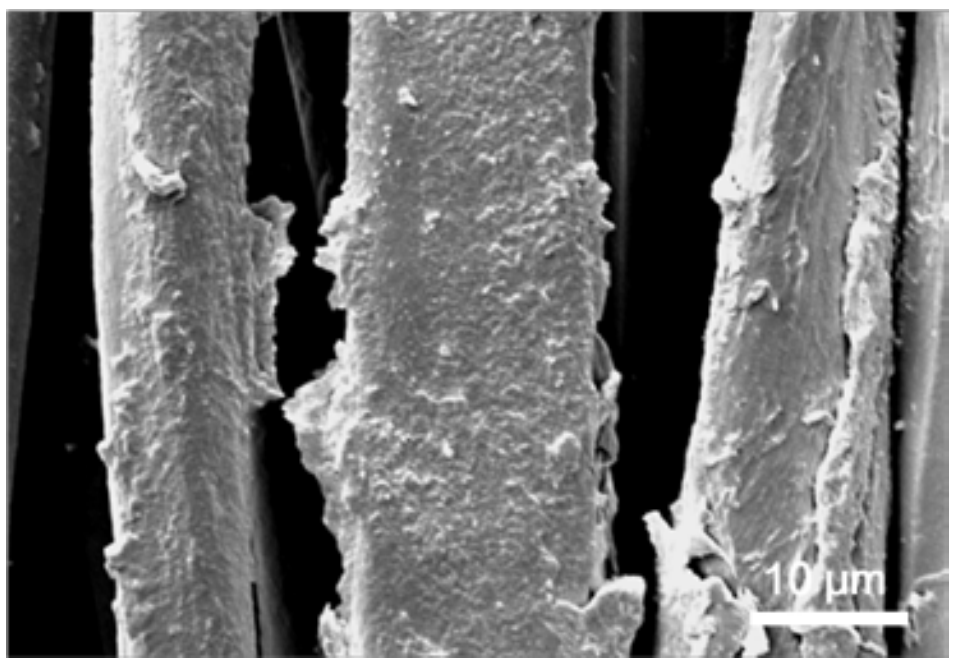

**Figure 43** SEM image of the G-fabric, showing the rough fish-scale-like surface coating.

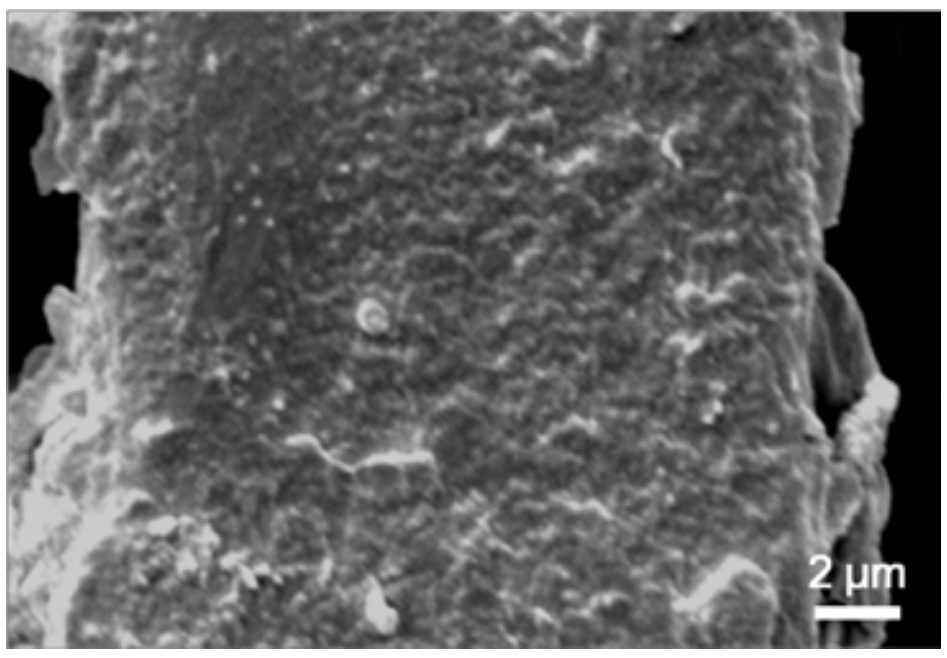

**Figure 44** Magnified SEM image of G-fabric's surface, where the orientation of the fish-scale-like structure can be observed.

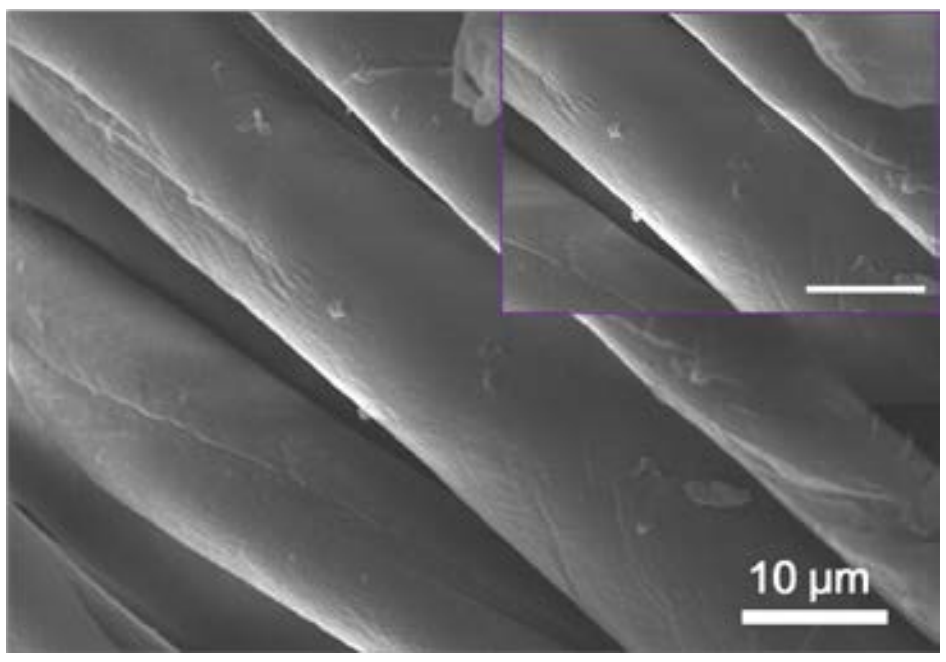

**Figure 45** SEM image of bare cotton fabric, where the smooth fiber surface without coating is observed. Insets shows the magnified view of the fiber surface.

## 4.3 Tuning the surface wettability of G-fabric

The surface wettability of G-fabric was characterized by the contact angle measurement. Droplets with a preset volume of 2 μL were casted on the G-fabric surface and their contact angles were analyzed by the OCA-15 contact angle measurement and contour analysis system. A representative image showing the 2 μL water droplet sitting on the G-fabric surface was presented in Figure 46. In this case, a contact angle of 138.2°±5.1° was observed, indicating the substantial hydrophobicity of the fabric surface with graphene coating. It is noted that the WCA of graphite is in the range of 80°-90°, which is generally considered hydrophilic.[102] However, the hydrophobicity observed for the graphene coating can be ascribed to the heterogeneity of the as-obtained graphene. As previously mentioned, GO was not fully reduced in the coating process, which made the as-obtained graphene a heterogeneous state between the oxygen intercalated carbon plane and the pure carbon plane. The replacement of the C=O group by C-H group during the hydrazine reduction may substantially lower the surface energy of the as-obtained graphene, while the intensive reactions on the atomically thin graphene layer would abruptly raise the roughness even in the molecular level.[103-104] These effects during the hydrazine reduction process thus may generate the unexpected hydrophobicity of the graphene coating. On the other hand, the mass loading of the graphene coating on the fabric is rather low, with a measurable value of ~0.5 wt%, indicating the formation of a very thin coating layer on the fabric surface, which is more susceptible to form the nano-wrinkles. Nonetheless, the G-fabric became more and more hydrophilic upon repeating the reduction process. The phenomenon is predictable since the heterogeneity of the as-obtained graphene would be gradually reduced during the repeatedly reduction, and homogeneous atomically smooth carbon basal plane was eventually formed, which demonstrated similar hydrophilicity to the graphitized carbon.

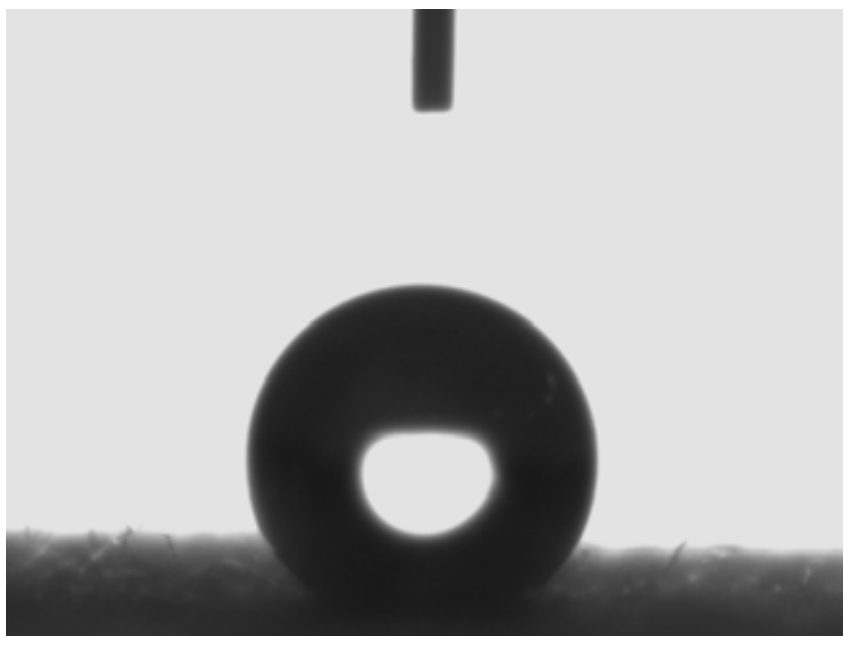

**Figure 46** Photograph showing a 2 μL water droplet sitting on the G-fabric surface.

The relationship between the hydrophobicity of G-fabric and the time of reduction was also investigated. Intriguingly, it was found that the WCA on G-fabric quickly raised above 135° after a short reduction duration of 10 minutes, and remained stable between 135° and 140° even the time of reduction was increased to 5 hours (Figure 47). It thus indicated that the reaction between GO and hydrazine quickly reached its equilibrium on the fabric surface at 80°C for 10 minutes, as further increasing the reaction time did not significantly alter the surface wettability. The reason for this notable phenomenon can be possibly attributed to both the excess amount of hydrazine used and the low loading of graphene on the fabric, which dramatically increase the reduction rate in the given condition.

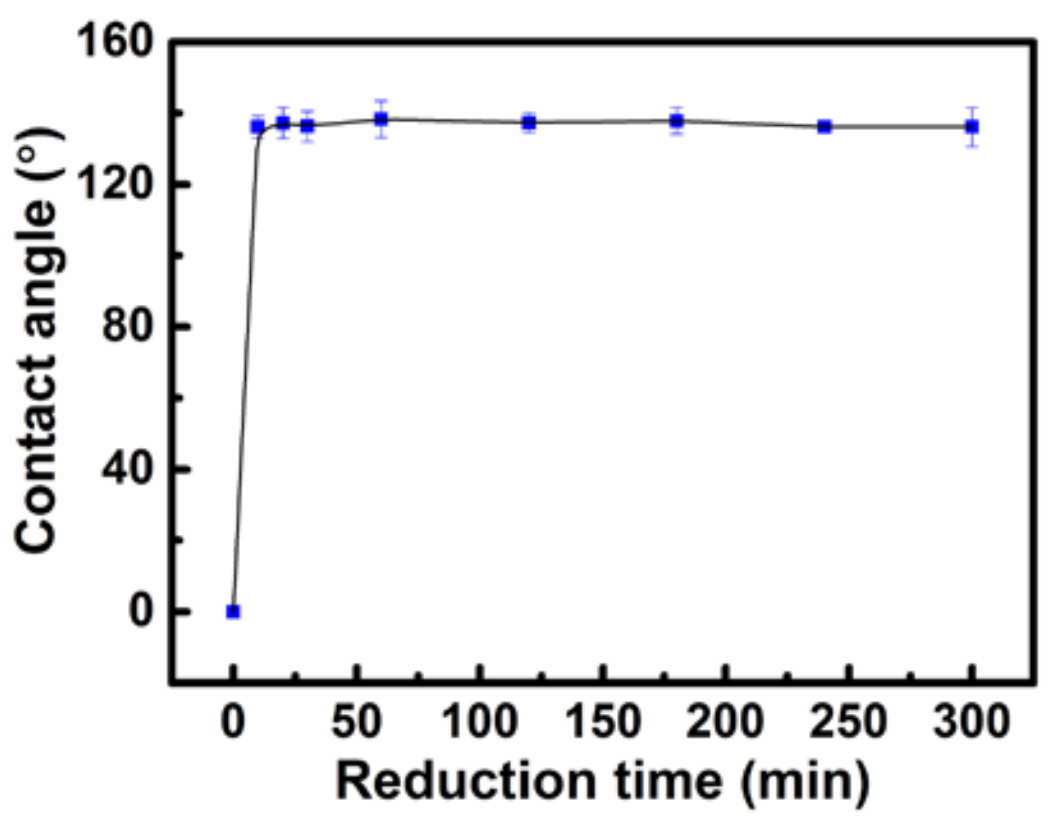

**Figure 47** The relationship between contact angle and reduction time in the fabrication process of G-fabric.

Tunable wettability can be achieved on G-fabrics by a simple absorption process. In this regard, small molecules, including water and acetone, were used as the external stimuli to trigger the surface wettability change of G-fabrics. In a typical process, the G-fabric was immersed and sonicated in either water or acetone, or their mixtures, for a short duration (~30 min). After drying, these treatments eventually resulted in the enhanced hydrophilicity or hydrophobicity on the G-fabric, respectively. These changes were reflected by the corresponding WCA on the G-fabrics, as shown in Figure 48. It can be found that the WCA increased to 140.1°±3.6° as G-fabric was sonicated in pure acetone, indicating the lowering of surface energy and enhancement in hydrophobicity by the acetone absorption. On the contrary, as the acetone was replaced by water, the G-fabric exhibited a zero contact angle after sonication, indicating the conversion from hydrophobic to hydrophilic by the water absorption. Further manipulating the volume ratio between acetone and water may result in a series of intermediate states between the case of pure acetone and pure water, as indicated by the changes in WCA (Figure 48). Nonetheless, it is speculated that this novel surface wettability change is induced by three major factors, including: (i) the evaporation-induced surface roughness increase on graphene; (ii) different configurations of the molecules adopted in absorption: in the case of acetone, it absorbed onto graphene by pointing the methyl groups outwards; (iii) amplification of the surface changes by the high surface area of graphene. Similar phenomena were also observed on graphene coated gold, aluminum, and highly ordered pyrolitic graphite (HOPG) surfaces.[105]

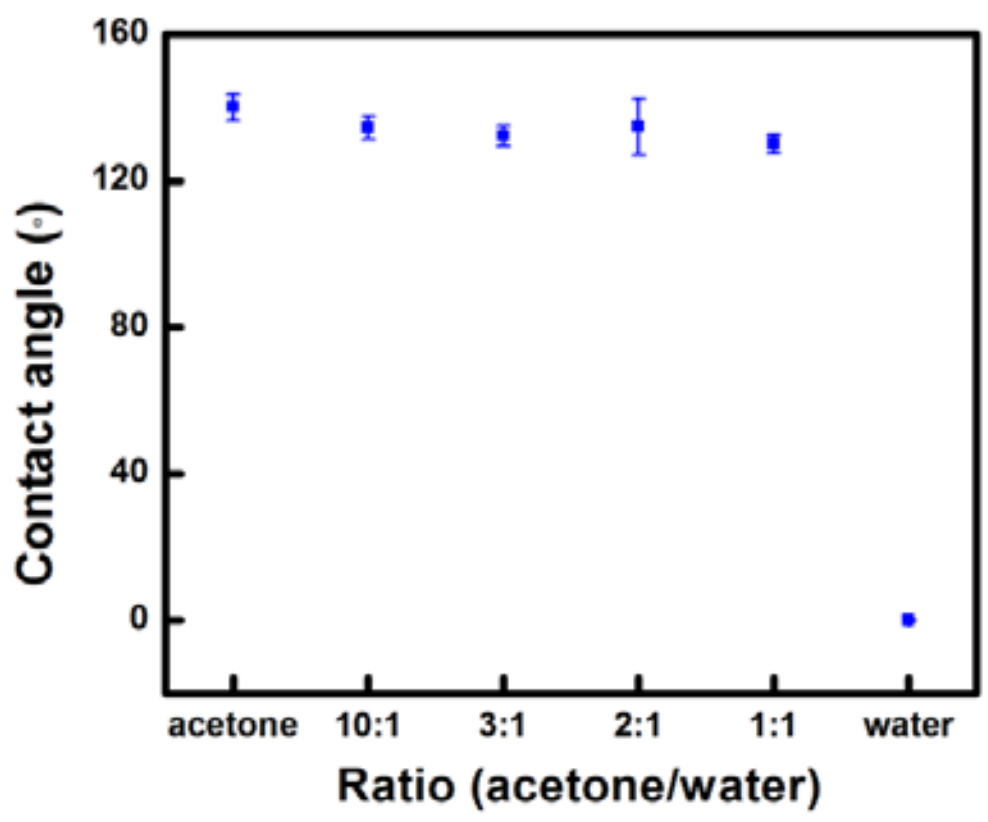

**Figure 48** Changes of water contact angles on G-fabric by sonication in the water-acetone mixture with different water/acetone volume ratios.

# Chapter 5. Surface energy engineering on G-fabric surface using titania

## 5.1 Introduction

Even though tunable surface wettability can be obtained on G-fabric surface by the absorption of different molecules, the process is still relatively sluggish without appropriate reversibility. On the other hand, tuning the surface wettability over the full spectrum between superhydrophobic region and superhydrophilic region is also hard to achieve by the absorption of small molecules. Last but not least, the small molecules are easy to desorb from the fabric surface upon washing or mechanical abrasion, which make them rather impractical for the fabrication of robust SHS. In order to obtain full-spectrum tunable wettability on SHS which is of practical significance, the technique used and process involved should be simple, while the tuning function is smart and stimuli-responsive. To satisfy these goals, a screening over the materials was conducted to locate the most suitable catalog. With both easy-processability and photoactivity, the transition metal oxides (e.g. titania, tungsten oxide, vanadium pentoxide, zinc oxide) caught the eyes as a result of their abundance and low toxicity.[106] Among the metal oxides, titania ($TiO_2$) is well-known for its photoactivity under ultra-violet (UV) light, which exhibits a reversible conversion from hydrophobic to superhydrophilic upon UV exposure. The UV superhydrophilic conversion of $TiO_2$ has been widely observed on the thin films, composites, and nanostructures of titania.[107-109] However, it is rather intriguing to observe this phenomenon on the hydrophobic graphene surface.

## 5.2 Synthesis of titania sol-gel with homogeneous nanoparticle size distribution

The concept of "dual roughness" was applied to realize the superhydrophobicity on the $TiO_2$ coated graphene (GT) surface. The $TiO_2$ was made into the high density nanoparticles (NPs) which were hierarchically deposited onto the single graphene layer coated on the cotton fabric surface, forming the nanostructured rough surface texture. The $TiO_2$ NPs were synthesized using a thermal hydrolysis method with titanium isopropoxide as the precursor. After hydrolysis and the subsequently aging in the ambient, a transparent pale sol-gel of $TiO_2$ NPs can be obtained, which is also denoted as the "titania nanosol".

TEM and high-resolution TEM (HRTEM) were used to characterize the as-obtained $TiO_2$ sol-gel. It can be observed from the TEM image that the $TiO_2$ sol-gel is composed of $TiO_2$ NPs with sizes ranging from 10 to 30 nm (Figure 49). According to the TEM image, it was observed that the $TiO_2$ NPs were agglomerated together and formed large particulate clusters. However, the boundaries of the single $TiO_2$ NP can still be clearly observed in the clusters, indicating the good crystallinity and stability of the as-synthesized $TiO_2$ NPs. The crystalline structure of the $TiO_2$ NPs was further investigated by HRTEM. Under the HRTEM observation (Figure 50), the lattice structures of the $TiO_2$ NPs can be clearly identified, and a lattice spacing of 0.35 nm was obtained, which corresponded to the (101) reflection of anatase $TiO_2$. Even though the lattice boundaries were randomly distributed between the $TiO_2$ NPs, the lattice structure was well-preserved in each single crystal (Figure 50). The crystalline structure of the $TiO_2$ NPs was further confirmed by SAED, as the diffraction rings corresponding to the (101), (211), (004), (200), (105), (204), (220) can be clearly indexed, which can be unambiguously attributed to the lattice reflections of anatase $TiO_2$ (Figure 51). The XRD spectrum showed in Figure 52 was also coincident with the TEM & HRTEM observation, as the diffraction peaks were in good agreement with the anatase $TiO_2$ (JCPDS 21-1272).

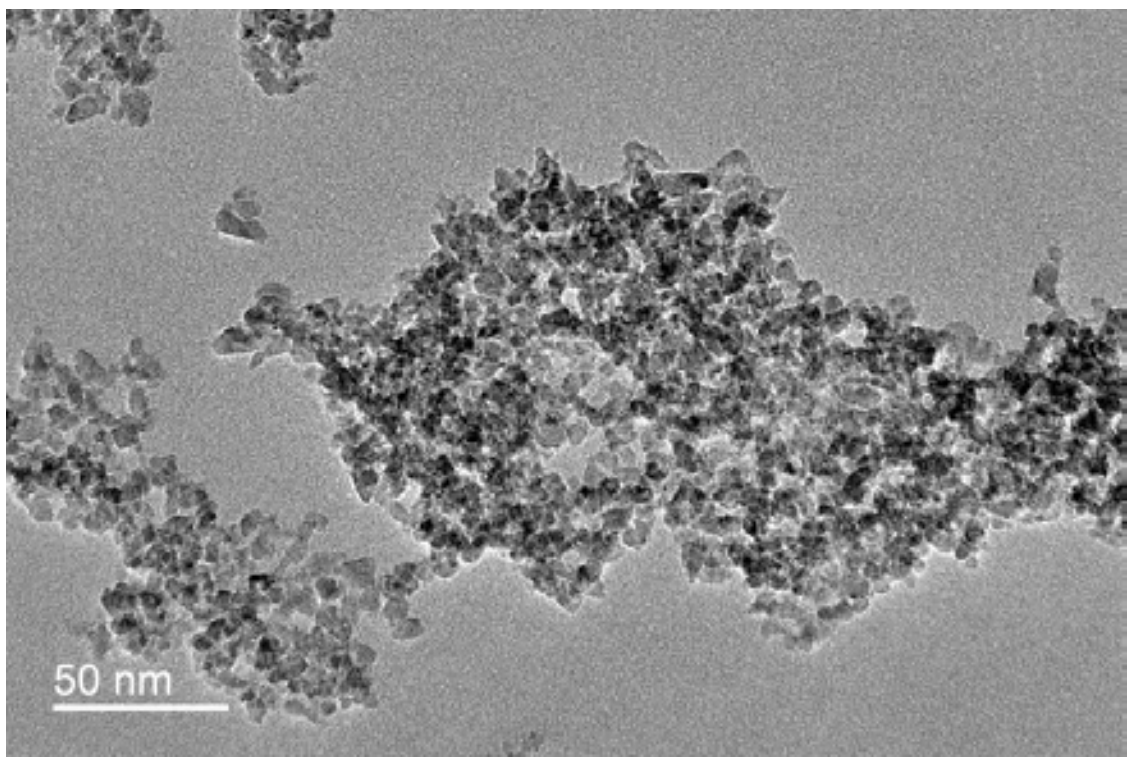

**Figure 49** TEM images of the as-obtained $TiO_2$ sol-gel.

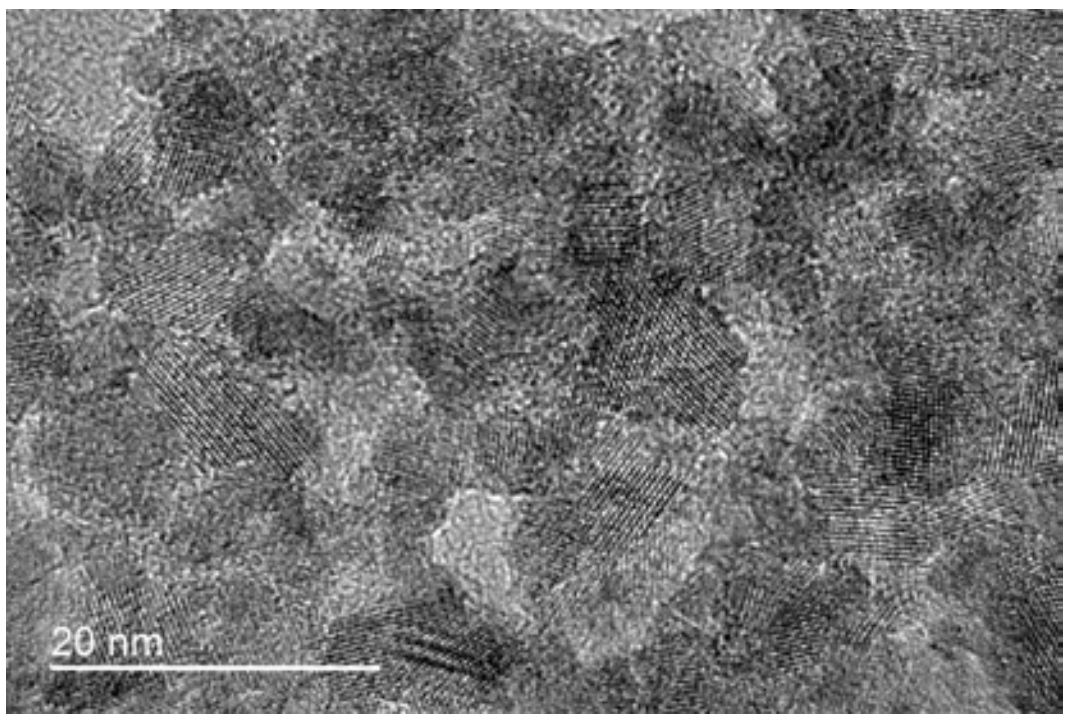

**Figure 50** HRTEM image of the $TiO_2$ sol-gel showing the lattice structure of the $TiO_2$ NPs.

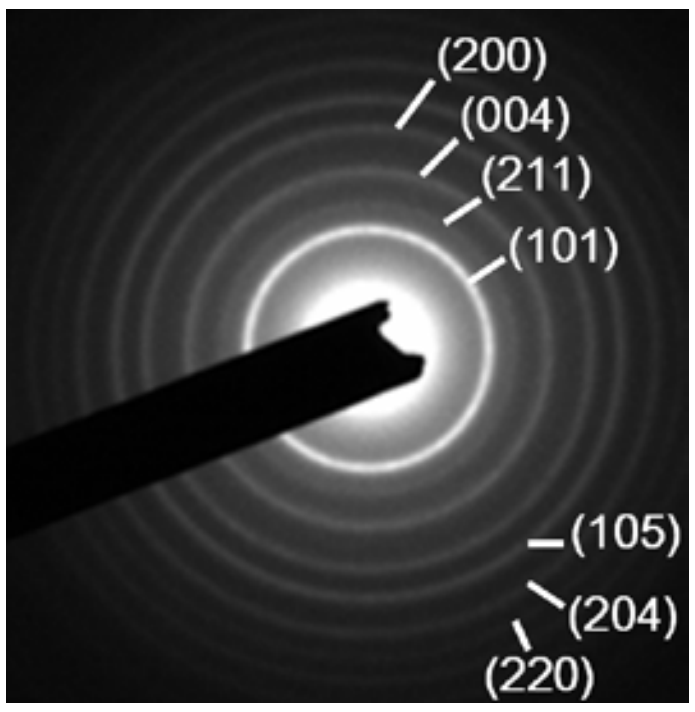

**Figure 51** SAED pattern of the $TiO_2$ NPs shown in the TEM image.

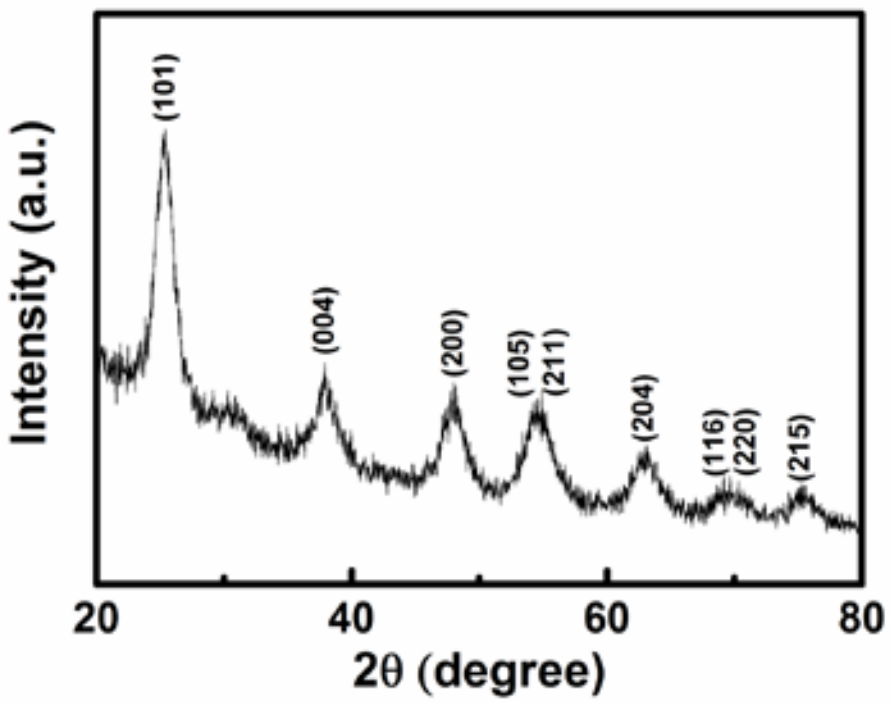

**Figure 52** XRD spectrum of the as-obtained $TiO_2$ sol-gel.

## 5.3 Fabrication of titania nanofilm on G-fabric

The as-synthesized $TiO_2$ NPs were subsequently applied onto G-fabrics using a simple dip-pad-dry-cure method. Briefly speaking, the G-fabric was firstly dipped in the 1% $TiO_2$ sol-gel solution for 5 minutes, and then passed through a vertical padder with an applied pressure of 2 kg $cm^{-2}$ in medium speed, in order to remove the excess liquid absorbed. The $TiO_2$ sol-gel treated fabric was then dried at 80°C for 10 minutes and subsequently cured at 160°C for 3 minutes. After cooling in the ambient, the hierarchical $TiO_2$ NP/graphene coated cotton fabric (GT-fabric) was obtained. It was speculated that the $TiO_2$ sol-gel condensed to form a nanofilm consisting of $TiO_2$ NPs on the G-fabric's surface. A schematic illustration regarding the preparation process of GT-fabric was shown in Figure 53. The successful formation of the $TiO_2$ coating on G-fabric was confirmed by EDX mapping, as the elementary distributions of Ti and O were found over the observation window (Figure 54). The surface morphology of the as-prepared GT fabric was characterized by SEM. According to the as-obtained SEM image (Figure 55), the surface feature of GT-fabric was substantially different from G-fabric, as a thin film with bright contrast can be observed on top of the fabric surface, which was distributed continuously on the individual fibers. This thin film is formed from the condensation of the $TiO_2$ sol-gel during the curing process, resulting in the packing of the $TiO_2$ NPs and the formation of the thin solid film. The thin film composed of the $TiO_2$ NPs was visually smooth and continuous on the longitudinal direction of the fibers (Figure 56). The magnified SEM image revealed that the thin film was composed of densely-packed few layers of $TiO_2$ NPs, by which the preferential packing direction was following the local curvatures of the fiber, indicating the formation of a nanoscale thin film (Figure 57). The formation of $TiO_2$ nanofilm on the cotton fabric's surface in a thermal process has also been reported previously.[110] In that case, the $TiO_2$ NPs were directly deposited on the cotton fabric without the intermediate graphene layer, and substantial hydrophilicity was demonstrated.

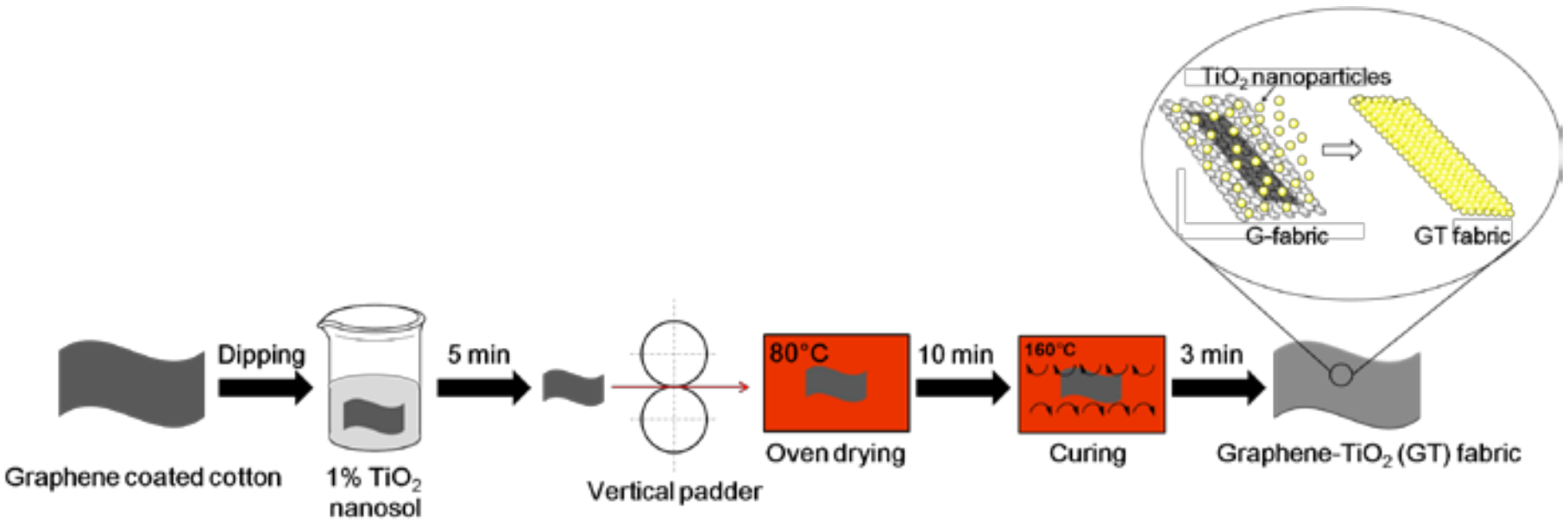

**Figure 53** Schematic illustration of the GT-fabric preparation process.

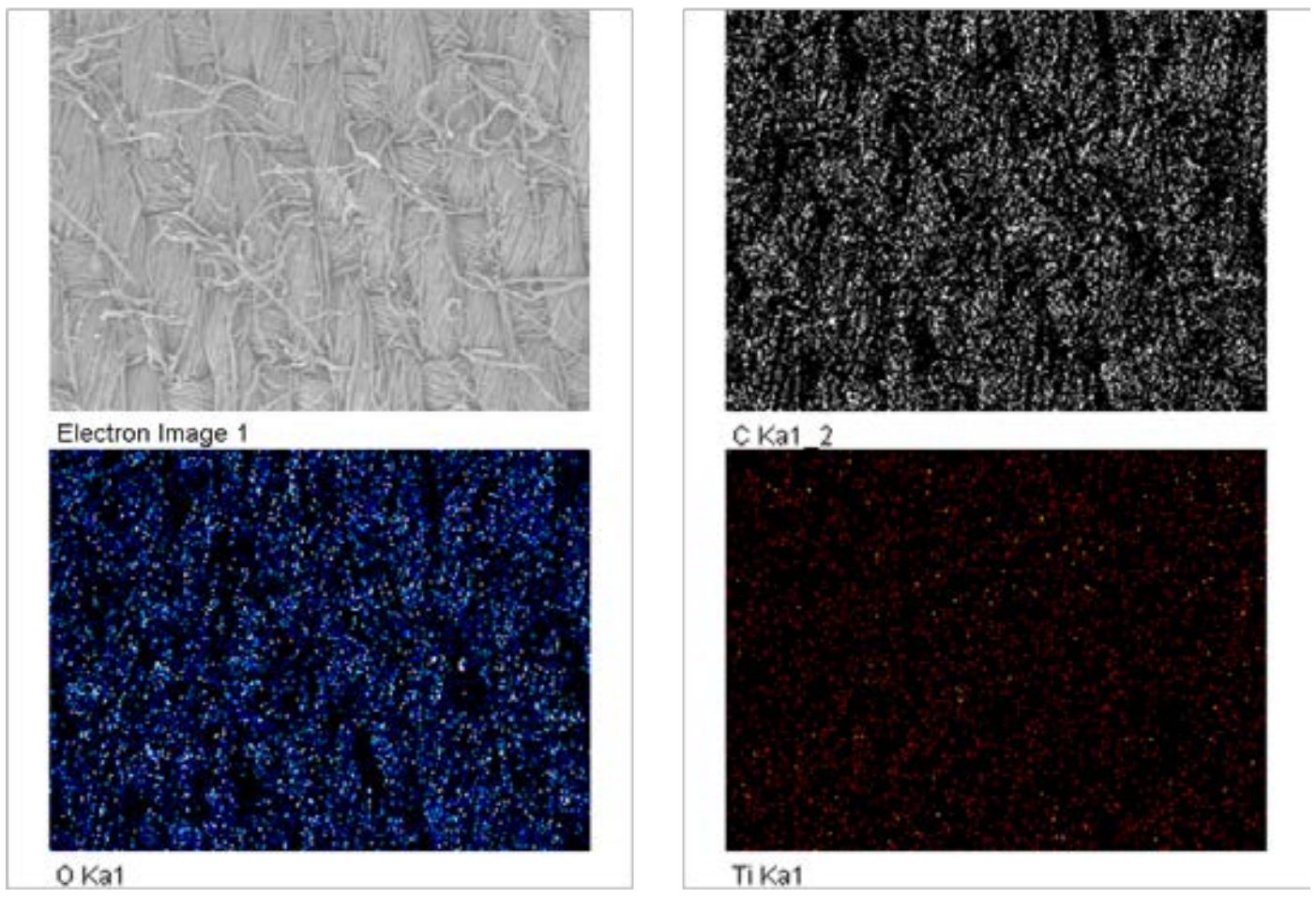

**Figure 54** EDX mapping of the selected area on GT-fabric (electron image 1) for the elements of carbon (C), oxygen (O) and titanium (Ti).

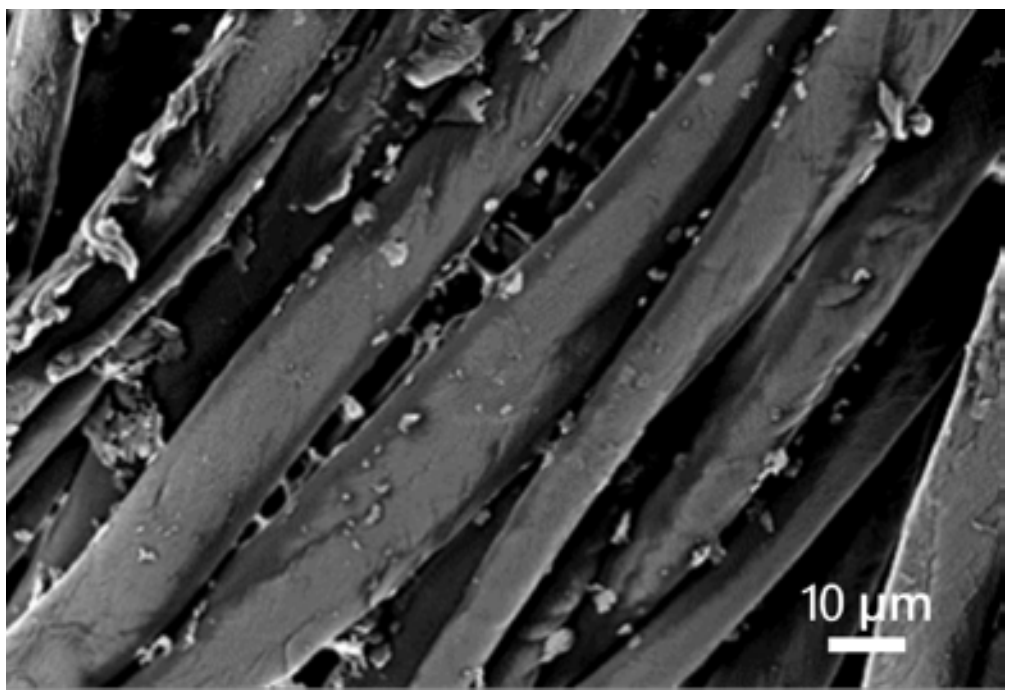

**Figure 55** SEM image of the as-prepared GT fabric, where the titania nanocoating with brighter contrast can be clearly observed.

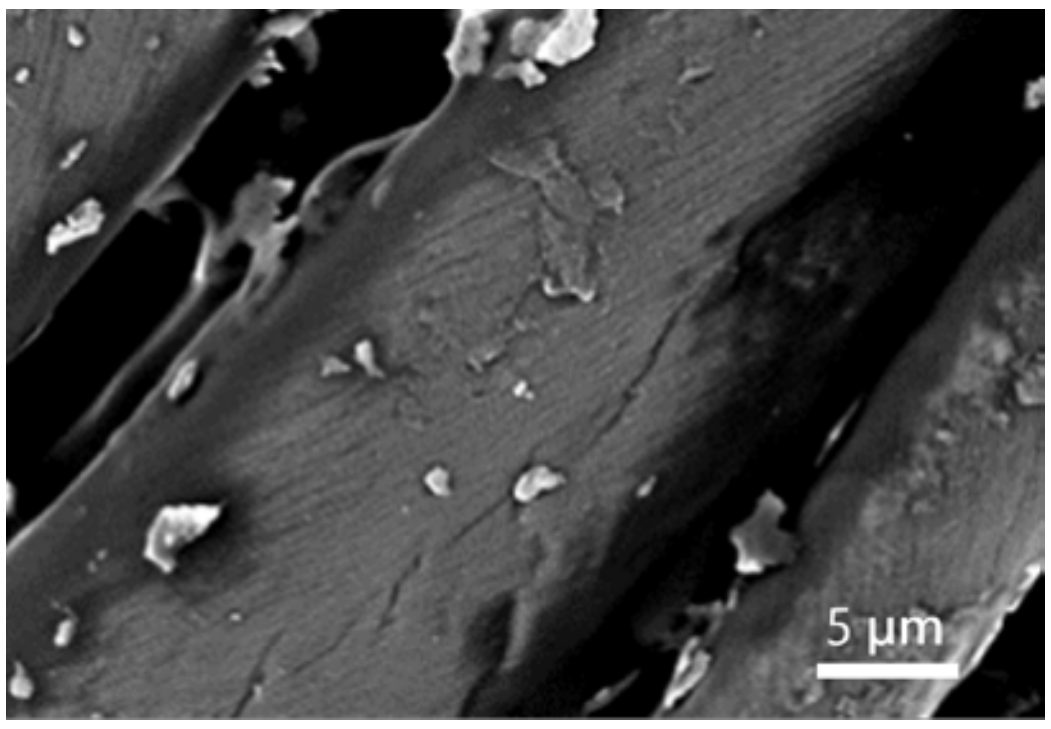

**Figure 56** Magnified SEM image showing the morphology of the titania nanocoating on the G-fabric surface.

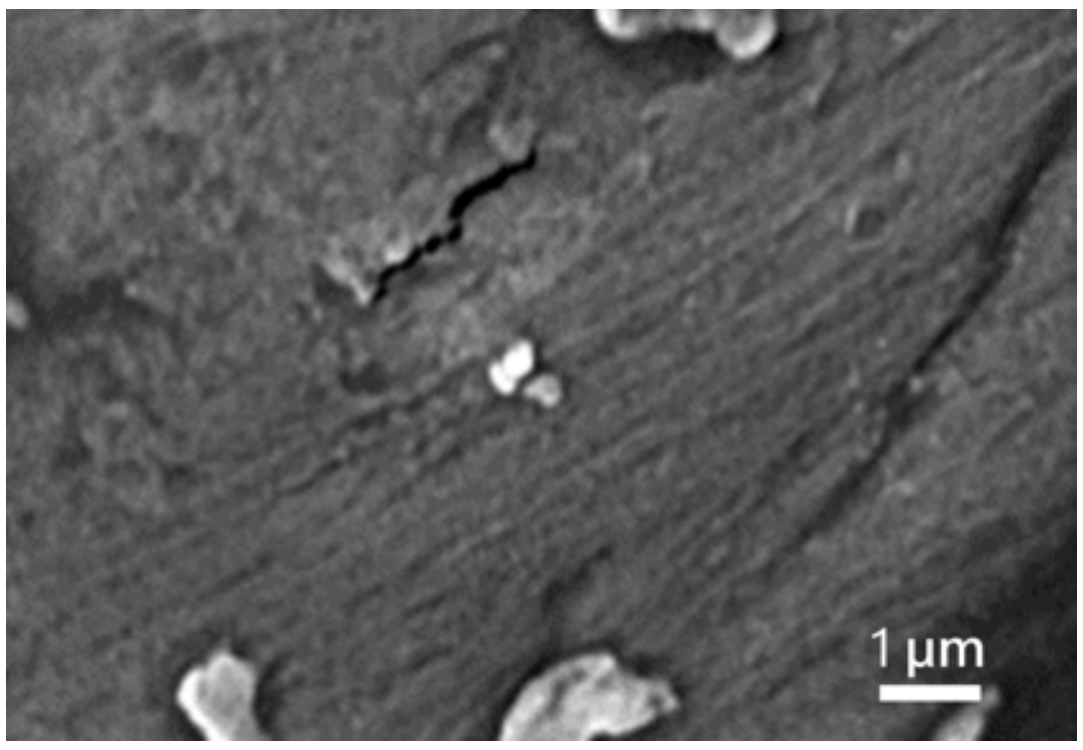

**Figure 57** Magnified SEM image showing the morphological detail of titania nanofilm, which is composed of densely packed $TiO_2$ NPs.

## 5.4 Tuning the surface wettability of GT-fabric

Stimuli-responsive full-spectrum tuning of the surface wettability can be achieved on GT-fabric as a result of the unique hierarchical graphene/$TiO_2$ surface nanocoating. Due to the "dual roughness" feature provided by the nanotextures formed by the packing of $TiO_2$ NPs and the crumpling of the graphene sheet; and the microtextures formed by duplicating the intrinsic curvatures of the cotton fibers, the GT-fabric exhibits substantial superhydrophobicity with high WCA and relatively low SA. Based on the sessile drop measurement, a WCA of 151° can be observed on the GT-fabric using 2 μL droplets, indicating the characteristic angle of Cassie state (Figure 58). On the other hand, a SA of 45° can be observed with a droplet size of 30 μL. The relatively large SA on GT-fabric can be attributed to the presence of $TiO_2$, which is intrinsically hydrophilic and can form hydrogen bonds with water via surface oxygen groups.[111] The molecular bonding between $TiO_2$ and water thus generates higher adhesion which may require more energy for de-pinning. However, the SA on GT fabric can be further improved by the addition of a low surface energy barrier layer between $TiO_2$ and the impinging water, which is also worthy of investigation in the future. By utilizing the UV superhydrophilic conversion on $TiO_2$, the superhydrophobic GT-fabric can be tuned into superhydrophilic by simple UV radiation, while intermediate states can also be facilely achieved by using different radiation time. The WCAs on GT-fabrics were measured after being UV-treated for 1, 2 and 3 hours, respectively (Figure 58). The decreasing trend is obvious since the WCAs change from 132.4°, 99.6°, to 0° as the UV time increases from 1, 2, to 3 hours. It thus indicates the elevation of surface energy on GT-fabric upon UV exposure. However, according to the relatively smooth decreasing trend represented by the curve, the reduction in surface energy can be interpreted as a stimuli-responsive UV-dependent process, which can be utilized as the essential tool for the surface energy engineering. Interestingly, different wettability characteristics can be observed on the two sides of the GT-fabric, as superhydrophilicity is obtained on the side facing UV radiation while superhydrophobicity is maintained on the opposite side. Furthermore, the WCAs on the UV-opposing side remain almost stable with increasing UV radiation time, indicating the UV penetration depth is considerably smaller than the thickness of the GT fabric. However, the observed surface energy gradient is essential for the realization of the surface energy engineering on GT-fabrics, which also serves as the very basis for advanced microfluidic applications.

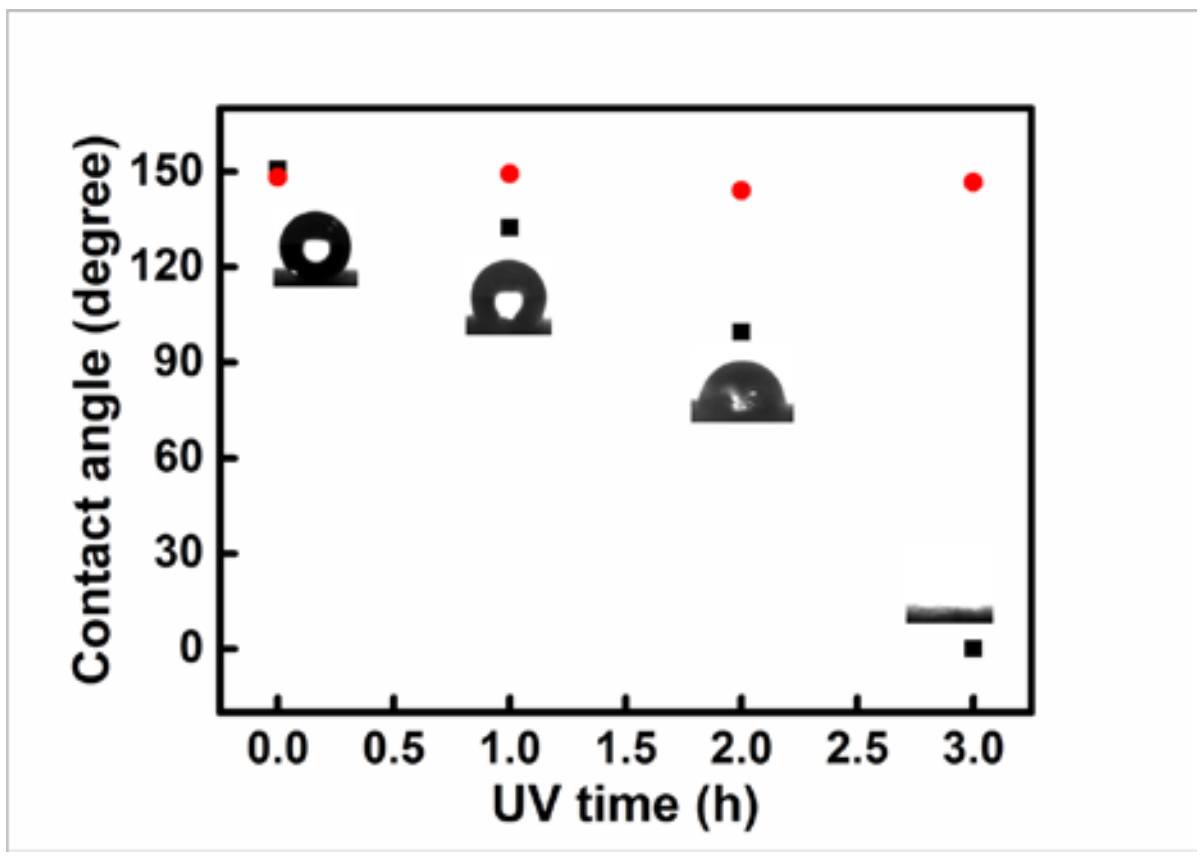

**Figure 58** Contact angle vs. UV radiation time on the GT-fabric's surfaces. The black squares represent the contact angles on the fabric side facing UV radiation while the red circles represent the contact angles on the opposite side.

# Chapter 6. Advanced applications of superhydrophobic hierarchical graphene/$TiO_2$ surface

## 6.1 Introduction

Upon achieving the full-spectrum wettability tuning of GT-fabric surface through the fabrication of hierarchical graphene/$TiO_2$ NP nanocoating, multifunctionality can be implemented on GT-fabric upon controlling the wetting, adhesion, and transport phenomenon on the fabric's surface. By introducing opposite wettability on the two faces of GT-fabric, directional water transport with controlled speed can be realized in the fabric's axial direction. This novel transport phenomenon can spontaneously alter the mass distribution of water within the fabric matrix and may have potent applications in controlling the microenvironment at the fabric/skin interface. On the other hand, by the selection of proper droplet size, the water droplets can be stored on the GT-fabric surface via a so-called impregnated Cassie state, which exhibits both high contact angle and high adhesion. By utilizing this effect, the GT-fabric can be transformed to a versatile platform for microfluidic manipulation, and novel manipulative functions, e.g. droplet storage, transport, and gas sensing, are achieved on the GT-fabric based microfluidic device. Last but not least, owing to its full-spectrum tunable wettability, GT-fabric also demonstrates high separation speed to the oil/water mixture and a wide-range adaptability to oil/water mixtures with different surface tensions.

## 6.2 Directional water transport

Smart directional water transport can be achieved on the GT-fabric by engineering the surface energy gradient in the fabric axial direction. Utilizing the stimuli-responsive property of the graphene/$TiO_2$ nanocoating, the surface energy gradient along the fabric axial direction can be tuned from the low to high levels depending on the time of UV radiation. In the case of high surface energy gradient, the deposited water droplet will be spontaneously transported from the superhydrophobic side of the fabric to the superhydrophilic side, preferentially in a relatively high speed fully driven by the gradient. This novel phenomenon is significant for the micro-environmental control in high humidity condition, which can extensively enhance the moisture drainage and creates the comfort feeling. The directional transport is also of great usage for the controlled collection and release of microfluids, which may have wide-spread applications in biological fluid handling.

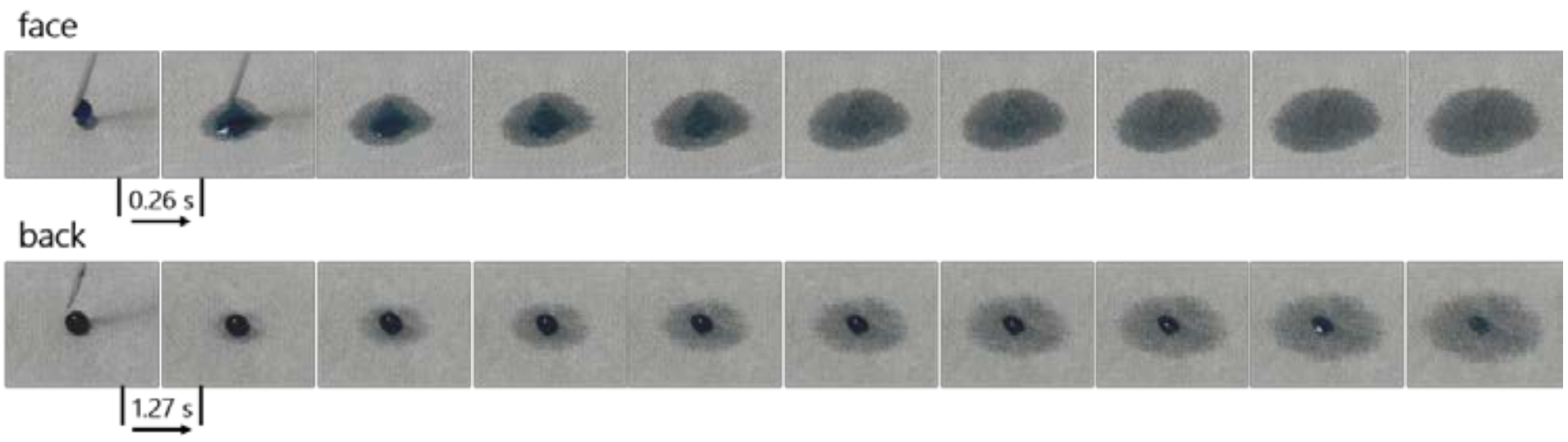

**Figure 59** Sequential frames showing the different wetting phenomena on two sides of 3h-UV GT-fabric. The water droplet (10 μL) instantly spread on the face side (UV-facing) on the fabric while it maintained the spherical shape on the back side (UV-opposing) and gradually transported to the face side.

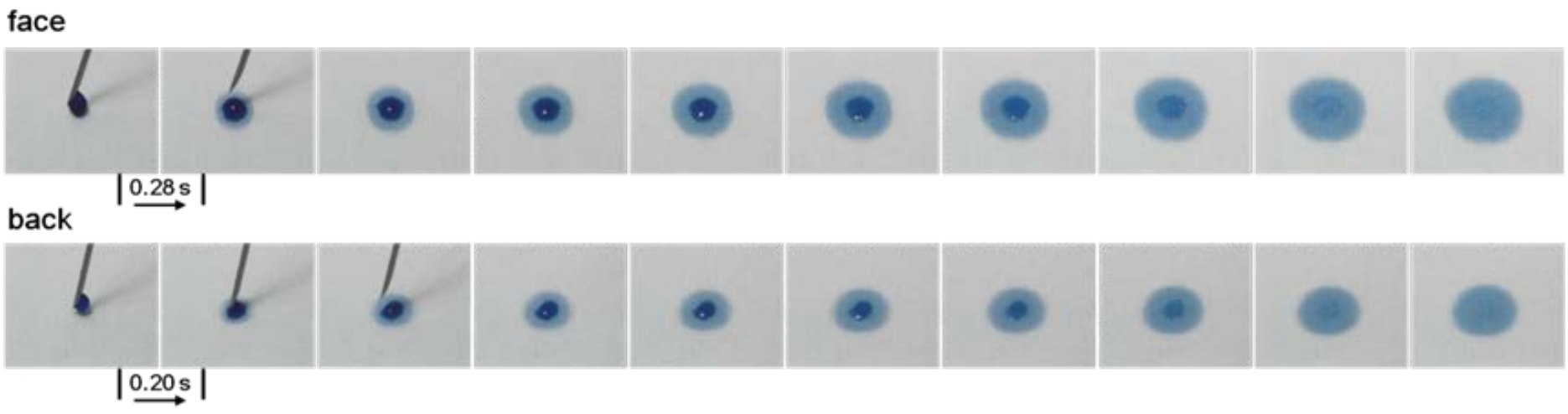

**Figure 60** Sequential frames showing the wetting phenomena on bare cotton fabric. Identical water spreading was observed on both sides of the fabric.

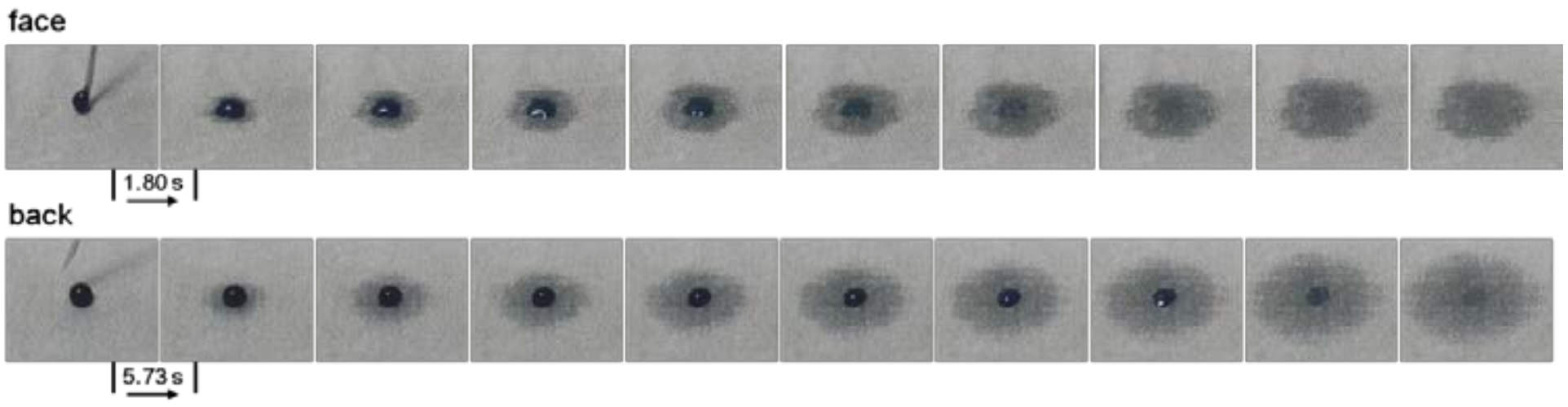

**Figure 61** Sequential frames showing the difference in wetting phenomenon on the 2h-UV GT-fabric.

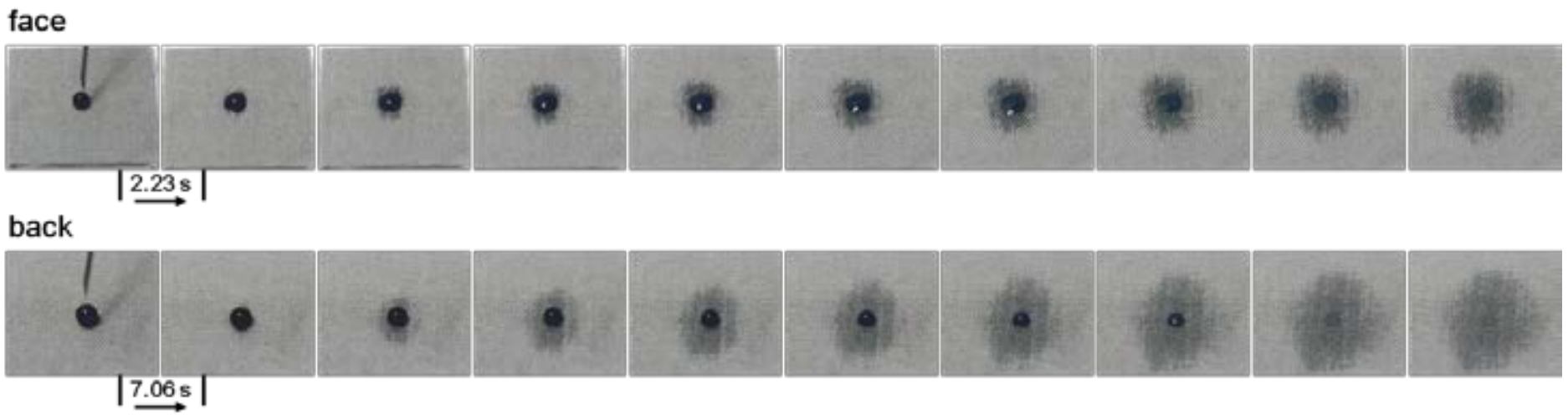

**Figure 62** Sequential frames showing the difference in wetting phenomenon on the 1h-UV GT-fabric.

The transport phenomenon was observed by directly casting of a 10 μL water droplet on the 3h-UV treated GT-fabric. Since the substantial surface energy gradient has been established, different water transport dynamics can be observed on the two sides (UV-facing & UV-shading) of the fabric (Figure 59). As the water droplet was deposited on the superhydrophilic side (UV-facing), it instantly spread on the surface while the shape of droplet changed from spherical to an extended liquid film, indicating a drastic change on the contact angle driven by the capillary force. This fast spreading process thus indicated the established superhydrophilicity on the UV-facing side, while significant different phenomenon was observed on the other side. To speak more specifically, the deposited water droplet was found to gradually wick from the superhydrophobic side to the superhydrophilic side, and the droplet was observed to be remained spherical for a lengthy duration in this process. The strongly impeded spreading and the slowly decreasing contact angle thus indicated the superhydrophobic characteristics of the UV-opposing side. However, as previously mentioned, a surface energy gradient was established between the two sides of GT fabric by UV radiation. As a result, directional capillary force was generated by the surface energy gradient, and the water is tended to transport from the more hydrophobic surface to more hydrophilic surface through directional wicking, which is the case for the droplet deposited on the UV-opposing side of GT-fabric. Different from the normal spreading process on the untreated cotton fabric (Figure 60), strong controls can be implemented on the speed and direction of the mass transport on GT-fabric. By controlling the intensity of the established surface energy gradient, the mass transport speed of water on GT-fabric can be facilely tuned from high to low. A representative example was provided by the transport phenomena observed on the GT-fabric treated by 1, 2 and 3 hours' UV radiation. The sequential frames of the corresponding mass transport phenomenon on the GT fabric treated by 2 hours' and 1 hour's UV radiation are show in Figure 61 and Figure 62, respectively. Both the fastest spreading and transport processes are observed on the 3h-UV treated GT-fabric, and both of the spreading and transport speeds are following the order of 3h UV treated > 2 h > 1h, which further prove the proportional relationship between the mass transport speed and the surface energy gradient. The mass transport phenomenon on GT-fabric was then analyzed quantitatively by measuring the time required for the disappearance of a 10 μL water droplet on the hydrophobic and hydrophilic sides of the UV treated fabric, respectively. As a result, the time required for the transport process was measured to be 39.7±7.8 s, 37.7±10.7 s, and 12.5±5.2 s for the GT-fabrics treated by 1, 2 and 3 hours' UV radiation. On the other hand, the time required for the spreading process was measured to be 26.3±4.2 s, 14.9±3.1 s, and 3.3±2.0 s for the 1, 2 and 3 hours' treated GT-fabrics (Figure 63). A clear trend of decrease can be observed for both the transport and spreading time with increasing time of UV treatment. However, the changes in spreading time exhibit a more linear trend which can be fitted linearly into: $y = 37.8 - 11.5x$. It thus further confirms the underlying linearity between the surface hydrophilicity and UV time (Figure 64). Clear time gaps can be observed between the spreading time and transport time, which can be a useful indication of the intensity of surface energy gradient. By interpreting the changes in time gaps of different UV time, an inversely proportional correlation was found between the time gaps and the UV time, indicating a lower time gap may represent a higher surface energy gradient.

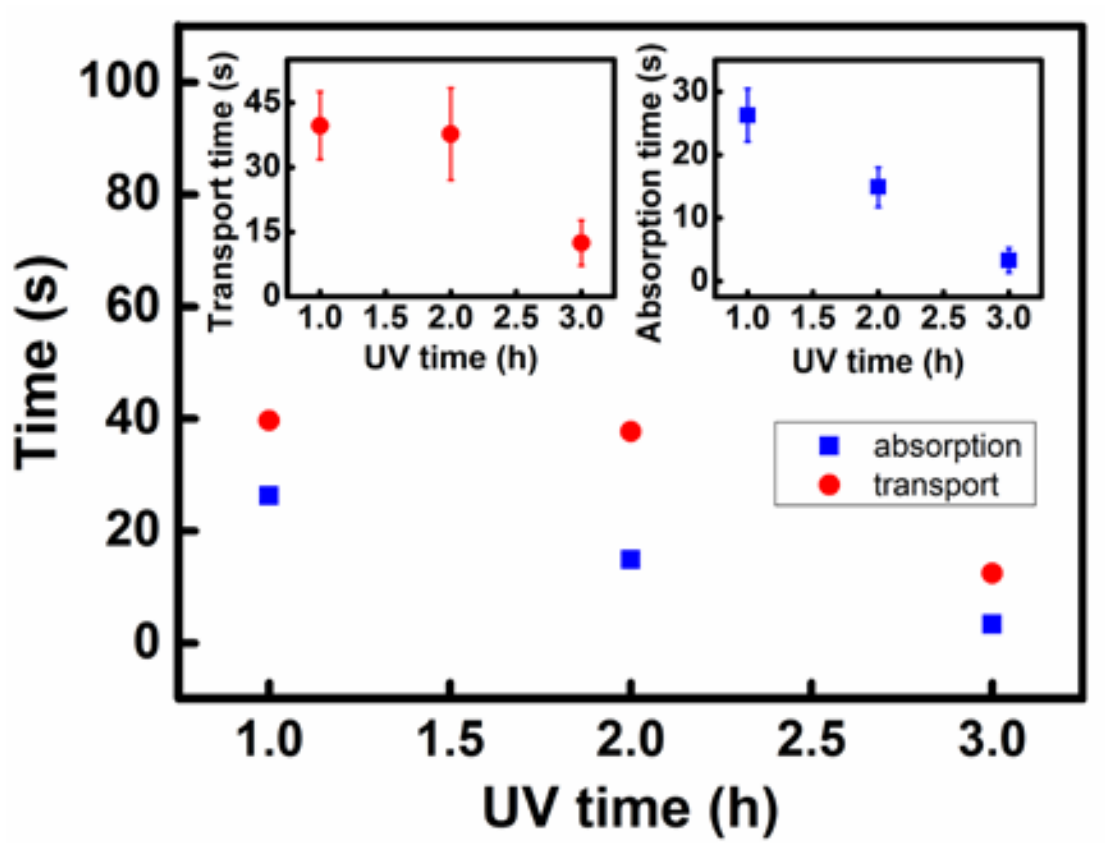

**Figure 63** The relationship between the spreading and transport speeds on two sides of GT-fabric treated by different UV radiation time, i.e. 1, 2 and 3 hours. The points at the x-axis of 1.5 h correspond to the case of bare cotton fabric.

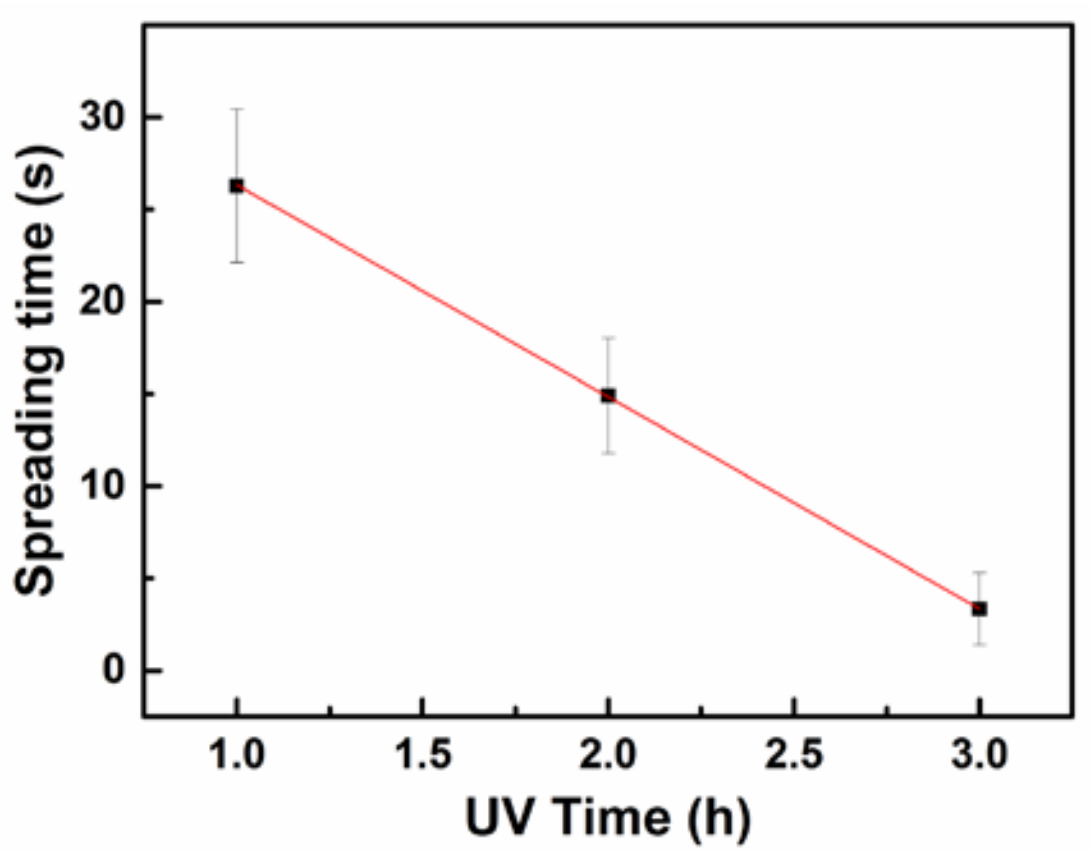

**Figure 64** Linear curve fitting of the spreading time on GT-fabric as a function of UV radiation time.

The surface energy engineering on GT-fabric by UV radiation was also found to reactivate the internal capillary networks of the cotton fibers, indicating by the gradually increasing wicking distance on the fabric with longer UV treatment (Figure 65). The wicking test was conducted according to AATCC TM 197-2011. Almost zero wicking distance was observed for the 30 min-UV treated GT-fabric, as the internal capillary network was still sealed by the superhydrophobic graphene/$TiO_2$ nanocoating. Gradually increasing wicking distances were observed as the GT-fabric was treated by longer time of UV radiation, i.e. 1, 2 and 3 hours, which indicated the gradual reactivation of the internal capillary networks by the superhydrophilic conversion of the nanocoating. Even though superhydrophilicity was observed on the 3h-UV treated GT-fabric, its wicking distance was still substantially lower than bare cotton fabric. This phenomenon can be attributed to the penetration depth of the incident UV light in the fabric. As the UV intensity received by the coated fibers decreased with incremental penetration depth, the fibers located underneath the fabric surface would receive less radiation while the fibers on the surface would take more. This in turn would generate a heterogeneous hydrophilic conversion within the GT-fabric matrix, as a higher hydrophilic conversion rate can be obtained near the fabric surface and a lower rate is expected underneath, and so a surface energy gradient is achieved. As a result, a substantial portion of fibers below the fabric surface layer may still be hydrophobic even the surface layer is turned superhydrophilic, and the remained hydrophobic fibers will restrict the capillary network of the fabric to a great extent, which significantly reduces the wicking distance. From this point of view, the time of UV radiation should be extended beyond 3 hours in order to fully activate the internal capillary network of the fabric; or both sides of the fabric should be exposed to the UV radiation for the same duration.

By the creation of the unique hierarchical graphene/$TiO_2$ nanocoating, full-spectrum wettability tuning can be achieved on cotton fabric in a simple and smart manner. The novel surface coating spontaneously adjusts its wettability to the incident UV light and water transport with speed and direction control can be readily achieved. Compared to the G-fabric substrate, both the function and performance of GT-fabric have been promoted to a higher level, which further demonstrate the powerful concept of "function-led design". However, the beauty of function-led design does not only lie in the promotion of the existing functions, but also the generation of multivariant new features and applications through the integration of the materials and structures, presumably from their extensive interfacial areas and interactions.

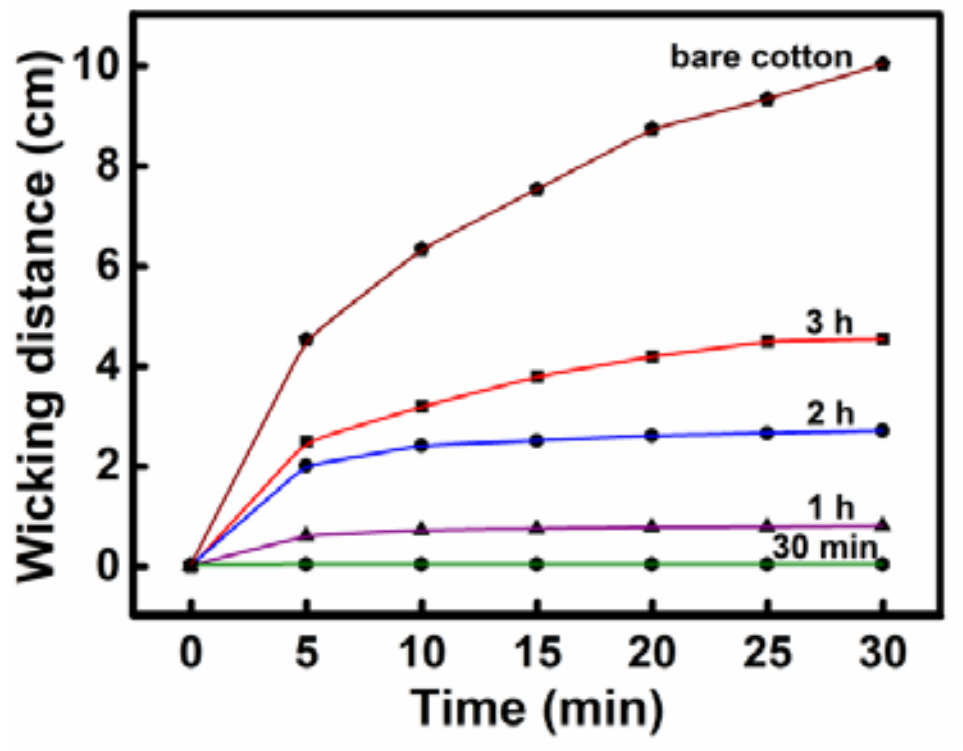

**Figure 65** Wicking distance vs. test time on bare cotton fabric and GT-fabrics treated by different UV exposure duration.

## 6.3 Microfluid manipulation

Besides directional water transport, the GT-fabric can also be used as a versatile platform for microfluid manipulation owing to its superhydrophobicity and tunable adhesion. Micro-droplet storage can be directly implemented on GT-fabric without further surface modification. As shown in Figure 66, a 4×4 array of 10 μL water droplets was directly casted on GT-fabric surface. The array can remain stable on the GT-fabric surface during the experimental observation without penetrating into the fabric or cross-contaminated with each other. High spatial storage stability was also demonstrated by GT-fabric regardless of the tilting angle. The droplet array was observed to pin firmly on the fabric surface without visible deformation even at tilting angles of 90° and 180°, indicating the high adhesion state formed between the water droplets and the fabric. This phenomenon can be attributed to the formation of a metastable partial impregnated Cassie state between the droplet and the fabric surface. In this case, both high WCA and high hysteresis are obtained on the three-phase contact line, as a portion of the air-cushion are still remained.

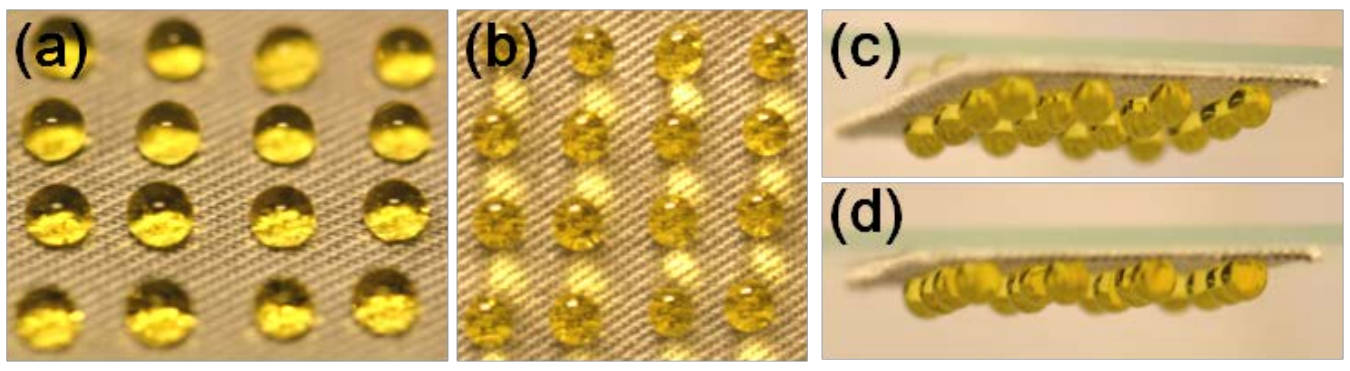

**Figure 66** (a) Photographs of a 4×4 water droplet array casted on the surface of GT-fabric. (b, c, d) The droplet array remained stable on GT-fabric surface at the tilting angles of 90° (b) and 180° (c, d).

Utilizing the special hydrophobic state of GT-fabric, novel rotating droplet transfer can be realized between the fabric and the target substrates. Figure 67 shows the transfer of a 3×3 array of 10 μL water droplets from GT-fabric to the silicon wafers. Upon rotating the GT fabric attached to one end of a glass stick, the droplets in the array were subsequently transferred to the silicon wafer surface, forming a new 3×1 array of larger drops. This unique motion of droplet manipulation is accomplished by the characteristic high WCA and high hysteresis on the GT-fabric surface, and only trace amount of droplet residues was found to remain on the fabric surface. The GT-fabric can also be utilized as a multifunctional sensing platform for various gas molecules. By casting the micro-droplet containing specific indicators, different gas sensing arrays can be fabricated on GT-fabric, including ammonia ($NH_3$), hydrochloric acid (HCl), and ethylene diamine (EDA), as shown in Figure 68. Both qualitative and quantitative sensing can be achieved on GT-fabric. In the case of $NH_3$, the presence of the gas molecule is indicated by the color change of the droplets from nearly colorless to dark blue. On the other hand, the concentration of the $NH_3$ gas is indicated by the depth of the blue color, and a linear increase in the color depth can be observed with increasing $NH_3$ concentrations (Figure 69).

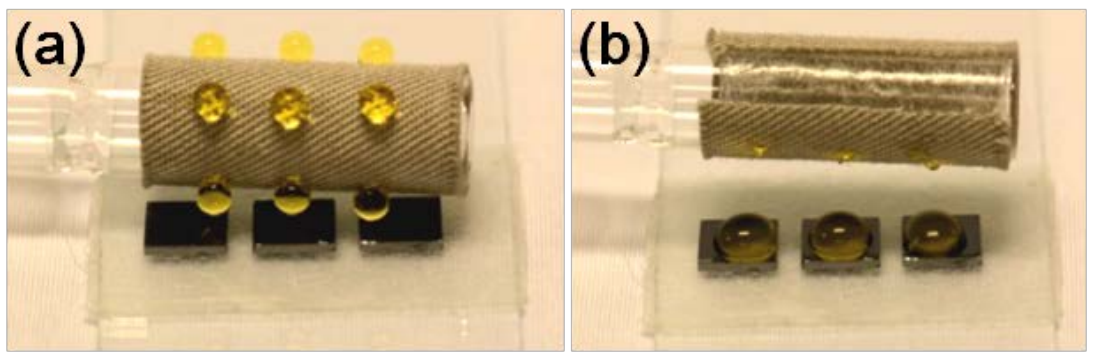

**Figure 67** (a) Rotating droplet transfer between the GT-fabric with a 3×3 water droplet array and the silicon substrate. (b) A 3×1 droplet array on silicon surface is eventually formed.

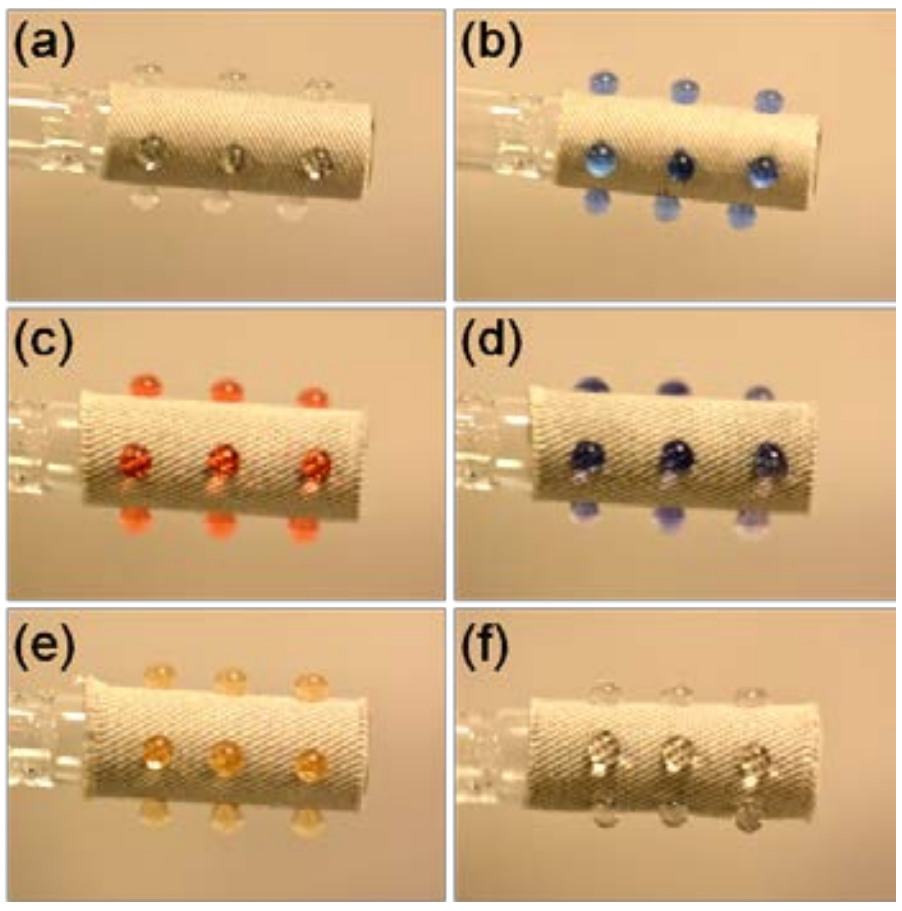

**Figure 68** (a-f) Different sensing arrays prepared by water droplet loaded with different indicators. The color changes of the droplet arrays indicate the presence of specific gas molecules, i.e. ammonia (a, b), HCl (c, d) and EDA (e, f).

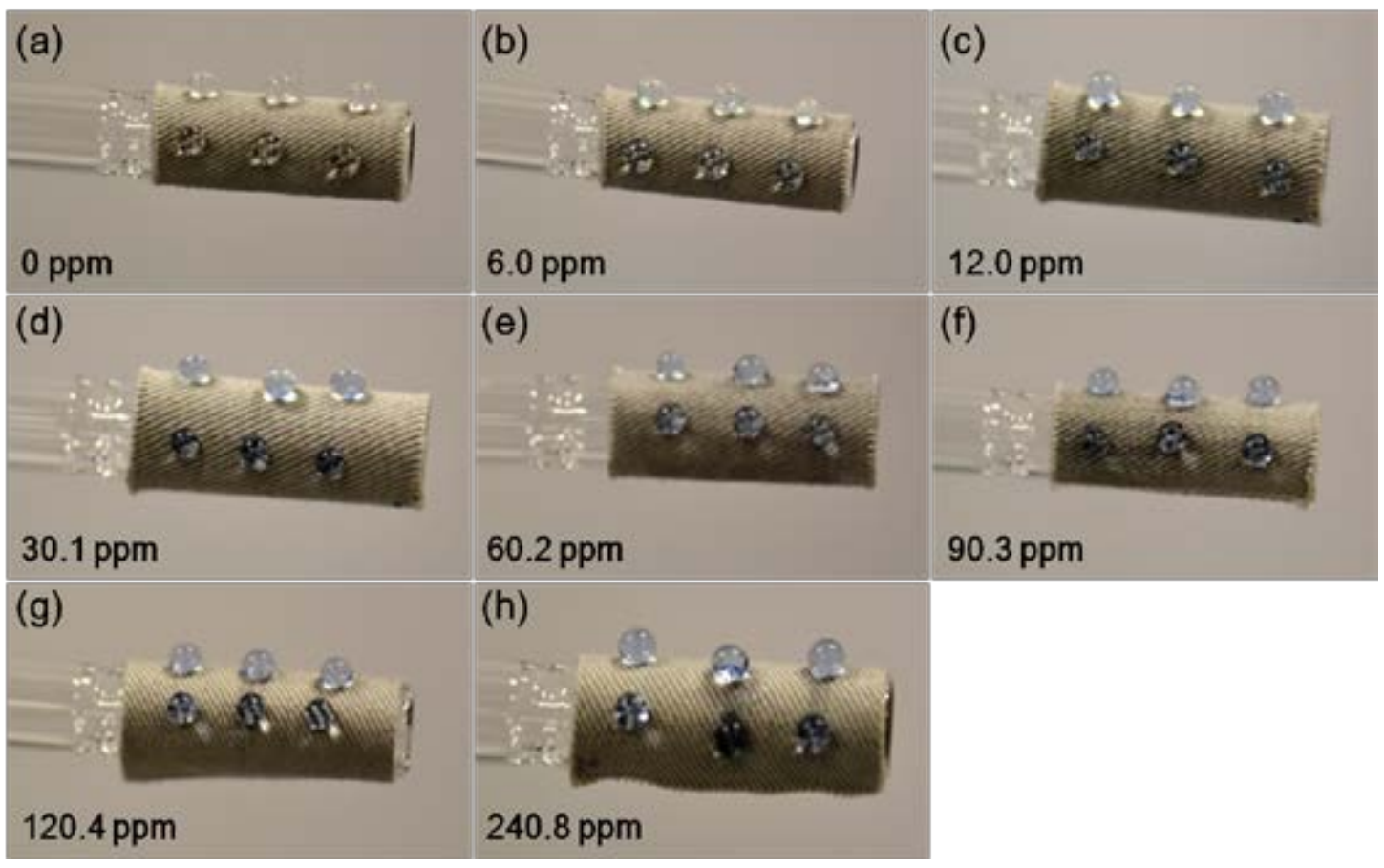

**Figure 69** (a-h) Sequential images showing the chromatic changes (from nearly colorless to deep blue) of the gas sensing array correlate to the concentrations of the ammonia.

Anti-staining property is also observed on the GT-fabric surface as a result of its superhydrophobicity. Solid stains can be easily removed from GT-fabric surface regardless of the small particle sizes. To demonstrate the anti-staining effect on GT-fabric, a small quantity of carbon nanotubes (CNTs) was spread on the fabric surface, which was subsequently removed by an extremely small quantity of water (Figure 70). After the removal of the CNTs, there are no either solid or liquid residues left on the fabric surface, indicating its excellent anti-staining property.

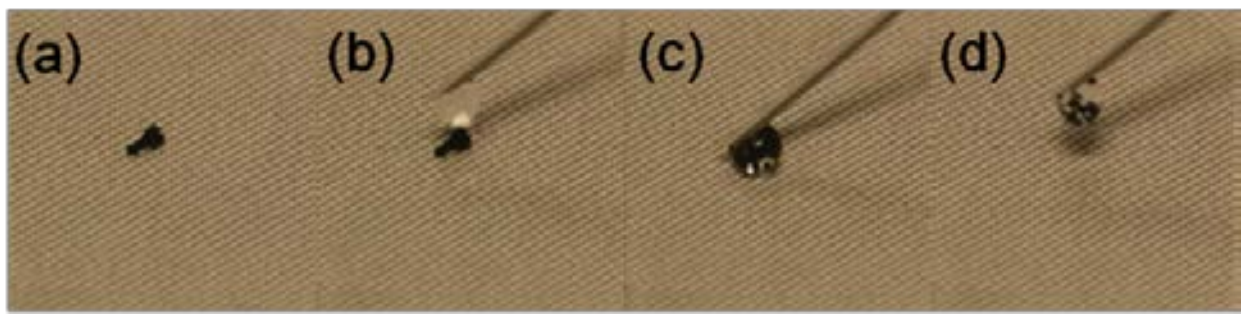

**Figure 70** (a-d) Sequential images showing the removal of the CNT staining from GT-fabric surface by a single water droplet.

## 6.4 Oil-water separation

As a result of the full-spectrum tunable wettability and microporosity, rapid oil-water separation can also be achieved by using GT-fabric as the filtration membrane. The issue of oil-water separation is widely encountered in many different engineering fields, ranging from environmental protection, water treatment, cooling system, to food industry.[112-115] The development of high performance membrane for oil-water separation is highly desired in the industries. Considering the scalability of the fabrication process and cost-effectiveness of the membrane material, GT-fabric is well-suited for meeting the requirement of rapid oil-water separation, owning to its extremely tunable wettability and high porosity with nano-pore distribution.

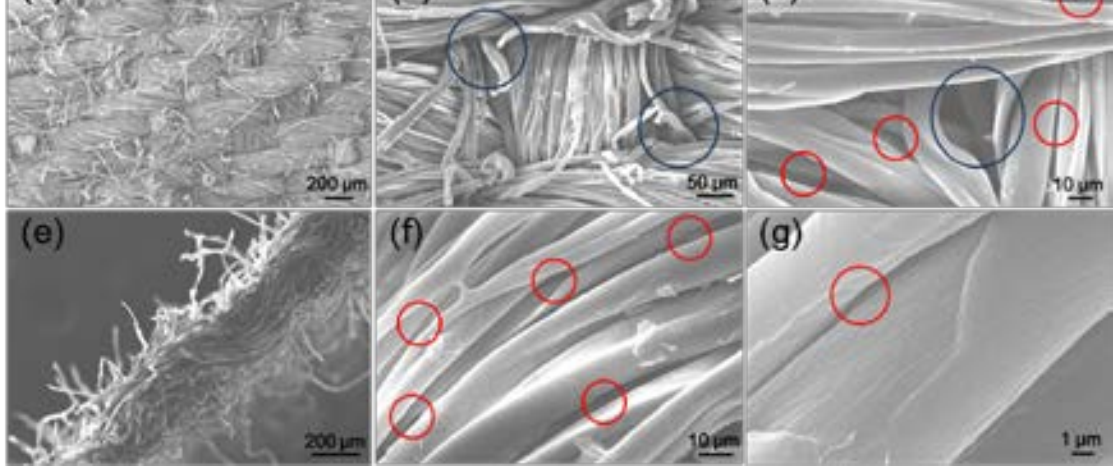

**Figure 71** SEM images showing the pore distribution on GT-fabric surface from top-view (a, b and c), and the pore distribution in the cross-sectional area of GT-fabric (e, f and g) with the blue circles and red circles to indicate the inter-yarn pores and the inter-fiber pores, respectively.

The pore structure of GT-fabric was characterized by SEM and the results were shown in Figure 71. From the top view of GT-fabric (Figure 71a), the closely woven structure consisting of inter-winded yarns and fibers can be clearly observed. Magnification view of the fabric surface indicated that majorly the inter-yarn pores (blue circle) and inter-fiber pores (red circle) were existed on the fabric (Figure 71b and c). The inter-fiber pores on the fabric can be observed more comprehensively from the fabric cross-sectional view (Figure 71e), as micro-channels are found to exist among the adjacent fibers (Figure 71f and g). By counting several tens of the pores, the size distribution was found to be between 21-56 μm for the inter-yarn pores and between 0.2-6 μm for the inter-fiber pores. The water-chloroform mixture was used to demonstrate the oil-water separation performance of GT-fabric. Due to the superhydrophobicity of the fabric, the water phase will be prevented from penetrating while the chloroform is free to flow through. The high porosity of GT-fabric thus resulted in a high separation speed featured by a flow rate of 3162 ± 128 L $m^{-2}$ $h^{-1}$. Sequential frames showing the separation of a 300 mL water-chloroform mixture (v/v=1:1) are presented in Figure 72. It can be observed that the separation process is completed within 30 seconds, indicating the high effectiveness of GT-fabric in separating the oil-water mixture. The separation efficiency was calculated to be 99.2% by carefully measuring the weight change of chloroform before and after the separation process. Besides high speed and high efficiency, the GT-fabric is also suitable for the separation of different types of oil-water mixtures. The mixtures of water/hexane, water/dodecane, and water/hexadecane were also successfully separated by the GT-fabric (Figure 73). For the oil-water mixture with an oil phase density lower than water, further improvement on the separation speed can be implemented by tuning the surface wettability of GT-fabric from superhydrophobic to superhydrophilic. In this case, since the water phase is heavier than the oil phase, it would penetrate through the superhydrophilic GT fabric while the oil is impeded, resulted in a fast and effective gravity-driven separation process.

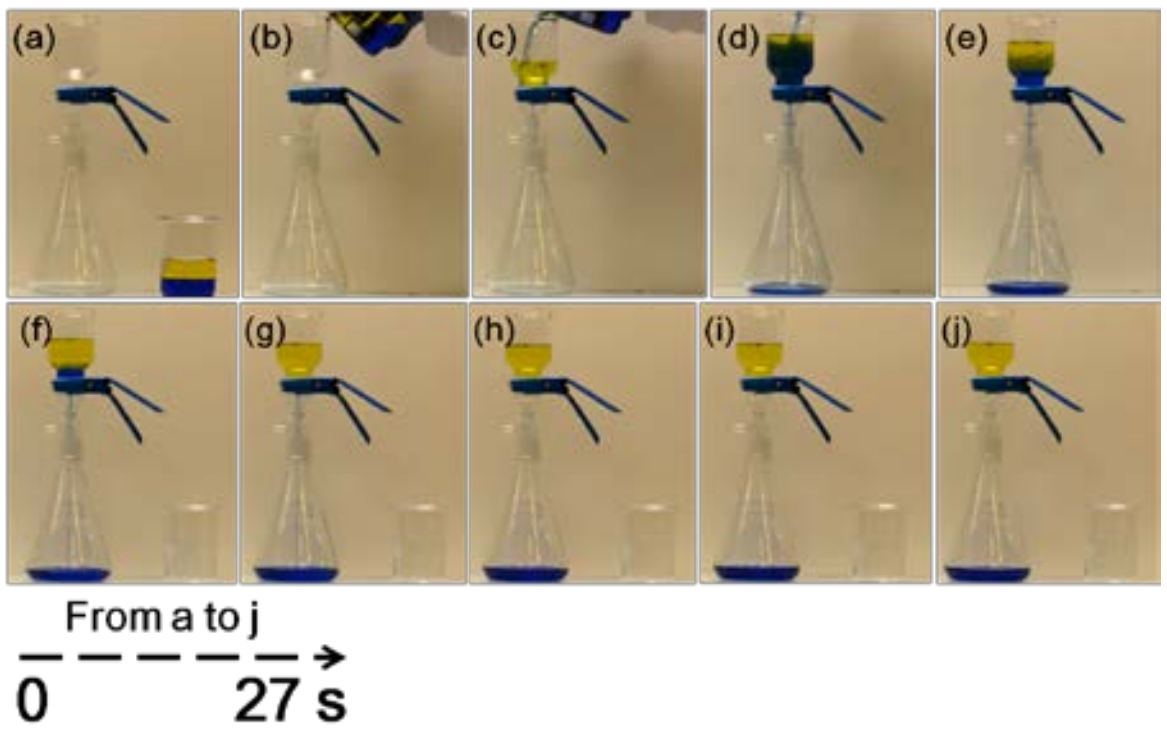

**Figure 72** (a-j) Sequential frames showing the separation process of a water-chloroform mixture by GT-fabric.

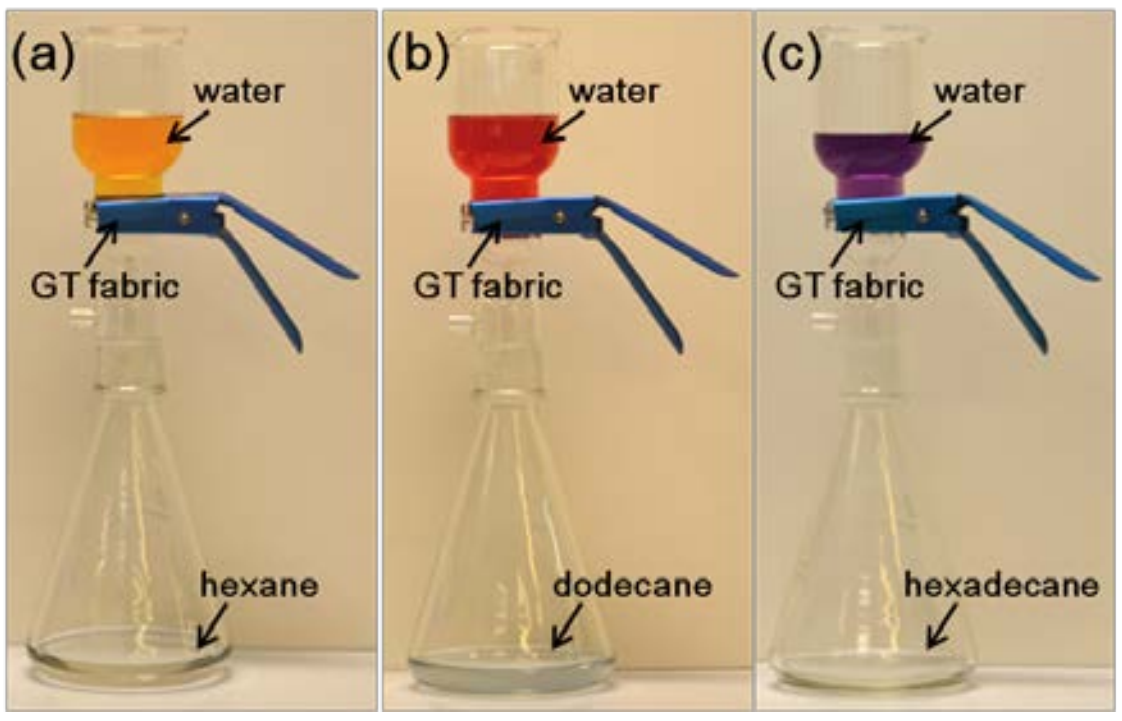

**Figure 73** (a-c) The successful separation of different oil-water mixtures by GT-fabric, i.e. water/hexane (a), water/dodecane (b), and water hexadecane (c).

The high performance in oil-water separation demonstrated by GT-fabric inevitably reveals the advanced multifunctionality of GT-fabric originated from the smart graphene/$TiO_2$ nanocoating. It thus envisages that by using the concept of function-led design in the engineering of smart SHS, not only the desired functions can be achieved through careful selection of materials and structures, but also value-added new functions can be generated and extended beyond the existing functions. Nonetheless, the true power of function-led design may lie in the realization of high-end integrated functions beyond the scope of design, which is intriguing and worthy of thinking. Since the stimuli-responsive and multivariant advanced functions of GT-fabric have been studied comprehensively, further investigations will be focused on the reasons for the smart responsiveness showed by the GT-fabric, which emphasize the interfacial interactions on the $TiO_2$ surface, and at the interface between graphene and $TiO_2$.

# Chapter 7. Mechanism study on $TiO_2$ superhydrophilic conversion and graphene/titania interfacial interaction

## 7.1 Introduction

The superhydrophilic conversion of $TiO_2$ surface under UV radiation is worthy of investigation since it acts as the very basis for the full-spectrum wettability tuning on GT-fabric. A more in-depth understanding on the conversion mechanism can be used to create new engineering methods to manipulate the conversion process, resulted in increased conversion speed or controllable degree of conversion. A novel research approach is used for investigating the superhydrophilic conversion mechanism on $TiO_2$ surface. The physical structure changes of the $TiO_2$ under UV are characterized from the macroscopic level (thin film) to the microscopic level ($TiO_2$ nanocrystal), and a new mechanism is proposed by interpreting the internally linked experimental evidences. On the other hand, the role of graphene in the superhydrophilic conversion on the GT surface is also evaluated. As a result, enhanced conversion rate is obtained in the presence of graphene, which can be attributed to the accelerated electron-hole pair separation on $TiO_2$ through the interfacial charge transfer interaction between the $TiO_2$ and graphene.

## 7.2 Macroscopical surface change of titania during UV radiation

The macroscopic surface change of titania was observed on a thin film casted from the as-synthesized $TiO_2$ sol-gel. The $TiO_2$ sol-gel was diluted by D.I. water and stirred to achieved homogeneous. Afterwards, approximately 0.05 mL of the $TiO_2$ sol-gel dispersion was casted onto a pre-cleaned silicon wafer. After drying at ambient, a visually smooth pale $TiO_2$ thin film was formed on the wafer surface, which was used for further characterization. In order to measure the macroscopic change on the $TiO_2$ surface upon UV, the thin film was exposed to 3 hours' UV radiation to become superhydrophilic and its surface morphology change was subsequently characterized by SEM. The as-obtained SEM images are shown in Figure 74. According to the SEM images, the flat thin film surface is divided into many fragments by the visible cracks, which then mark the boundaries between the fragments (Figure 74a). These irregular cracks are possibly induced by the internal stress of the thin film during the drying process. However, larger gaps can be observed between the fragments in the pristine thin film, according to the magnified SEM images (Figure 74b, c). After being radiated by UV for 3 hours, the large, loosely packed fragments on the surface of the $TiO_2$ thin film transformed into smaller, compact fragments (Figure 74e), and the gaps between the fragments were significantly reduced (Figure 74f, g), indicating a curing effect induced by UV radiation. Tiny $TiO_2$ NPs in the fragments can be observed upon further increasing the magnification of the SEM images. However, the $TiO_2$ NPs showed different contrast under the electron beam before and after the UV treatment (Figure 74d, h), indicating the possible micro-structural change, which should be measured in a smaller scale.

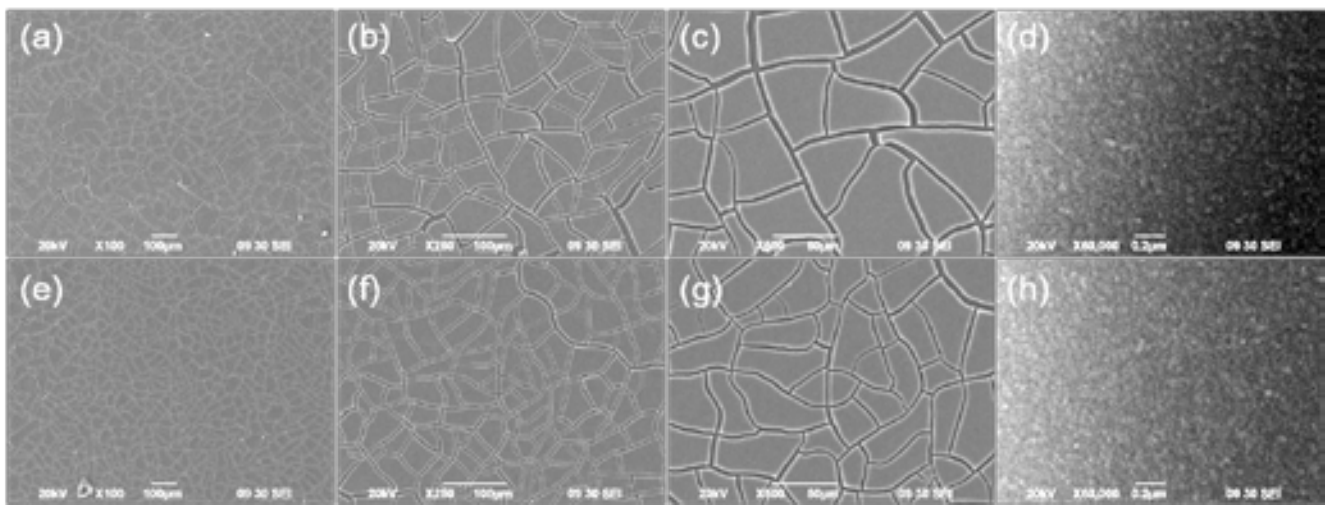

**Figure 74** SEM images showing the surface morphologies of the $TiO_2$ thin film before (a-d) and after (e-h) UV radiation for 3 hours.

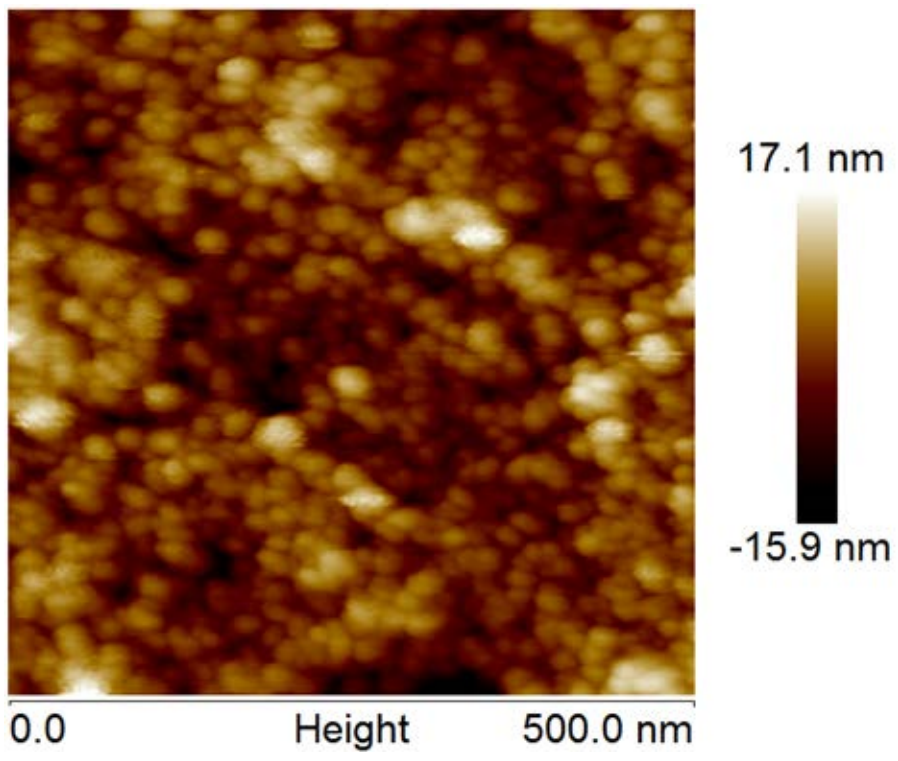

**Figure 75** AFM image of the $TiO_2$ thin film surface. The topographical structure of the $TiO_2$ NPs can be clearly observed at the nanoscale.

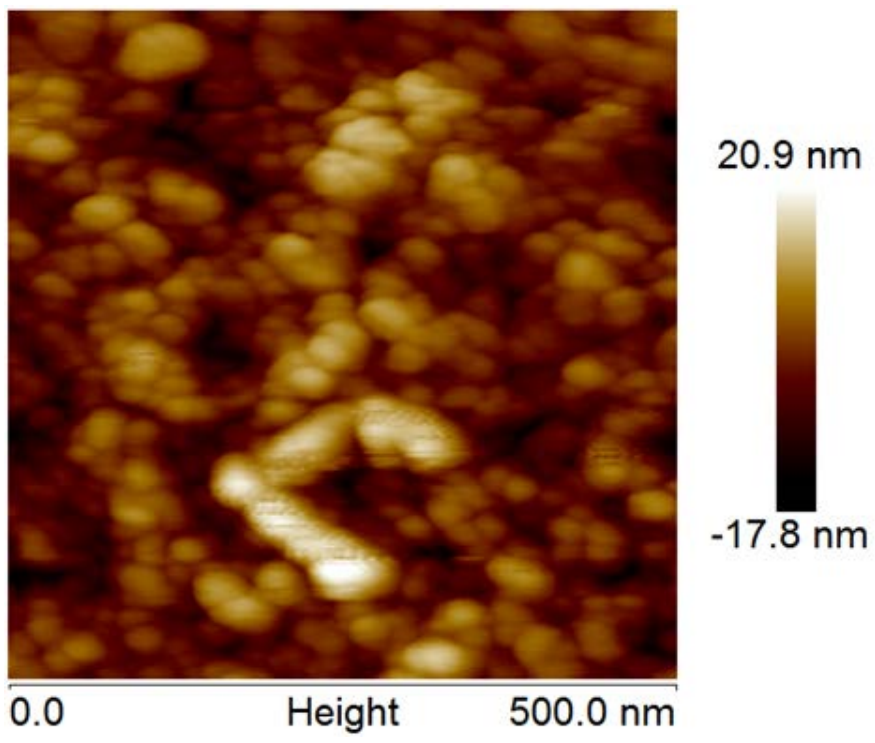

**Figure 76** AFM image of the 3h-UV treated $TiO_2$ thin film. Clear topographical structure changes can be observed as compared to the pristine film. The $TiO_2$

NPs collide with each other to form larger clusters.

Atomic force microscopy (AFM) was subsequently applied to investigate the microstructural change on the $TiO_2$ surface during UV radiation. The sample for AFM was prepared by firstly drop-casting 0.05 mL of the graphene dispersion obtained from the fabric reduction process; and the subsequently drop-casting of the 10-times-diluted titania nanosol (0.05 mL) on top of the graphene layer, which mimicked the hierarchical GT coating on the fabric surface. For the $TiO_2$ surface before UV radiation, it is majorly composed of $TiO_2$ NPs with the sizes between 12 and 39 nm. A roughness factor $R_q$=4.65 nm was also obtained by calculating the surface topography of the $TiO_2$ thin film based on the AFM image (Figure 75). However, a significant change on the NP sizes was observed in the AFM image of 3h-UV treated $TiO_2$ thin film, as the sizes increased to between 24 and 31 nm (Figure 76). The $R_q$ was also increased to 5.34 nm on $TiO_2$ surface after 3h-UV radiation. Nonetheless, the increase in the NP size indicates a homogenization and secondary crystal growth processes induced by UV treatment, by which the NP fusion and collision may occur. This homogenization process in the microscale is also coincident with the curing of the fragment boundaries in the macroscale. The fusion and re-growth of the NPs can be observed more clearly in the topographical phase images, as the pristine high density sharp phase edges collided into flattened, bigger structures after the UV treatment (Figure 77a, b). By re-constructing the AFM images into 3D simulations, a clear fused and flattened surface with expanded structural unit volume can be observed after UV (Figure 78b), as compared to the sharp and extruded surface microtexture before the UV treatment (Figure 78a). However, the fusion and collision of $TiO_2$ NPs during UV radiation should induce by the change of their crystalline structure, accompanied with the atomic rearrangement at the angstrom level ($10^{-10}$ m). To gain a comprehensive roadmap of the superhydrophilic conversion process, it is necessary to further obtain the information in the extremely small scale.

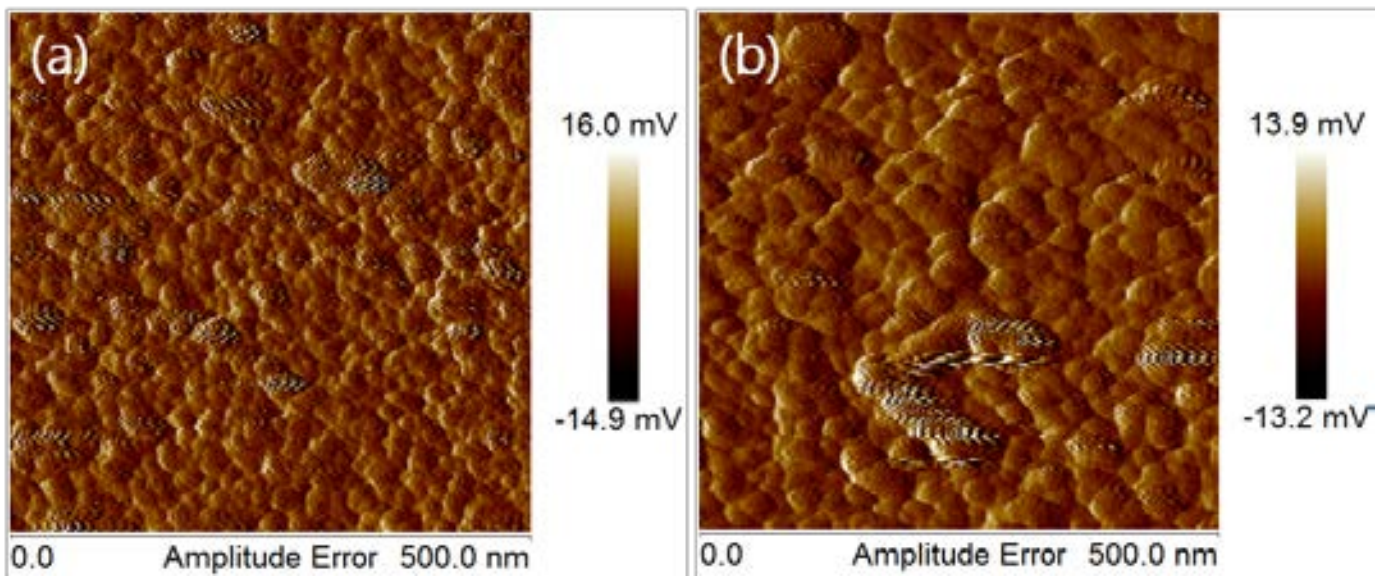

**Figure 77** Phase images of the pristine (a) and the 3h-UV treated (b) $TiO_2$ thin film.

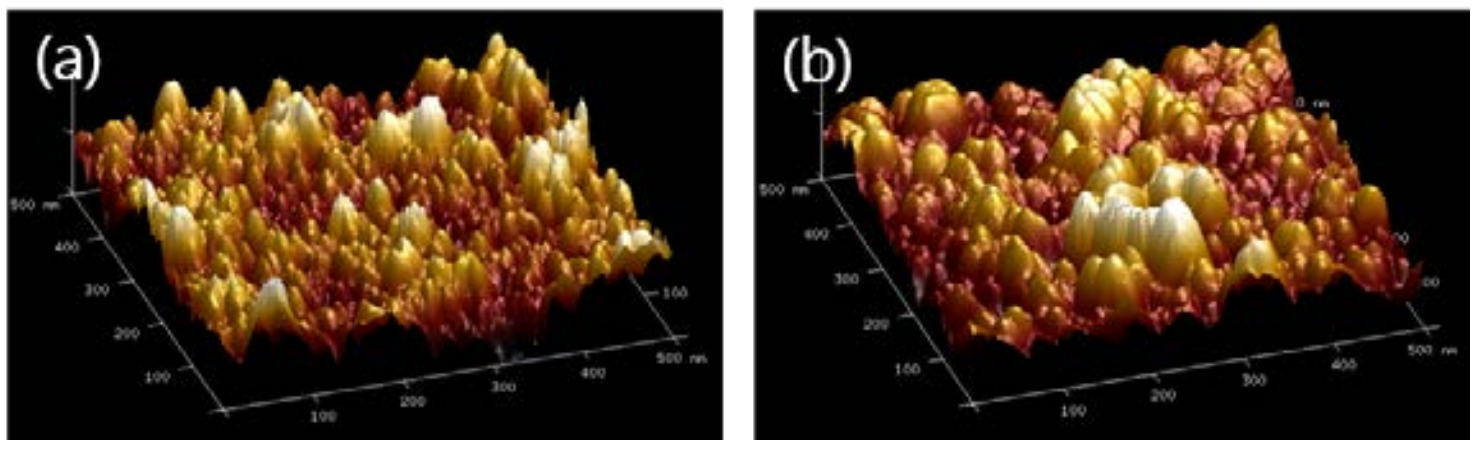

**Figure 78** 3D topographical model of the pristine (a) and 3h-UV treated (b) $TiO_2$ thin film surface. Clear fusion and flattening of the surface microstructures can be observed after UV radiation.

## 7.3 Microscopical surface change of titania during UV radiation

HRTEM was used to characterize the change on $TiO_2$ surface during UV radiation at the atomic scale. The lattice structure of the $TiO_2$ NPs can be clearly observed in the as-obtained HRTEM image (Figure 79). Information regarding the crystalline facet of the NPs was obtained by analyzing the lattice spacing of the nanocrystals in the HRTEM image. For the pristine $TiO_2$ NPs, a uniform lattice spacing of 0.35 nm was measured by counting tens of the nanocrystals, corresponding to the (101) lattice reflection of anatase $TiO_2$ (Figure 80). It thus indicated that the NPs were in the anatase phase with {101} facet exposed on the surface, which was coincident with the previous characterizations (e.g. TEM & XRD). The nanocrystals of $TiO_2$ NPs were also found to orient along the [001] direction, as indicated by the lattice angle of 68.3° between the [001] and [101] direction, and the high intensity of (004) diffraction in the XRD spectrum. However, after being exposed to UV radiation, a drastic change on the lattice structure of $TiO_2$ nanocrystals was observed (Figure 81). By carefully analyzing the crystalline structure in the HRTEM image, a lattice spacing of 0.23 nm can be observed for the nanocrystals of $TiO_2$ NPs, corresponding to the (001) lattice reflection of anatase $TiO_2$ (Figure 82). It thus revealed a substantial lattice structure transformation from {101} to {001} on $TiO_2$ surface during UV radiation. Nonetheless, this atomic scale rearrangement may eventually result in the macroscopic changes as-observed, including the fusion and flattening of the surface microtexture, and the curing of the cracks on the $TiO_2$ thin film.

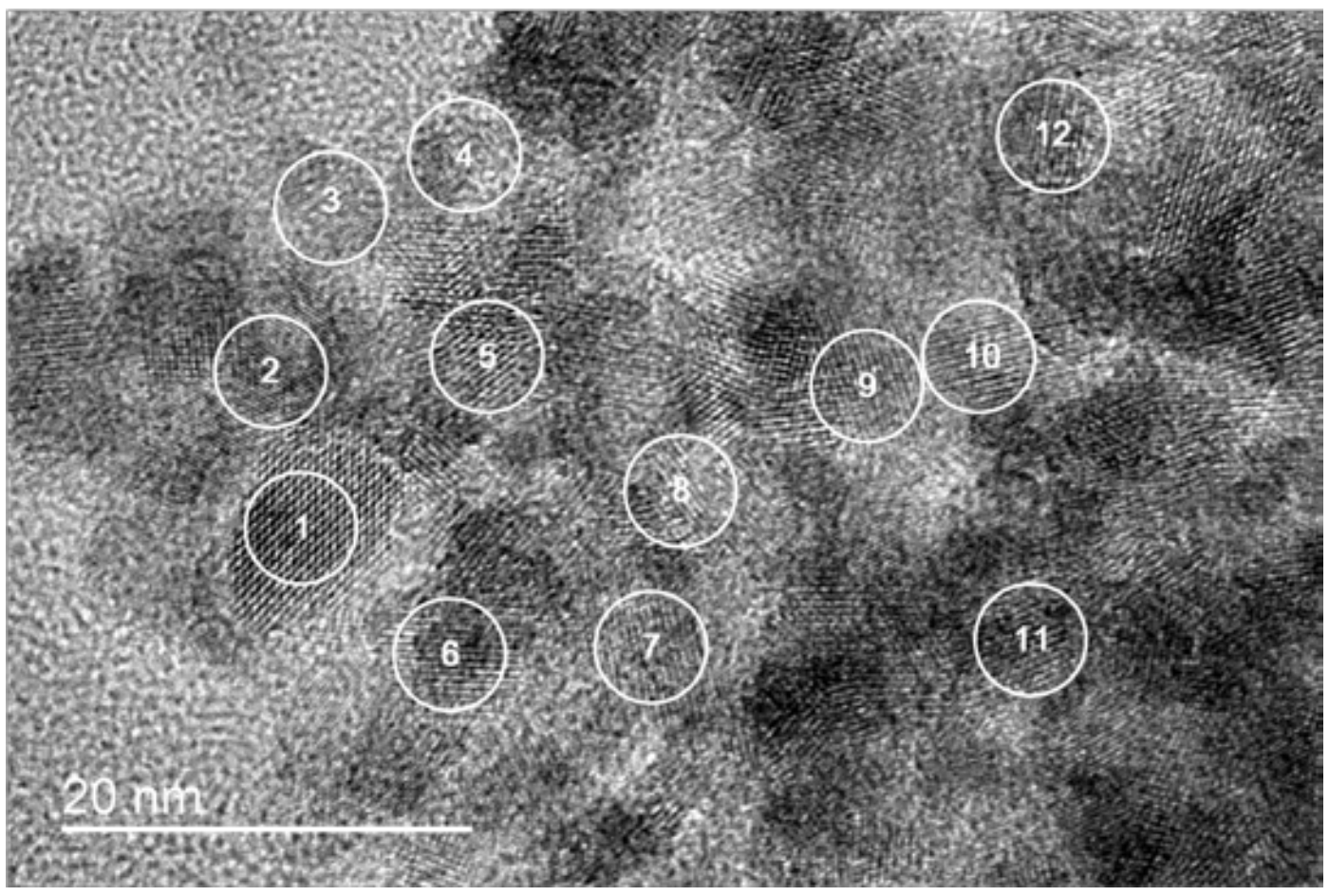

**Figure 79** HRTEM image of the $TiO_2$ NPs in the $TiO_2$ thin film, the lattice structure of the individual NP can be clearly observed. The numbers indicate the NPs used for the counting purpose.

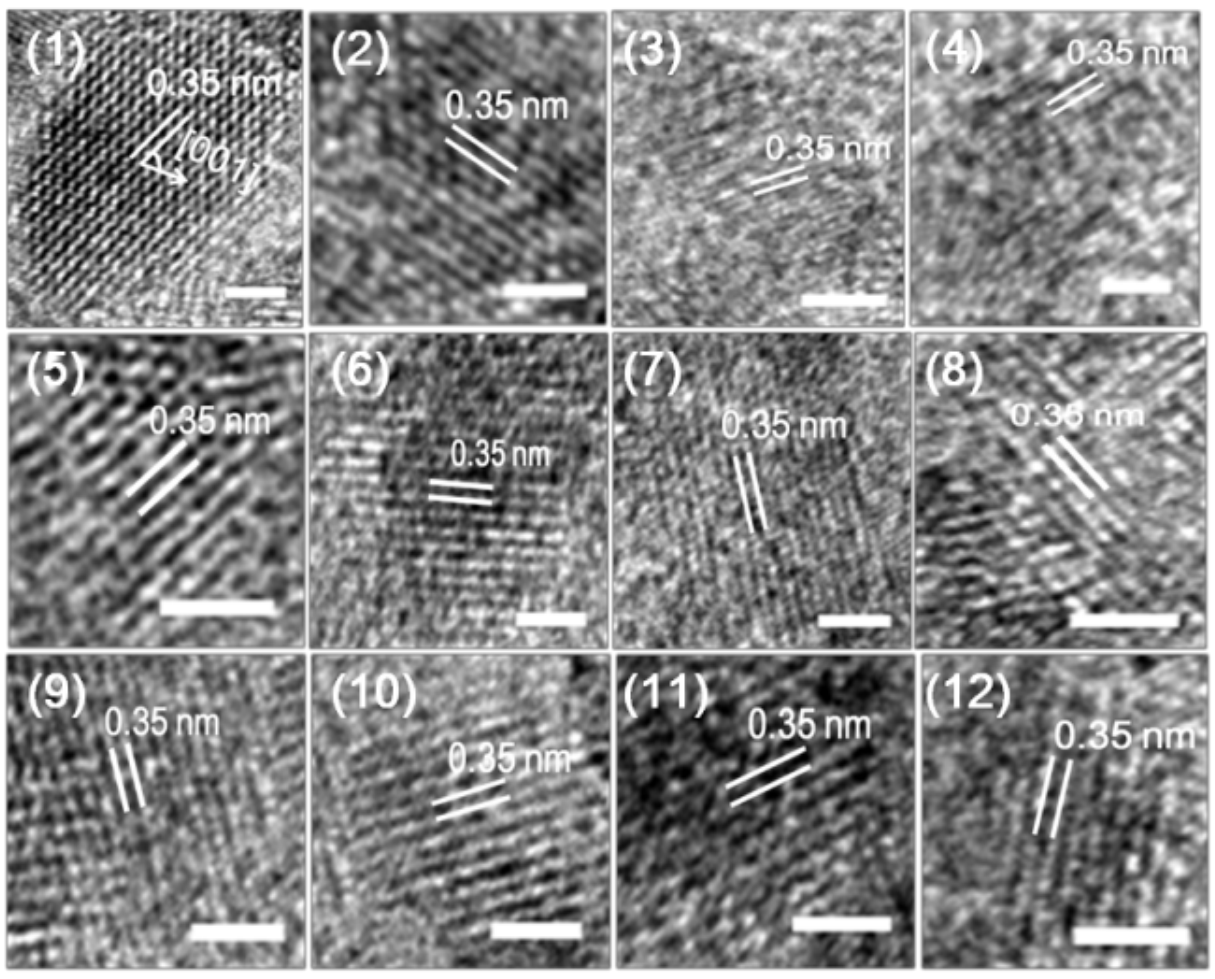

**Figure 80** Magnified HRTEM images of the $TiO_2$ NPs with the lattice spacing indexed. The numbers correspond to the $TiO_2$ NPs counted in Figure 79.

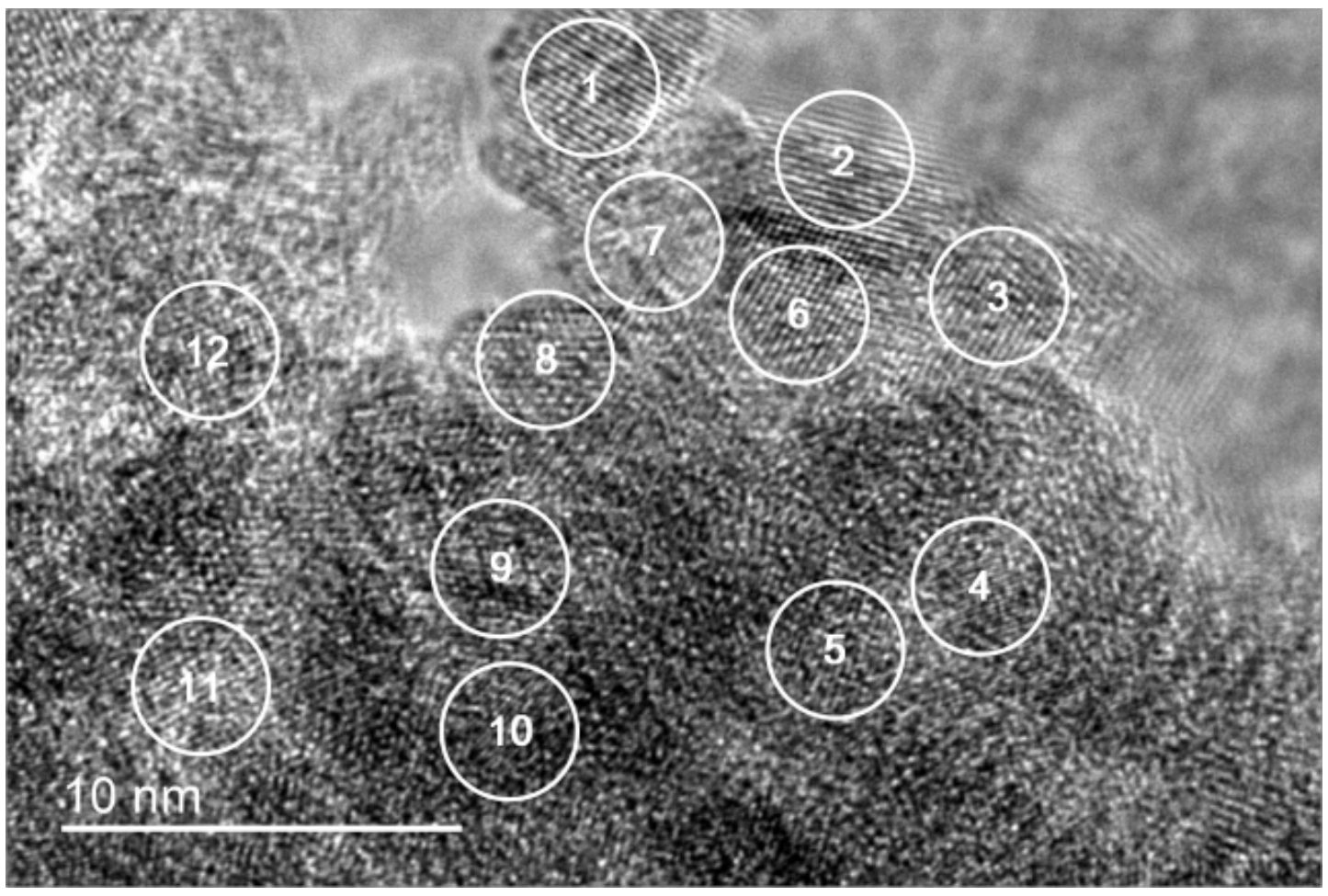

**Figure 81** HRTEM images showing the $TiO_2$ NPs in the 3h-UV treated $TiO_2$ thin film. The lattice structure of the NPs is also clearly observed and the NPs used for the counting purpose are indexed by the numbers.

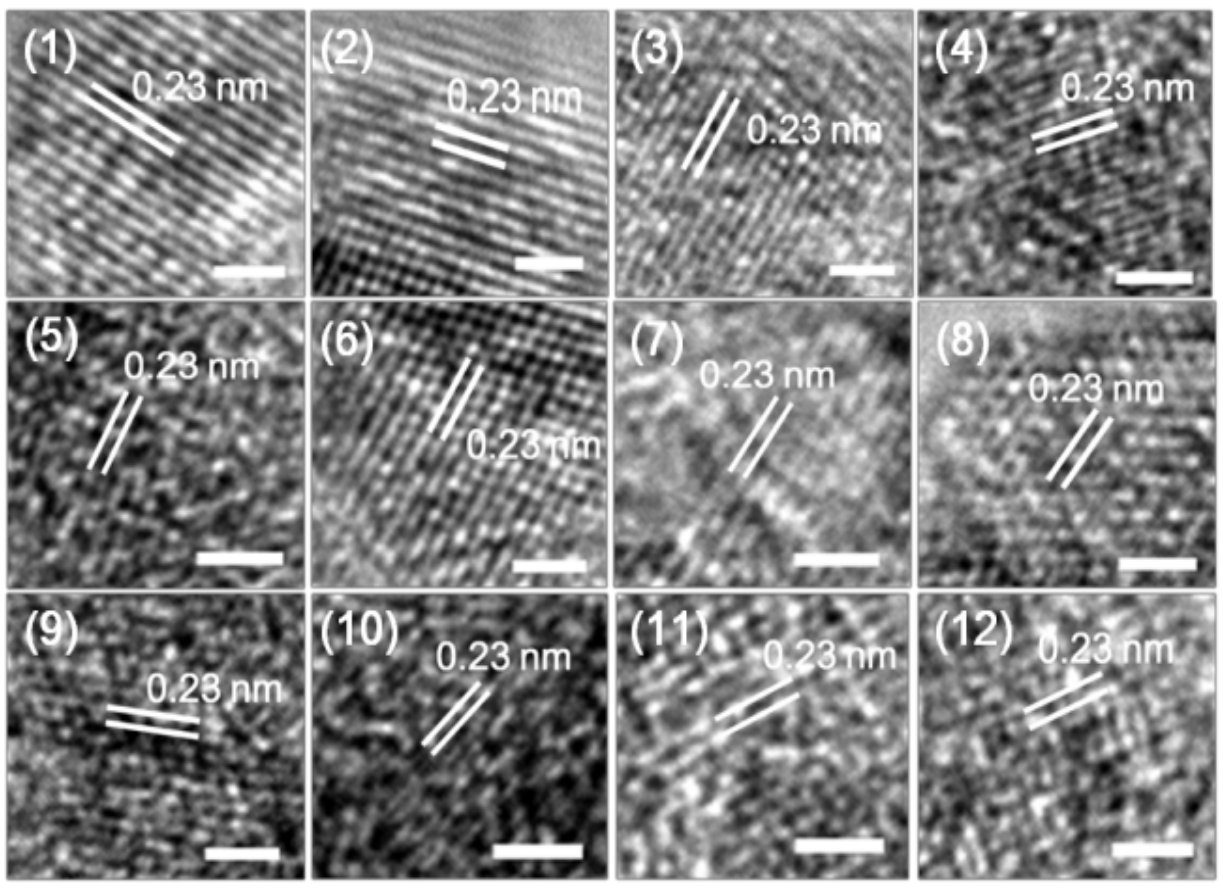

**Figure 82** Magnified TEM images showing the lattice spacing of the $TiO_2$ NPs. The numbers in the image correspond to the $TiO_2$ NP counted in Figure 81.

Based on these discoveries, the unseen process of the superhydrophilic conversion on $TiO_2$ can be even traced down to the sub-atomic level. As the $TiO_2$ nanocrystal was radiated by UV, extensive electron-hole recombination happened on {101} facet, which resulted in substantial heat release to induce the partial melting of the nanocrystal.[116] This particle melting may subsequently trigger the atomic rearrangement on the nanocrystal through a solid-liquid process, which generates more thermodynamically stable {001} facet under this circumstance. It was speculated that a heterojunction between {101} and {001} facets was formed on $TiO_2$ nanocrystals during UV radiation. The presence of this heterojunction significantly enhanced the electron-hole pair separation on the $TiO_2$ surface. As the electron-hole pair was separated on the heterojunction, the electrons were moved to {101} facet and the holes were moved to {001} facet. As a result, the {001} facet may have higher oxidation potential while higher reduction potential is expected on {101} facet. A schematic illustration of the transformation process is shown in Figure 83. The holes on {001} facet can further break the covalent bonds between titanium atom and oxygen atom, resulting in surface active sites such as $Ti^{3+}$ and oxygen vacancy. The surface active sites further react with adsorbed water molecules to generate new hydroxyl groups on the $TiO_2$ surface, which significantly enhance the surface hydrophilicity. The generation of hydroxyl groups on $TiO_2$ surface during UV radiation was confirmed by X-ray photoelectron spectroscopy (XPS). The XPS spectra were taken on the pristine $TiO_2$ thin film and the thin film exposed to 3h-UV radiation (Figure 84a, b). The abundance of surface hydroxyl groups is indicated by the deconvoluted components of the O1s peak at 530.8 eV. As can be observed from the XPS spectra, the O1s peak is essentially deconvoluted into three sub-components, which correspond to the binding energy of lattice oxygen (O-Ti-O), surface hydroxyl group (Ti-OH), and adsorbed water molecule ($H_2O$). It can be clearly observed that both the intensities of O-Ti-O and $H_2O$ decrease after the UV treatment, while the intensity of Ti-OH increases. It thus indicates the possible reaction between the lattice oxygen and adsorbed $H_2O$, which produces new hydroxyl groups on $TiO_2$ surface.

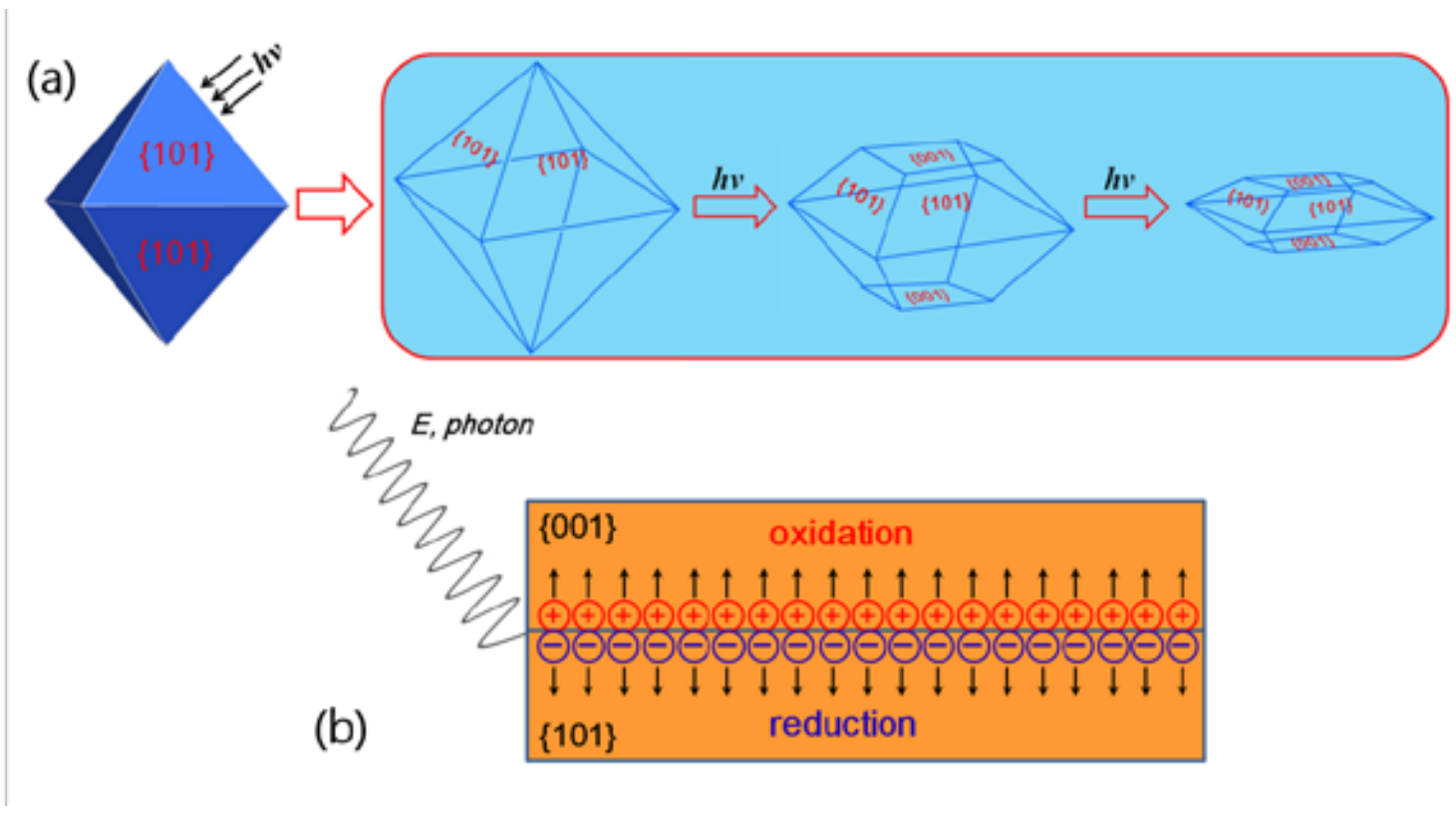

**Figure 83** Schematic illustration of lattice transformation on the anatase $TiO_2$ crystal surface induced by UV radiation. The upper graph (a) illustrates the structural change of the $TiO_2$ crystals during UV and the bottom graph (b) depicts the electron-hole pair separation at the {001}/{101} heterojunction.

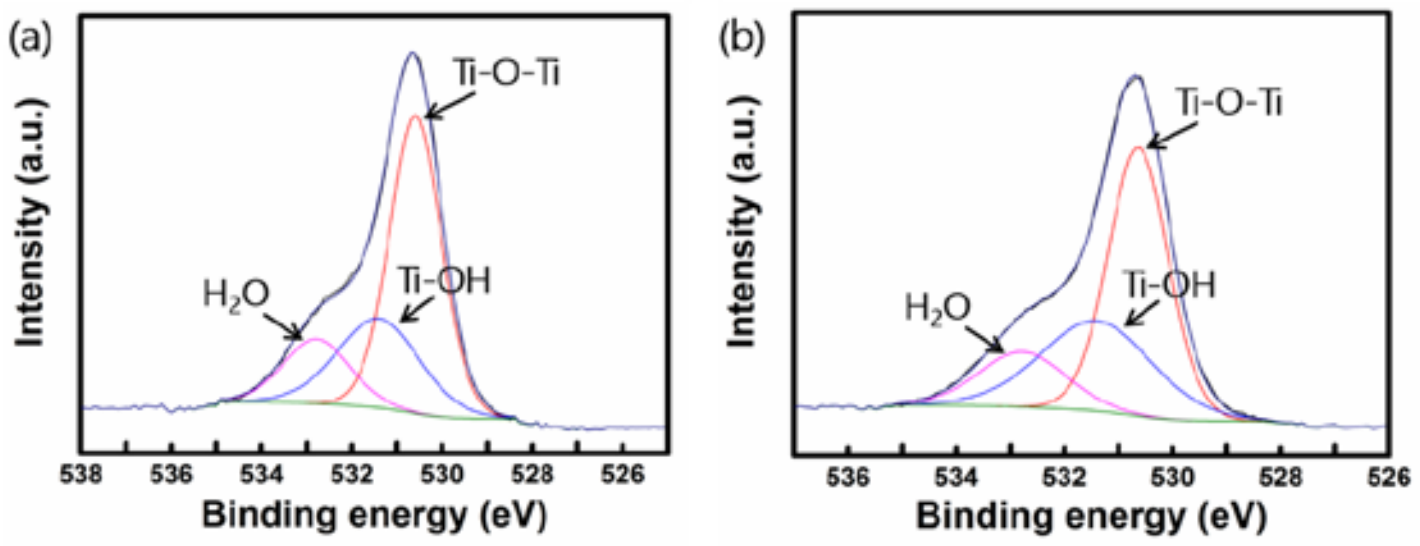

**Figure 84** O1s XPS spectra of the pristine (a) and 3h-UV treated $TiO_2$ thin film (b). The O1s peak is deconvoluted into three sub-components of the lattice oxygen (Ti-O-Ti), surface hydroxyl group (Ti-OH), and adsorbed water ($H_2O$).

## 7.4 Interfacial interaction between graphene and titania

It is envisaged that extensive interfacial interactions may exist between graphene and $TiO_2$ NPs as a result of their extremely small size and excellent electronic property. Extensive research effort has been devoted to the development of high performance graphene/$TiO_2$ composite system with various structures and forms. Based on the results of these research work, new types of interfacial interactions which deliver enhanced material functions have been widely observed.[117-121] In the present study, enhanced charge transfer interactions are believed to exist at the interface between the hierarchically nanostructured graphene and $TiO_2$. As a result of this enhanced charge transfer interaction, both the rate and extent of superhydrophilic conversion on $TiO_2$ NP surface are substantially improved, which constitutes the very basis for the stimuli-responsive full-spectrum wettability tuning on the GT fabric.

To evaluate the effect of graphene in superhydrophilic conversion, three different types of thin films, including hierarchical graphene/titania (GT), sole titania, and sole graphene, were casted on silicon wafer and their changes in water contact angles were measured before and after 3h-UV treatment (Figure 85). The pristine WCAs on the thin films are following the order of graphene > GT > $TiO_2$, which is reasonable since the CA of GT surface locates between graphene and $TiO_2$ (Figure 85a-c). However, the order becomes graphene > $TiO_2$ > GT as the thin films are exposed to UV radiation (Figure 85 d-e). In this regard, the GT surface showed both the highest rate and extent for superhydrophilic conversion, which were superior to the sole $TiO_2$ surface, indicating the substantial enhancement induced by graphene. Besides, a large increase in WCA was also observed on the sole graphene surface. Compared to the reduction of WCA on the GT and $TiO_2$ surface during UV, the increase of WCA on graphene surface can be possibly attributed to the increase in surface roughness, and the hydrophobicity on the graphene surface is further improved.[122] The recovery rate of the thin films under dark storage was also studied (Figure 86). It is not surprised to observe that the GT surface showed the highest recovery rate and the recovery efficiency, as it can fully restore the original WCA in 72 hours' dark storage, regardless of its lowest WCA after UV treatment. However, the sole $TiO_2$ surface showed much slower recovery rate and efficiency, which may require more than 96 hours to nearly restore the original state. Furthermore, the recovery curves of GT and graphene showed similar trends during dark storage, representing both high rates and efficiencies. It thus evidenced the essential role of graphene in accelerating the superhydrophilic conversion process.

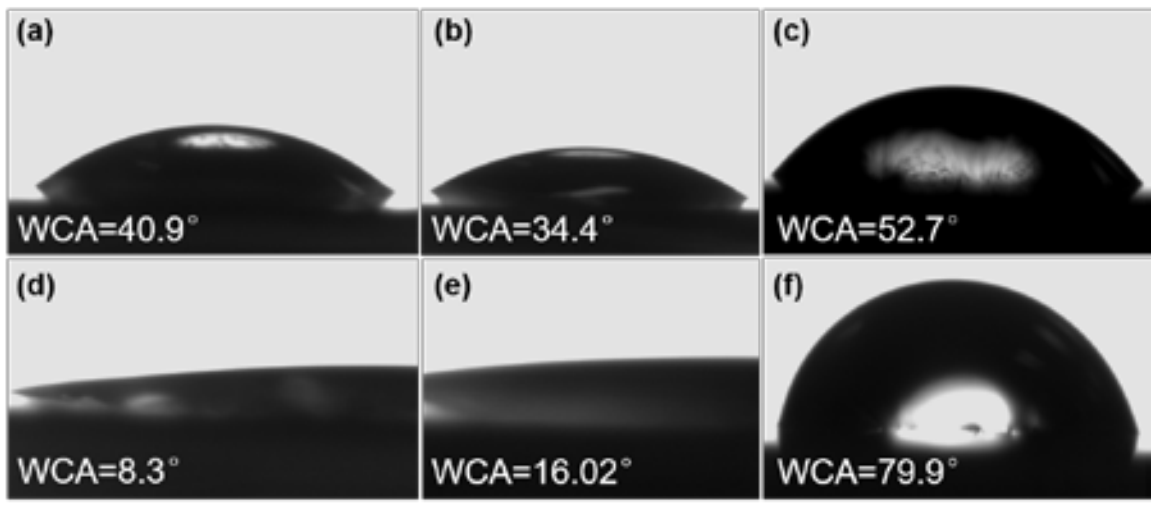

**Figure 85** Photographs showing the contact angle of a 1 μL water droplet on the GT (a, d), $TiO_2$ (b, e) and graphene (c, f) surfaces. The upper images show the water contact angles on the pristine surfaces while the bottom images show the water contact angles after 3 hours' UV radiation.

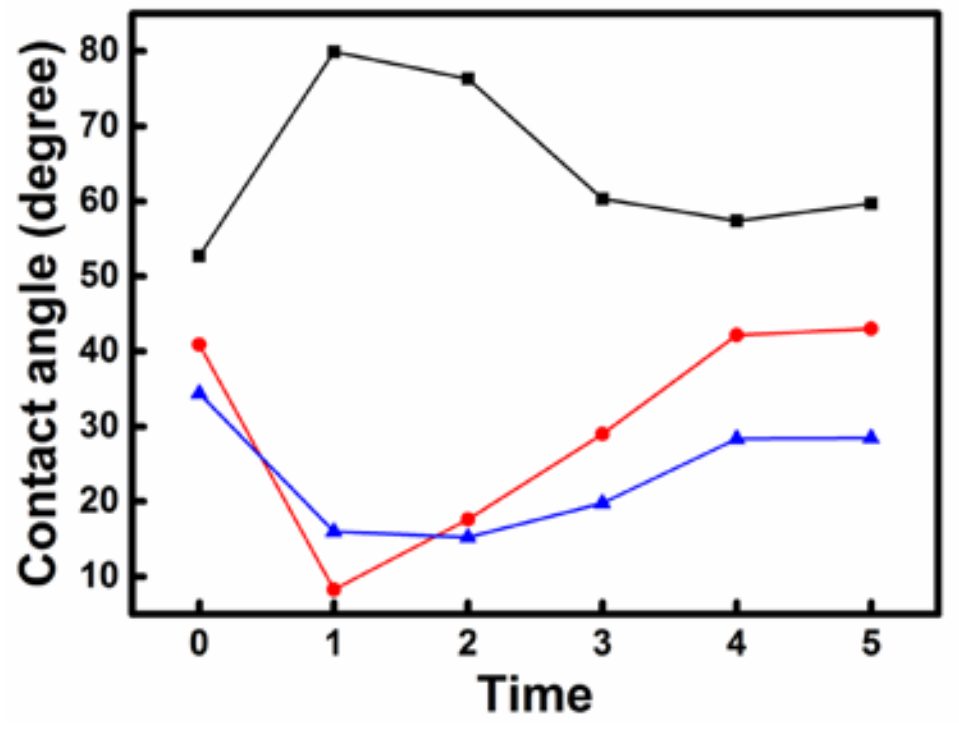

**Figure 86** The contact angle changes on the GT (red circle), $TiO_2$ (blue triangle) and graphene (black square) surfaces as a function of time. The point 0 at the x-axis represents the pristine state while point 1 represents the state after 3h-UV radiation. Point 2, 3, 4, 5 represent the dark storage time of 24, 48, 72, and 96 hours, respectively.

It is speculated that graphene enhances the superhydrophilic conversion on $TiO_2$ surface through interfacial charge transfer interaction. As a result of the low reduction potential, graphene can effectively function as an electron acceptor in the hierarchical GT system. In this case, the efficiency of electron-hole pair separation on the $TiO_2$ surface during UV will be considerably improved, as the generated electron is transferred to graphene, the equilibrium of the separation reaction will be shifted to the right-hand side, which favors a higher separation rate. The enhanced electron-hole pair separation on $TiO_2$ surface thus accelerates the superhydrophilic conversion process via faster lattice transformation and surface structure re-configuration. The enhanced lattice transformation is also indicated by the changes in XPS spectrum of GT surface when UV is applied (Figure 87). As can be seen from the O1s XPS spectra of GT surface, a higher intensity of Ti-OH can be obtained after the UV treatment (Figure 87b), as compared to the pristine GT surface (Figure 87a). This phenomenon thus reveals a higher rate of formation for the surface hydroxyl groups on GT surface, which is associated with a faster lattice transformation process.

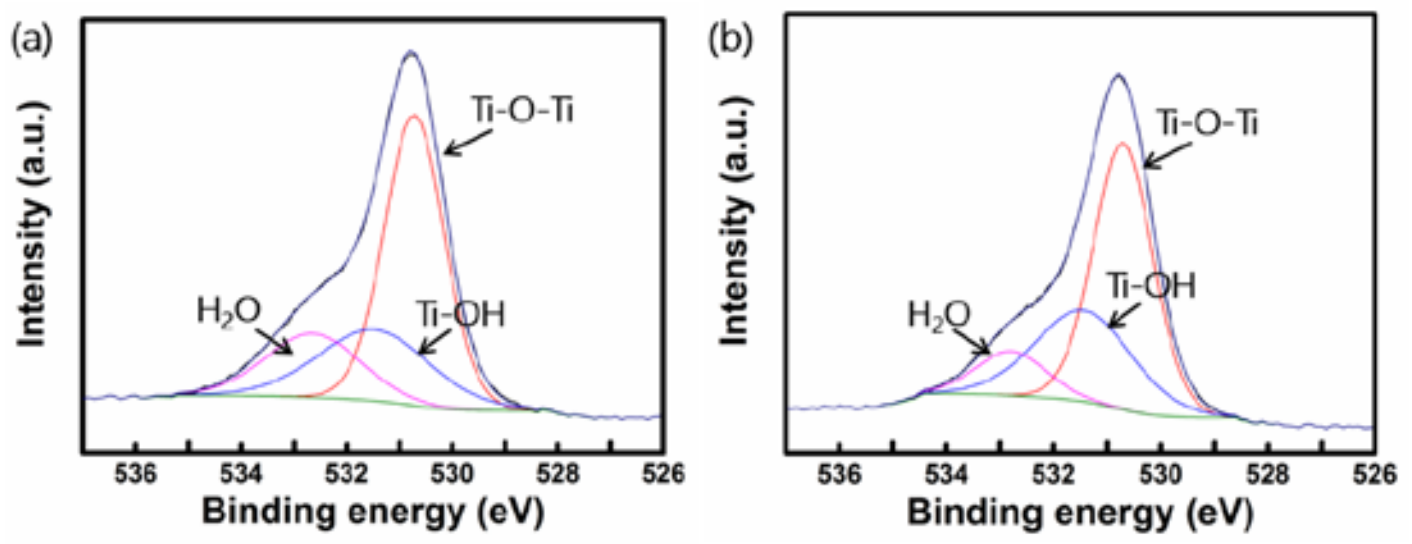

**Figure 87** O1s XPS spectra of the pristine (a) and 3h-UV treated (b) GT surfaces. The O1s peak is deconvoluted into three components of lattice oxygen (Ti-O-Ti), surface hydroxyl group (Ti-OH) and adsorbed water ($H_2O$).

To make a summary, the underlying processes of superhydrophilic conversion on $TiO_2$ surface were investigated comprehensively both in the macroscopic and microscopic levels. A new mechanism regarding the superhydrophilic conversion is proposed, which highlights the micro to macro structural changes on $TiO_2$ surface during UV radiation. It is speculated that the interfacial charge transfer interaction between graphene and $TiO_2$ enhances the electron-hole pair separation on the heterojunction between {101} and {001}, which subsequently accelerates the lattice transformation of the {101} facet into {001} facet. Hydrophilic surface hydroxyl groups are generated during the transformation and they also induce structural reconfiguration at the larger scale. The corresponding fusion and flattening are observed on the surface microstructures of $TiO_2$ thin film, and a micro-crack healing process is also observed, resulting in a roughened, superhydrophilic $TiO_2$ surface with zero water contact angle.

# Chapter 8. Conclusion and future work

In conclusion, novel graphene coated cotton fabric (G-fabric) has been synthesized using a simple dip-coating method. The hydrophilic cotton fabric turned hydrophobic with extremely low graphene loading (less than 0.5 wt%), and tunable wettability can be obtained by the absorption of small molecules, i.e. water and acetone, onto the G-fabric surface. To further improve the precision and speed of the wettability tuning on the G-fabric surface, the concept of function-led design is utilized to direct both the material and structural engineerings that necessary to achieve the targeted functions. The as-developed hierarchical graphene/titania nanofilm coating with dual roughness on cotton fabric (GT-fabric) showed unique full-spectrum tuning of wettability with precise control over the water/substrate interactions such as transport, adhesion and spreading. Based on the full-spectrum tuning of wettability on GT-fabric, three innovative functional aspects can be realized, including:

(i) Spontaneous directional water transport in the axial direction of the fabric, which can be subsequently utilized for the micro-environment control, such as humidity and moisture management. By changing the duration and intensity of the UV radiation, the water transport speed can be facilely tuned from low to high on GT-fabric, which indicates its application potential in control release and microfluidic device.

(ii) Multivariant microfluidic manipulation can be achieved on GT-fabric

surface by tuning the water adhesion. By carefully selecting the size of droplet, micro-droplet arrays demonstrating high adhesion on GT-fabric can be obtained. The droplet arrays are stable regardless the configuration of GT-fabric. Dynamic manipulative motions, such as droplet transfer and droplet sensing can also be realized, which demonstrate the potential for no-loss droplet transfer and high chromatic gas sensing.

(iii) Excellent oil-water separation performance with both high speed and separation efficiency can also be achieved on GT-fabric. As a result of the superhydrophobicity and high porosity of GT-fabric, the oil phase can pass through it with high flow rate and low pressure requirement, which can be purely gravity-driven. In the case that the oil phase has a lower density than water, the GT-fabric can be switched to the superhydrophilic state, which grants the water passage and impedes the oil penetration.

The novel function-led design concept not only realizes the full-spectrum wettability tuning on cotton fabric but also generates multiple new functions for various innovative applications. It thus envisages that the function-led criterion may bring about the unprecedented functional aspects beyond the scope of design, as a result of the extensive interfacial interactions between the integrated materials and structures. However, new theories and mechanisms can also be generated from the function-led design process, from which the normal phenomenon can be viewed in a different and intriguing angle.

The superhydrophilic conversion on titania surface is also studied by using GT-fabric as the model system. The stable superhydrophobicity and high conversion speed on the GT-fabric surface are found to correlate with the interfacial charge transfer interaction between graphene and $TiO_2$ NPs. The charge transfer interaction thus accelerates the electron-hole pair separation on $TiO_2$ surface during UV radiation and subsequently triggers a unique lattice transformation from {101} to {001} on $TiO_2$. As being viewed from microscopically to macroscopically, the lattice transformation is correlated to the microstructural fusion and flattening, which subsequently induce the micro-crack healing process on the $TiO_2$ thin film surface, and generate a higher surface roughness. The generation of hydrophilic hydroxyl groups on {001} facets during lattice transformation and the enhanced surface roughness both increase the hydrophilicity of the $TiO_2$ NPs at the microscale, which are eventually magnified and turned into the as-observed superhydrophilicity at the macroscale.

On the other hand, the function-led design directs the specific fabrication process of the SHS with desired functions, from the steps of material selection, structural configuration, to function realization. However, since the function-led design is a typical reverse-engineering process and its ultimate target is to achieve the desired functions, the innovative functions may serve as the initiatives for the function-led design. In the field of SHS, the state-of-the-art developmental trends in the functional aspects may include:

(i) The fabrication of SHS with extremely super-repellent property, or omniphobicity. Compared to the water-repellent SHS, the super-repellent SHS can effectively de-wet from both water and low surface energy oil, even fluorocarbon. It can also remain non-wetting in the extreme application environment, such as in high pressure, high temperature, and high humidity conditions.

(ii) The fabrication of SHS with stimuli-responsive, tunable wettability. Tunable wettability is essential for the SHS to switch their functions under different application conditions, which enables smart features on SHS such as directional spreading, tunable adhesion, and no-loss transport. The SHS with programmable wettability is also essential to achieve specific high-end applications, which require the integration and cooperation between the hydrophilic and hydrophobic regions, such as mist collection.

(iii) The fabrication of robust SHS under dynamic interfacial interaction. As the real-field applications require robust SHS which is stable and durable in the conditions provided, new types of SHS which not only obtain the superhydrophobicity under static water contact, but also possess the preferred liquid-solid interactions under dynamic interactions are urgently needed. These innovative types of SHS may possess new functions such as accelerated droplet bouncing, droplet trampolining in reduced pressure, droplet mobility during vaporization, self-healing and highly resistant to the mechanical deformation.